\documentclass[11pt]{article}

\usepackage[dvipsnames,table]{xcolor} 
\usepackage{comment,soul}
\usepackage[hidelinks]{hyperref}
\usepackage{enumitem}
\usepackage{amsmath,amssymb,epsf,cite,graphicx,subfigure}
\usepackage{amsmath,braket,tikzsymbols}
\usepackage[bbgreekl]{mathbbol}
\usepackage{comment}
\usepackage{mathtools}
\usepackage{cancel}

\usepackage{blindtext}

\numberwithin{equation}{section}
\newcommand{\nn}{\nonumber}

\newcommand{\beq}{\begin{equation}}
\newcommand{\eeq}{\end{equation}}

\newcommand{\tr}{\text{tr}}

\newcommand{\cO}{\mathcal{O}}

\newcommand{\rmT}{\mathrm{T}}

\newcommand{\mc}[1]{\multicolumn{1}{|c}{#1}}

\usepackage[mathscr]{euscript}

\def\lie{\pounds}
\def\dow{\partial}

\def\df{\mathrm{d}}
\def\SO{\mathrm{SO}}
\def\scX{\mathscr{X}}
\def\scB{\mathscr{B}}
\def\scK{\mathscr{K}}
\def\lb{\left(}
\def\rb{\right)}
\def\half{\frac12}

\def\eqb{\text{eqb}}
\def\ideal{\text{ideal}}
\def\hs{\text{hs}}
\def\nhs{\text{nhs}}

\def\eff{\text{eff}}

\let\Im\relax

\DeclareMathOperator{\Im}{Im}

\usepackage{cleveref}

\crefname{equation}{eq.}{eqs.}
\crefname{section}{section}{sections}
\crefname{result}{result}{results}
\crefname{appendix}{appendix}{appendices}
\crefname{figure}{figure}{figures}
\crefname{table}{table}{tables}

\usepackage{hhline}
\usepackage{accents}
\newcommand{\ut}[1]{\underaccent{\tilde}{#1}}

\def\tmu{\tilde\mu}
\def\tvarpi{\tilde\varpi}

\newcommand\gray[1]{{\color{Gray}#1}}

\def\fD{\mathfrak D}

\begin{document}
\def\arraystretch{1.3}
\baselineskip=15.5pt

\begin{center}
{\LARGE \bf Dipole superfluid hydrodynamics II}
\vskip 1cm

\textbf{
Akash Jain$^{1,2,a}$, 
Kristan Jensen$^{3,b}$, Ruochuan Liu$^{3,c}$ and 
Eric Mefford$^{3,d}$}

\vspace{0.5cm}

{\small ${}^1$Institute for Theoretical Physics, University of Amsterdam, 1090
  GL Amsterdam, The Netherlands \vspace{.3cm}}

{\small ${}^2$Dutch Institute for Emergent Phenomena, 1090 GL Amsterdam, The Netherlands \vspace{.3cm}}

{\small ${}^3$Department of Physics and Astronomy, University of Victoria, Victoria, BC V8W 3P6, Canada\\}

\vspace{0.5cm}

{\tt \small ${}^a$a.jain@uva.nl,}
{\tt  \small ${}^b$kristanj@uvic.ca,}
{\tt \small ${}^c$liur@uvic.ca,}
{\tt \small ${}^d$ericmefford@uvic.ca\\}

\medskip

\end{center}
\thispagestyle{empty}

\vskip1cm

\begin{center} 
{\bf Abstract}
\end{center}
We present a dissipative hydrodynamic theory of ``s-wave dipole superfluids'' that arise in phases of translation-invariant and dipole-symmetric models in which the U(1) symmetry is spontaneously broken. The hydrodynamic description is subtle on account of an analogue of dangerously irrelevant operators, which requires us to formalize an entirely new derivative counting scheme suitable for these fluids. We use our hydrodynamic model to investigate the linearized response of such a fluid, characterized by sound modes $\omega \sim \pm k - ik^2$, shear modes $\omega\sim-ik^2$, and magnon-like propagating modes $\omega \sim \pm k^2 - ik^4$ that are the dipole-invariant version of superfluid ``second sound'' modes. We find that these fluids can also admit equilibrium states with ``dipole superflow'' that resemble a polarized medium. Finally, we couple our theory to slowly varying background fields, which allows us to compute response functions of hydrodynamic operators and Kubo formulas for hydrodynamic transport coefficients.

\hspace{.3cm}

\newpage

\tableofcontents

\section{Introduction}
\label{sec:intro}

In this work we study finite-temperature transport in quantum mechanical systems with a conserved U(1) charge together with conserved dipole moment. Such systems have been of recent theoretical interest as prototypes of ``fracton'' physics, wherein elementary and isolated charges are immobile by symmetry~\cite{Chamon:2004lew, 2011AnPhy.326..839B, Haah:2011drr, Vijay:2015mka,Nandkishore:2018sel,Seiberg:2020bhn, Vijay:2016phm, Williamson:2016jiq, You:2018oai, Slagle:2017wrc,Seiberg:2020wsg, Pretko:2018jbi, Pretko:2020cko}. They are also of experimental interest in the context of tilted optical lattices~\cite{OpticalLattice1}, (2+1)-dimensional elasticity~\cite{Pretko:2017kvd, 2018PhRvL.120s5301P, Nguyen:2020yve, Gromov:2017vir}, and vortices in a superfluid~\cite{Doshi:2020jso}. 

This manuscript continues our previous work~\cite{Jain:2023nbf}, where we began developing the hydrodynamic description of translationally-invariant systems with conserved dipole moment; see also~\cite{Gromov:2020yoc, Grosvenor:2021rrt, Glorioso:2021bif, Armas:2023ouk, Glodkowski:2022xje, Glorioso:2023chm} for related work. At finite temperature, these systems tend to spontaneously break the dipole symmetry~\cite{Stahl:2021sgi,Lake:2022ico,Jensen:2022iww}, generating either a ``p-wave dipole superfluid'' phase where the dipole symmetry is spontaneously broken but the U(1) symmetry is intact, or an ``s-wave dipole superfluid'' phase where the U(1) symmetry is spontaneously broken and, on account of translation symmetry, the dipole symmetry is spontaneously broken as well. In either case, one is describing a gapless phase of matter with a low-energy hydrodynamic description at finite temperature. There is evidence that there is no hydrodynamic transport in a translationally-symmetric phase where the dipole symmetry remains spontaneously unbroken~\cite{Jensen:2022iww, Jain:2023nbf}. The broad goal of these works is to systematically build the hydrodynamic description for these exotic phases of matter, so as to arrive at robust predictions for the hydrodynamic modes and response functions, independent of microscopic (and often incalculable) details.

Our discussion in~\cite{Jain:2023nbf} focused primarily on p-wave dipole superfluids, where the low-energy variables include a dipole Goldstone $\phi_i$, in addition to the conventional fluid degrees of freedom. The main technical ingredient was a novel derivative counting scheme appropriate for systems in such a phase, characterized by a dynamical exponent $z=2$. This is consistent with the absence of sound modes in p-wave dipole superfluids, and results in dissipative transport coefficients appearing at leading order in the gradient expansion, as also observed in~\cite{Glodkowski:2022xje}. 
In this work, we shift our attention to s-wave dipole superfluids. In addition to a careful treatment of the appropriate power counting, another complicating factor in this analysis is the presence of the fluid mechanical analogues of ``dangerously irrelevant operators'' in quantum field theory~\cite{Amit:1982az}. These operators are accompanied with transport coefficients that, formally, appear at higher orders in the gradient expansion, but effectively contribute to the hydrodynamic mode spectrum and response functions at lower orders, and in this work, at leading order.\footnote{Previous discussions of dangerously irrelevant operators in hydrodynamics can be found in \cite{PhysRevLett.123.141601, Davison:2018nxm,Davison:2022vqh}. In those works higher gradient transport coefficients effectively appear at lower (but not leading) order in the gradient expansion. Here, on account of dipole symmetry, the terms we discuss are of the same importance as the ideal part of the constitutive relations.} These dangerously irrelevant operators are not merely formal properties of these fluids, but instead imply an important physical consequence, namely sound modes, which do not exist in the p-wave phase.

A preliminary version of s-wave dipole superfluid dynamics has already appeared in our previous work~\cite{Jain:2023nbf}; see also~\cite{Armas:2023ouk}. In addition to the vector dipole Goldstone $\phi_i$, we introduced a scalar U(1) Goldstone $\phi$ to accommodate the spontaneously broken U(1) global symmetry. We also noted that the dipole Goldstone is generically massive with an on-shell relation $\phi_i = - \partial_i \phi + \ldots$ that fixes the vector Goldstone $\phi_i$, leading to a low-energy effective description purely in terms of the scalar Goldstone $\phi$ and the ordinary hydrodynamic degrees of freedom. Up to a Josephson condition for the scalar Goldstone $\phi$, our preliminary treatment simply replaced $\phi_i$ in the constitutive relations, retaining the same $z=2$ power counting scheme as in a p-wave phase.
However there is a seeming inconsistency between this power counting and the linearized spectrum of fluctuations, which includes propagating sound modes and magnon-like modes
\begin{align}
\label{E:firstModes}
    \omega_{s,\pm} = \pm v_{s} k + \ldots~, \qquad 
    \omega_{m,\pm} = \pm v_m k^2 + \ldots~,
\end{align}
where both $v_s$ and $v_m$ are real. As we explain below, the latter are the dipole superfluid analogue of the ``second sound'' modes familiar from ordinary dipole-non-invariant superfluid dynamics; see e.g.~\cite{Bhattacharya:2011tra}. This leads to a puzzle: the sound mode suggests a power counting with $z=1$, while the magnon-like second-sound mode suggests one with $z=2$. If we adopt the former counting, it requires a complete reorganization of the gradient expansion and thus hydrodynamics, while the latter implies the existence of ``dangerously irrelevant operators'' giving rise to the normal sound mode, which also forces a careful reconsideration of the gradient expansion.

In addition to the preliminary investigations in \cite{Jain:2023nbf}, some aspects of s-wave dipole superfluid dynamics were also studied in~\cite{Armas:2023ouk}. Their ``pinned s-wave fracton superfluid'' contains the massive vector dipole Goldstone $\phi_i$ in addition to the massless scalar U(1) Goldstone $\phi$, while their ``U(1) fracton superfluid'' phase arises as a low-energy description after $\phi_i$ has been integrated out. The authors in~\cite{Armas:2023ouk} focused primarily on the ideal limit, with some gradient corrections, and we find these results to be in agreement with ours where they overlap.

In this work we systematically uncover a power counting appropriate to these fluids, agnostic of the ambiguous dynamical scaling exponent. With it in hand, it is tedious but straightforward to construct the ensuing hydrodynamics; we discuss the ideal s-wave dipole superfluids, together with the complete set of dissipative and non-dissipative leading gradient corrections allowed by the Second Law of thermodynamics. We also present a precise all-order road-map between the two-Goldstone and single-Goldstone formulations of s-wave dipole superfluid dynamics. As in our previous work~\cite{Jain:2023nbf}, we couple these fluids to slowly varying external fields, which allows us to deduce hydrodynamic response functions including the linearized spectrum of fluctuations. We also report the Kubo formulae for various transport coefficients, in hopes of aiding the experimental detection of these fluids.
In the remainder of this Introduction, we summarize our power counting scheme as well as some physical properties of these fluids. 

\paragraph*{Power counting:}

In an ordinary relativistic or non-relativistic fluid, the fluid velocity $u^i$ is a dimensionless parameter that characterizes the equilibrium state. The situation is more complicated for a dipole and translationally-invariant system, where the symmetry algebra naively forbids equilibrium states of nonzero velocity. Already in a p-wave dipole superfluid, one finds that $u^i\sim \cO(\dow^1)$ and thus carries dimensions of inverse length~\cite{Jain:2023nbf, Armas:2023ouk}. In the s-wave phase, we find that the velocity picks up an anomalous dimension, scaling as $u^i\sim \cO(\dow^{\varepsilon})$. Physically, we infer the existence of a length scale that does not decouple from the low-energy hydrodynamic description. When the dipole Goldstone $\phi_i$ is parametrically light, we can think of this scale as being inherited from the associated mass gap.
We have not found a principle that fixes the anomalous dimension $\varepsilon$. We only find that it must lie within $0\leq \varepsilon\leq 1$ for the hydrodynamic derivative expansion to be well-posed. Within our power counting scheme, time derivatives are treated separately from spatial derivatives, with $\partial_t \sim \cO(\partial^z), \partial_i\sim\cO(\dow^1)$; the ambiguity in the dimension of $u^i$ also leads to an ambiguity in the dynamical exponent $z$, and we find $z=1+\varepsilon$ in the range $1\leq z \leq 2$. 

To accommodate this non-trivial scaling we introduce a momentum scale $\ell\sim\cO(\dow)$ in the hydrodynamic description, taking $u^i\sim \ell^{\varepsilon}$ and $\partial_i \partial_j \phi\sim \ell^{1-\varepsilon}$,  and write down the constitutive relations as an expansion in the powers $\ell$ as well as gradients. For any choice of $\varepsilon$, we find that there are two hydrodynamic analogues of dangerously irrelevant operators in the equation of state of the fluid: $\vec u^2\sim\ell^{2\varepsilon}$ and $(\vec\nabla^2\phi)^2\sim\ell^{2-2\varepsilon}$. 
The corresponding transport coefficients appear through inverse powers in the parameters $v_s$ and $v_m$ in the sound and magnon-like second-sound modes respectively in~\cref{E:firstModes}. In fact, both of these terms in the constitutive relations come from operators in the effective action describing the hydrostatic behavior of the phase; see~\cite{Banerjee:2012iz,Jensen:2012jh,Bhattacharyya:2012xi} for the construction of hydrostatic effective actions for ordinary dipole-non-invariant fluids. As a result they are not merely the hydrodynamic analogues, but dangerously irrelevant operators in the conventional sense of the term.

We should note that this issue with the derivative counting scheme is rather formal in nature, akin to an accounting scheme for organizing the constitutive relations. In practice, the physical results such as the mode spectrum, response functions, and Kubo formulae, are agnostic to the choice of scheme as long as we work up to sufficiently higher orders in the derivative expansion. To this end, we implement an ``overcomplete'' derivative counting scheme that is designed to be consistent with any $0\leq\varepsilon\leq1$. To avoid confusion with the more naive counting of derivatives, we introduce a square bracket notation that we call ``$[m]$th derivative order.''
To wit, a generic term in the hydrodynamic constitutive relations has the derivative scaling $\cO(\partial^{n+a\varepsilon+b(1-\varepsilon)})$, for integers $n,a,b\geq 0$. Here $n$ counts explicit powers of derivatives, $a$ the powers of $u^i$, and $b$ the powers of $\dow_i\dow_j\phi$. The terms with $a=b$ should be counted trivially as $[n+a]$-derivative order. For $a>b$, the lowest integer derivative order these terms can contribute to is $(n+b)$ for $\varepsilon=0$, so we count these as $[n+b]$-derivative order. Similarly for $a<b$, the lowest integer derivative order these terms can contribute to is $(n+a)$ for $\varepsilon=1$, so we count these as $[n+a]$-derivative order.

In this manuscript we account for all transport coefficients to linear order in fluctuations at $[0]$th and $[1]$st derivative order, and a few interesting $[2]$nd order coefficients. We include a host of nonlinear terms in a Mathematica file attached to this submission. 

\paragraph*{Linearized spectrum:}

Having developed a consistent derivative expansion, we then analyze the linearized dispersion relations and response functions. In $d$ spatial dimensions, working with an equilibrium state with temperature $T$, chemical potential $\mu$, and zero ``dipole superflow'' $\dow_i\dow_j\phi = 0$, we find a pair of normal sound modes $\omega_{s,\pm}$, $d-1$ copies of a shear mode $\omega_D$, and a pair of magnon-like second-sound modes $\omega_{m,\pm}$. Their dispersion relations are schematically given as
\begin{align}
\begin{split}
    \omega_{s,\pm} 
    &= \pm v_s k - \frac{i}{2} \Gamma_s k^2
    + \mathcal{O}(k^3)~, \\
    \omega_D 
    &= -iD k^2+\mathcal{O}(k^4)~, \\
    \omega_{m,\pm} 
    &= \pm v_m k^2 
    + \mathcal{O}(k^4)~.
    \label{eq:spectrum-intro}
\end{split}
\end{align}
The definitions of the various parameters appearing here in terms of thermodynamic and transport coefficients can be found in \cref{sec:dispersion}. In particular, we note that the magnon-like mode only receives ``subdiffusive'' dissipative corrections at order $ik^4$; this is a consequence of dipole symmetry and similar physics was also found in p-wave dipole superfluids~\cite{Jain:2023nbf}. Owing to a careful consideration of background fields in the hydrodynamic description, we are also able to compute the linear response functions of hydrodynamic observables in an s-wave dipole superfluid, ultimately allowing us to derive Kubo formulas for various transport coefficients appearing in the hydrodynamic model. These have been presented in \cref{sec:response}.

To understand why the magnon-like modes in \cref{eq:spectrum-intro} are the analogue of second-sound, let us look at the zero temperature limit of the theory. In this limit, it is reasonable to expect that one gets a single component dissipationless fluid described by the Goldstone mode for which we may write a Wilsonian effective action. The simplest such action allowed by symmetry is
\begin{equation}
    S = \half \int {\rm d} t\,{\rm d}^{d}x 
    \left( \dot{\phi}^2 
    - a\, (\vec\nabla\phi)^2
    - b\, (\vec\nabla^2\phi)^2\right)\,,
\end{equation}
plus gradient and nonlinear corrections. For a dipole-non-invariant superfluid, the parameter $a\neq 0$, which results in the usual ``second sound'' mode $\omega \sim \pm k$. However, the kinematic constraint of the dipole symmetry in an s-wave dipole superfluid imposes $a=0$, which softens the dispersion of the second sound mode to $\omega \sim \pm k^2$. At finite temperature, the Goldstone mode mixes with the other hydrodynamic degrees of freedom and furthermore is attenuated by dissipative effects. However, evidently, both of these effects work in such a way as to preserve this quadratic dispersion along with subdiffusive attenuation.

The hydrodynamics of s-wave dipole superfluids also admit states with nonzero dipole superflow, characterized by a spatially-quadratic profile of the U(1) Goldstone, $\dow_i\dow_j\phi = \xi^0_{ij} \neq 0$. These states are reminiscent of superflow from ordinary dipole-non-invariant U(1) superfluids, in that they feature a dissipationless dipole flux. We find that states with isotropic dipole superflow $\xi^0_{ij} = \xi_0\delta_{ij}$ are qualitatively similar to the no-superflow states, except for certain shifts of thermodynamic parameters. However, states with anisotropic superflow can lead to genuinely distinct physical signatures. In particular, the fluid sound modes and shear modes non-trivially couple among themselves and even cause the shear modes in different transverse spatial directions to carry different diffusion constants. Interestingly, we find that the magnon-like modes remains largely unperturbed by dipole superflow, with the effects only appearing at the subdiffusive $ik^4$ level. Similar states also arise in p-wave dipole superfluids, characterized by a spatially-linear profile of the dipole Goldstone, $\dow_j\phi_i = \xi^0_{ij}$, except that the superflow tensor $\xi^0_{ij}$ is also allowed to have antisymmetric components; see~\cite{Jain:2023nbf} for more discussion in this regard.

Ordinary dipole-non-invariant superfluids allow for superflow states up to a certain critical superflow velocity, beyond which the second sound becomes perturbatively unstable~\cite{landauBook2,Gouteraux:2022qix,Arean:2023nnn}. With that in mind, one might wonder if dipole superfluids have an upper bound on superflow as well. Our hydrodynamic theory allows for the existence of such an upper bound along loci in the phase diagram where the squares of speed of sound $v_s^2$ or second sound $v_m^2$, or attenuation coefficients $\Gamma_s, D$, etc., change sign. The former would be analogous to a two-stream instability~\cite{twostream, twostream2} and the latter to the Landau instability of ordinary superfluids. However, it is unclear if such loci exist without more input from microscopic models. 

\paragraph*{States with nonzero fluid velocity:}

In ordinary dipole-non-invariant hydrodynamics, there are equilibrium states with nonzero velocity $u^i\neq0$, which labels the state in addition to the temperature $T$ and chemical potential $\mu$. The situation is quite different for a system with dipole symmetry. The dipole symmetry naively forbids a nonzero velocity at nonzero charge density, which makes intuitive sense because it forbids isolated charges from moving. In fact, there is a formal argument for this result from the point of view of the representation theory of dipole+U(1) symmetry together with translation invariance~\cite{Jensen:2022iww}. While p-wave dipole superfluids conform to this expectation, we find that s-wave dipole superfluids possess homogeneous equilibrium solutions with nonzero velocity, featuring zero net charge flux but nonzero net energy/heat flux. Physically, they can be interpreted as states where the two components of charge flux, fluid component and superfluid component, flow in opposite directions with no net flow of charge. Since p-wave dipole superfluids have spontaneously intact U(1) symmetry, there is no notion of a superfluid component and hence only states with zero velocity are allowed.

Interestingly, these exotic nonzero velocity states in s-wave dipole superfluids ultimately turn out to be a mirage: they are both thermodynamically and dynamically unstable. The origin of these instabilities can be traced back to the dipole symmetry. In an ordinary dipole-non-invariant fluid, states with nonzero velocity can alternatively be labeled by nonzero momentum density $\pi^i = \rho u^i$, where $\rho$ is the kinetic mass density, which protects them on account of momentum conservation. However in a dipole-invariant setting, while velocity is dipole-invariant, momentum density shifts under dipole transformations as $\pi^i \to \pi^i - q\psi^i$, where $q$ is the U(1) charge density and $\psi_i$ is the parameter of dipole transformation. Consequently, the momentum density of an s-wave dipole superfluid is given as $\pi^i = -q\dow^i\phi + \rho u^i$.
This means that it is possible for the system to transition from a nonzero velocity state to a zero velocity state without violating momentum conservation, by dumping the additional momentum into a spatially-linear profile for the U(1) Goldstone, $\phi = \rho/q\, u_i x^i$. The equilibrium states with zero velocity have lower canonical free energy compared to those with nonzero velocity, and thus are thermodynamically favored. Accordingly, we also find that the states with nonzero velocity are dynamically unstable. The dissipative corrections to the magnon-like second-sound modes from \cref{E:firstModes} acquire a sign-indefinite imaginary part at nonzero velocity, i.e. $\omega_{m,\pm} = \pm v_m k^2 + (I_m\pm iI'_m) (u\cdot k)k^2 + \ldots$, implying a dynamical instability.

\paragraph{Organization:}

The rest of the paper is organized as follows. In Section~\ref{sec:aristotlesRevenge}, we give a brief outline of the Aristotelian spacetime geometries to which these fluid are coupled; a longer treatise can be found in \cite{Jain:2021ibh}. In Section~\ref{sec:dipoleInvariant}, we discuss spontaneous breaking of U(1) and dipole symmetries, and outline the construction of dipole+U(1) invariant background fields, conserved currents, and the respective Ward identities. Technical aspects of this construction for p-wave dipole superfluids are discussed at length in \cite{Jain:2023nbf}. In Section~\ref{sec:hydrostatics}, we discuss the hydrostatic constitutive relations for s-wave superfluids based on the methods initiated in \cite{Banerjee:2012iz,Jensen:2012jh,Bhattacharyya:2012xi,Jensen:2013kka}. In Section~\ref{sec:dynamics}, we discuss non-hydrostatic corrections to the constitutive relations which satisfy the local Second Law of thermodynamics. In Section~\ref{sec:dispersionResponse}, we discuss the dispersion relations and a selection of response functions that arise from our construction, with and without equilibrium polarization. The paper is accompanied with a supplementary Mathematica notebook that contains the computational details of dispersion relations and response functions underlying Section~\ref{sec:dispersionResponse}, including more general results that are too technical to report in the main text.

\paragraph*{Note:}
When this paper was nearing completion,~\cite{Glodkowski:2024ova} appeared on arXiv which also aims to develop a hydrodynamic description of s-wave dipole superfluids. See the Discussion for some comments about the similarities and differences between our approach and theirs.

\section{Preliminaries}
\label{sec:background}

In this section, we review some preliminary material regarding dipole symmetry relevant for our work. We first review how to couple dipole-invariant field theories to a curved background spacetime and gauge fields in \cref{sec:aristotlesRevenge}. This allows us to systematically derive the Ward identities relevant for systems with dipole symmetry, including coupling to background sources. Then, in \cref{sec:dipoleInvariant}, we discuss the spontaneous breaking of dipole symmetry and its interplay with spontaneously broken U(1) symmetry.

\subsection{Sources, symmetries, and conserved currents}
\label{sec:aristotlesRevenge}

Dipole symmetry is compatible with time-translation, spatial-translation, and spatial-rotation symmetries, but not with spacetime boost symmetries; see e.g. \cite{Jain:2023nbf}. As a result, field theories realizing dipole symmetry can only be coupled to an Aristotelian spacetime background~\cite{Armas:2020mpr,deBoer:2020xlc,Bidussi:2021nmp,Jain:2021ibh}, which are generalizations of the Lorentzian pseudo-Riemannian spacetimes or the Galilean Newton-Cartan spacetimes to account for a preferred reference frame.\footnote{Another notable example are \textit{Carrollian spacetimes}, which are associated with the \textit{Carrollian symmetry algebra}, viz. the limit of the Poincar\'e symmetry algebra as the speed of light goes to 0; see e.g.~\cite{Ciambelli:2018wre, Ciambelli:2018xat, deBoer:2017ing}.}
In the following, we outline certain important features of Aristotelian spacetimes relevant for this work; more details can be found in \cite{Jain:2021ibh}.

\paragraph*{Aristotelian geometry:}

Instead of the symmetric metric tensor $g_{\mu\nu}$ used to describe Lorentzian spacetimes, an Aristotelian spacetime is described by a \textit{clock-form} $n_\mu$ representing the direction of time and a co-rank-1 degenerate symmetric tensor $h_{\mu\nu}$ representing the spatial metric. 
Since $h_{\mu\nu}$ has co-rank-1, there exists a unique vector field $v^\mu$ such that $v^\mu h_{\mu\nu}=0$, normalized as $v^\mu n_{\mu}=1$, which represents the preferred observer with respect to whom the notions of space and time are defined. Using these ingredients, we can also define an ``inverse'' spatial metric $h^{\mu\nu}$, satisfying $h_\mu^\nu \equiv h^{\mu\rho} h_{\rho\nu}=\delta^\mu_\nu - v^\mu n_\nu$ and $n_\mu h^{\mu\nu}=0$. Tensorial indices are raised and lowered by contractions with $h^{\mu\nu}$ and $h_{\mu\nu}$ respectively. In this language, a flat spacetime background is 
\begin{equation}
    n_\mu = \delta_\mu^t, \qquad 
    v^\mu = \delta^\mu_t, \qquad
    h_{\mu\nu} = \delta_\mu^i\delta_\nu^j \delta_{ij}, \qquad 
    h^{\mu\nu} = \delta^\mu_i\delta^\nu_j \delta^{ij}.
\end{equation}

Analogous to Lorentzian spacetimes, Aristotelian spacetimes are also equipped with a covariant derivative operator $\nabla_\mu$, with the connection\footnote{The action of the covariant derivative $\nabla_\mu$ on a rank-$(1,1)$ tensor $\mathcal{T}^\mu_{\;\;\nu}$ can be defined as
\begin{equation}
    \nabla_\rho \mathcal{T}^\mu_{\;\;\nu}= \partial_\rho \mathcal{T}^\mu_{\;\;\nu} + \Gamma^\mu_{\;\;\rho\sigma} \mathcal{T}^\sigma_{\;\;\nu} - \Gamma^\sigma_{\;\;\rho\nu} \mathcal{T}^\mu_{\;\;\sigma}~,
\end{equation}
and analogously for higher-rank tensors.}
\begin{equation}
    \Gamma^\lambda_{\;\;\mu\nu} = v^\lambda \partial_\mu n_\nu + \frac{1}{2} h^{\lambda\rho}\left( \partial_\mu h_{\nu\rho} + \partial_\nu h_{\mu\rho} - \partial_\rho h_{\mu\nu} \right).
\end{equation} 
The covariant derivative acts on the Aristotelian spacetime sources as follows
\begin{equation}
    \begin{split}
        \nabla_\mu n_\nu &=0~, \\
        \nabla_\lambda h^{\mu\nu}&=0~, \\
        \nabla_\lambda h_{\mu\nu}&=-n_{(\mu} \pounds_v h_{\nu)\lambda}~, \\
        h_{\nu\mu}\nabla_\lambda v^\mu &= \frac{1}{2} \pounds_v h_{\nu\lambda}~,
    \end{split}
\end{equation}
where $\lie_v$ denotes a Lie derivative along $v^\mu$. 
It is also useful to define a non-degenerate metric tensor $\gamma_{\mu\nu} \equiv n_\mu n_\nu + h_{\mu\nu}$, with inverse $\gamma^{\mu\nu} \equiv v^\mu v^\nu + h^{\mu\nu}$, and the associated spacetime volume factor $ \gamma \equiv \text{det}(\gamma_{\mu\nu})$. This leads to the identities
\begin{equation}
\begin{split}
\frac{1}{\sqrt{\gamma}}\partial_{\nu}\sqrt{\gamma}
&=\Gamma^{\mu}_{\;\;\mu\nu}+F^{n}_{\nu\mu}v^{\mu}~,
\\
T^{\lambda}_{\;\;\mu\nu}\equiv 2 \Gamma^{\lambda}_{\;\;[\mu\nu]}
&=v^{\lambda}F^{n}_{\mu\nu}~,
\end{split}
\end{equation}
where $F^{n}_{\mu\nu} \equiv (\df n)_{\mu\nu} = 2\partial_{[\mu}n_{\nu]}$. It is also useful to define a gradient operator $\nabla'_\mu = \nabla_\mu + F^n_{\mu\nu}v^\nu$, such that $\nabla'_\mu V^\mu = \frac{1}{\sqrt{\gamma}} \dow_\mu(\sqrt{\gamma}\,V^\mu)$ for any vector $V^\mu$.\footnote{Note that $\nabla'_\mu$ does not obey the product rule and instead satisfies $\nabla'_\mu({\cal T}^{\mu}_{~~\nu}{\cal Q}^\rho_{~\sigma}) = \nabla'_\mu{\cal T}^{\mu}_{~~\nu}{\cal Q}^\rho_{~\sigma} + {\cal T}^{\mu}_{~~\nu}\nabla_\mu {\cal Q}^\rho_{~\sigma}$, and similarly for higher-rank tensors.} Expressed in terms of the connection $\Gamma^\lambda_{~\mu\nu}$, the curvature tensor has the same form as Lorentzian spacetimes, i.e.
\begin{equation}
    R^\rho_{\;\;\sigma\mu\nu} 
    = \partial_{\mu} \Gamma^\rho_{\;\;\nu\sigma}
    - \partial_{\nu} \Gamma^\rho_{\;\;\mu\sigma}
    + \Gamma^\rho_{\;\;\mu\lambda} \Gamma^\lambda_{\;\;\nu\sigma} 
    - \Gamma^\rho_{\;\;\nu\lambda} \Gamma^\lambda_{\;\;\mu\sigma}\;.
\end{equation}
It can be checked that $n_\rho R^{\rho}_{\;\;\sigma\mu\nu} = 0$.

\paragraph{U(1) and dipole gauge fields:} 

In addition to the spacetime sources, we also include a background U(1) gauge field $A_\mu$ and a symmetric dipole gauge field $a_{\mu\nu}$ (s.t. $v^\mu a_{\mu\nu} = 0$). It is useful to combine these fields to define an effective dipole gauge field
\begin{align}
    A^\lambda_{~\mu} = 
    \biggl(
    n_\mu v^\rho F_{\rho\sigma}
    +\frac{1}{2}\lb
    h^{\rho}_\mu F_{\rho\sigma}
    + a_{\mu\sigma}
    \rb
    \biggr)h^{\sigma\lambda}~.
\end{align}
where $F_{\mu\nu} \equiv (\df A)_{\mu\nu} = 2\partial_{[\mu}A_{\nu]}$ is the background gauge field strength tensor. 
We also define the dipole field strength
\begin{align}
    F^\lambda_{~\mu\nu}
    &= \nabla_\mu A^\lambda_{\;\;\nu} - \nabla_\nu A^{\lambda}_{\;\;\mu} + F^n_{\mu\nu}v^\rho A^\lambda_{\;\;\rho} + 2n_{[\mu}A^\rho_{\;\;\nu]}\nabla_\rho v^\lambda~.
\end{align}
Note that $n_\lambda A^\lambda_{~\mu} = n_\lambda F^\lambda_{~\mu\nu} = 0$. For later use, we note the variation of $A^\lambda_{~\mu}$ in terms of other background fields is given as
\begin{align}
    h_{\nu\lambda}\delta A^\lambda_{~\mu} 
    &= 
    v^\rho F_{\rho(\nu} h_{\mu)}^\sigma \delta n_{\sigma}
    + \lb n_{\mu} v^{(\rho}A^{\sigma)}_{~~\lambda} h_\nu^{\lambda}
    - A^{(\rho}_{~~\mu} h_\nu^{\sigma)} \rb \delta h_{\rho\sigma}
    + \half h_\mu^\rho h_\nu^\sigma \delta a_{\rho\sigma}
    \nn\\
    &\qquad 
    + \half n_{\mu}v^\rho \delta F_{\rho\nu}
    + \frac{1}{2} h_\nu^\rho \delta F_{\mu\rho}~.
    \label{foot:aAtrans}
\end{align}

\paragraph{Symmetry transformations and Ward identities:}

Dipole-symmetric physical systems can be described by a generating functional $W = -i\ln{\cal Z}$, expressed as a functional of the background fields $n_\mu$, $h_{\mu\nu}$, $A_\mu$, and $a_{\mu\nu}$. The global spacetime and internal symmetries can be realized as the invariance of $W$ under ``spurionic'' background diffeomorphisms $\chi^\mu$, U(1) gauge transformations $\Lambda$, and dipole gauge transformations $\psi_\mu$ (s.t. $\psi_{\mu}v^{\mu}=0$). Denoting the symmetry parameters collectively as $\hat\scX = (\chi^\mu,\Lambda,\psi_\mu)$, their action on the background fields is defined as
\begin{subequations}
\begin{align}
\begin{split}
    \delta_{\hat\scX} n_\mu &= \lie_\chi n_\mu\,, 
    \\ 
    \delta_{\hat\scX} h_{\mu\nu} &= \lie_\chi h_{\mu\nu}\,, 
    \\
    \delta_{\hat\scX} A_\mu &= \lie_\chi A_\mu 
    + \partial_\mu\Lambda + \psi_\mu\,, 
    \\
    \delta_{\hat\scX} a_{\mu\nu} &= \lie_\chi a_{\mu\nu} 
    + h^{\rho}_\mu h^{\sigma}_\nu
    (\nabla_\rho \psi_\sigma + \nabla_\sigma \psi_\rho)~,
\end{split}
\end{align}
which further implies that
\begin{equation}
    \delta_{\hat\scX} A^\lambda_{~\mu} 
    = \lie_\chi A^\lambda_{~\mu} 
    + \nabla_\mu \psi^\lambda + n_\mu \psi^\nu \nabla_\nu v^\lambda\,,
\end{equation}
\label{eq:back-var}%
\end{subequations}
transforms as an effective dipole gauge field. 

Variations of the generating functional $W$ with respect to the background fields allow us to obtain the respective conserved currents: energy current $\epsilon^\mu$, momentum density $\pi^\mu$, symmetric stress-tensor $\tau^{\mu\nu} = \tau^{\nu\mu}$, U(1) current $J^{\mu}$, and symmetric dipole flux $J^{\mu\nu} = J^{\nu\mu}$, defined\footnote{These definitions are required to define the currents in addition to \cref{eq:PhysicalActionNoGoldstones} because the components of background fields are not all independent. For example, note that $v^\mu v^\nu \delta h_{\mu\nu}=0$, as required by $v^\mu v^\nu h_{\mu\nu}=0$, and hence \cref{eq:PhysicalActionNoGoldstones} by itself actually leaves the time component of $\pi^\mu$ undefined.}
such that $n_\mu\pi^\mu  = n_\mu\tau^{\mu\nu} = n_\mu J^{\mu\nu} = 0$. To wit,
\begin{align}
\begin{split}
    \label{eq:PhysicalActionNoGoldstones}
    \delta W 
    &= \int \df^{d+1}x\sqrt{\gamma}
    \Bigg[ 
    - \epsilon^\mu\delta n_\mu 
    + \biggl(v^{(\mu}\pi^{\nu)}+\frac{1}{2}\tau^{\mu\nu}\biggr)\delta h_{\mu\nu} + J^\mu\delta A_\mu 
    + J^{\mu\nu} h_{\nu\lambda}\delta A^\lambda_{~\mu}
    \Bigg] \\
    &= \int \df^{d+1}x\sqrt{\gamma}
    \Bigg[
    - \lb\epsilon^\mu
    + J^{\mu\nu}F_{\nu\rho}v^\rho\rb \delta n_{\mu}
    + \biggl(v^{(\mu}\pi^{\nu)}+\frac{1}{2}
    \lb \tau^{\mu\nu} - 2 A^{(\mu}_{~~\lambda} J^{\nu)\lambda}\rb \biggr)\delta h_{\mu\nu} \\
    &\hspace{26em}
    + J^\mu\delta A_\mu 
    + \half J^{\mu\nu} \delta a_{\mu\nu}
    \Bigg]~.
    \end{split}
\end{align}
Following the work of~\cite{Jain:2021ibh}, we have defined the conserved currents by using variations with respect to the effective dipole gauge field $A^\lambda_{~\mu}$ as opposed to $a_{\mu\nu}$, as they have comparatively more natural transformation properties under dipole transformations
\begin{align}
\begin{split}
    \epsilon^\mu &\to \epsilon^\mu + \left(2J^{\mu(\rho}h^{\sigma)\lambda}-J^{\rho\sigma}h^{\mu\lambda}\right)\psi_\lambda\frac{1}{2}\lie_v h_{\rho\sigma}\\
     \pi^\mu &\to\pi^\mu - \left(
     J^\nu n_\nu h^{\mu\lambda}-J^{\rho\mu}h^{\sigma\lambda}F^n_{\rho\sigma}\right)\psi_\lambda,\\
    \tau^{\mu\nu} &\to \tau^{\mu\nu}-2\psi^{(\mu} \nabla_\rho'J^{\nu)\lambda}+\nabla_\sigma'(\psi^\sigma J^{\mu\nu})~,
    \label{eq:current-trans}
\end{split}
\end{align}
while $J^\mu$ and $J^{\mu\nu}$ are dipole-invariant. 
The Ward identities for the symmetry currents that follow from \cref{eq:back-var} are
\begin{subequations}
\begin{align}
\begin{split}
  \nabla_\mu' \epsilon^\mu 
    &= -v^\mu  f_\mu 
    - (\tau^{\mu\nu} + \tau^{\mu\nu}_d)
    h_{\lambda\nu}\nabla_\mu v^\lambda~, \\ 
    \nabla_\mu'\lb v^\mu \pi^\nu 
    + \tau^{\mu\nu} + \tau^{\mu\nu}_d \rb
    &= h^{\nu\mu} f_\mu 
    - \pi_\mu h^{\nu\lambda}\nabla_\lambda v^\mu~, \\ 
    \nabla'_\mu J^\mu &=0~,\\
    \nabla_\mu'J^{\mu\nu} &= h^\nu_\mu J^\mu~,
\end{split}
\end{align}
where we have defined the force term $f_\mu$ and asymmetric dipole-stress $\tau^{\mu\nu}_d$ as
\begin{align}
\begin{split}
    f_\mu 
    &= -F^n_{\mu\nu}\epsilon^\nu 
    - h_{\mu\lambda}A^\lambda_{~\nu}J^\nu 
    + F^\lambda_{\;\;\mu\nu}h_{\rho\lambda}J^{\nu\rho} 
    + \half n_\mu \tau_d^{\rho\nu}\lie_v h_{\rho\nu}~,\\
    \tau_d^{\mu\nu} 
    &= - A^\mu_{~\rho}J^{\rho\nu}~.
    \label{eq:f-physical}
\end{split}
\end{align}
\label{eq:physical-Wardidentity}%
\end{subequations}
It can be explicitly checked that the dipole-transformations leave the Ward identities invariant. More details on the conserved currents and Ward identities of dipole-invariant field theories can be found in~\cite{Jain:2021ibh}; see also~\cite{Bidussi:2021nmp}.
The conservation equations \eqref{eq:physical-Wardidentity} together with the transformation properties \eqref{eq:current-trans} form the basis of our upcoming hydrodynamic discussion.

\subsection{Spontaneous breaking of dipole and U(1) symmetry}
\label{sec:dipoleInvariant}

We are interested in low-energy effective field theories featuring a conserved dipole moment. In our previous work~\cite{Jain:2023nbf} (see also~\cite{Jensen:2022iww}), we argued that for such a theory to admit any gapless excitations in the charge sector, the dipole symmetry must be spontaneously broken. This results in two non-trivial gapless phases for systems with dipole symmetry: a \emph{p-wave dipole superfluid phase} where the dipole symmetry is spontaneously broken but the U(1) symmetry remains unbroken, and a \emph{s-wave dipole superfluid phase} where both the U(1) and dipole symmetries are spontaneously broken. We have discussed the p-wave phase at length in~\cite{Jain:2023nbf}. In this paper, we will focus our attention on the s-wave dipole superfluid phase. 

Simultaneous spontaneous breaking of U(1) and dipole symmetries can generically be characterized by two Goldstone fields: a U(1) Goldstone field $\phi$ and a dipole Goldstone field $\phi_\mu$ (s.t. $v^\mu\phi_\mu = 0$), which transform non-linearly under global symmetries, i.e.
\begin{align}
\label{eq:goldstone-var}
\begin{split}
    \delta_{\hat{\scX}}\phi &= \lie_\chi\phi - \Lambda\\
    \delta_{\hat{\scX}}\phi_\mu &= \lie_\chi \phi_\mu - \psi_\mu.
\end{split}
\end{align}
In this phase, the generating functional $W$ can be expressed as a path integral of an effective action $S[\phi,\phi_\mu]$ over possible configurations of $\phi$ and $\phi_\mu$, i.e. $W = -i\ln\int[\df\phi][\df\phi_\mu]\exp(i S[\phi,\phi_\mu])$. The variations of $S[\phi,\phi_\mu]$ with respect to the background and dynamical Goldstone fields can be parametrized similar to \cref{eq:PhysicalActionNoGoldstones} as
\begin{align}
\begin{split}
    \label{eq:PhysicalActionVariation}
    \delta S
    &= \int \df^{d+1}x\sqrt{\gamma}\bigg[
        -\epsilon^\mu\delta n_\mu 
        + \biggl(v^{(\mu}\pi^{\nu)}
        +\frac{1}{2}\tau^{\mu\nu}\biggr)\delta h_{\mu\nu} 
        + J^\mu\delta A_\mu 
        + J^{\mu\rho} h_{\rho\sigma}\delta A^\sigma_{\;\mu}
    \\
    &\hspace{28em}
    + X\delta\phi
    + X^\mu\delta\phi_\mu\bigg].
    \end{split}
\end{align}
Since the dipole Goldstone satisfies $v^\mu\phi_\mu =0$, its time-component is not independent and thus we correspondingly choose $X^\mu n_\mu = 0$. The equations of motion associated with the Goldstones $\phi$ and $\phi_\mu$ are given as $X=0$ and $X^\mu = 0$ respectively.

As it turns out, the description of an s-wave dipole superfluid in terms of two Goldstone fields is redundant at low energies. When the U(1) symmetry is spontaneously broken on top of the spontaneously broken dipole symmetry, the dipole Goldstone $\phi_\mu$ generically gets Higgsed and acquires a mass $m$. This means that $\phi_\mu$ can be integrated out from the description at energy scales much smaller than $m$, leaving an effective theory expressed entirely in terms of the U(1) Goldstone $\phi$. The key observation in this regard is that we can combine $\phi_\mu$ and $\phi$ to construct a U(1)+dipole-invariant vector field
\begin{equation}
    B_\mu = \phi_\mu + \vec\xi_\mu~,
\end{equation}
where $\vec\xi_\mu = h^{\nu}_\mu\xi_\nu$ and
\begin{align}
    \xi_\mu = \dow_\mu\phi + A_\mu~,
\end{align}
is the U(1)-invariant superfluid velocity familiar from dipole-non-invariant superfluid dynamics.
This allows us to write down a U(1)+dipole-invariant term in the effective action taking the form
\begin{align} 
    \label{mass term for vector Goldstone}
    S\sim 
    -\int \df^{d+1}x\sqrt{\gamma}\,
    \frac{1}{2} m^2\,B_\mu B^\mu~,
\end{align}
which acts as a mass term for $\phi_\mu$ for some mass scale $m$. The equation of motion for $\phi_\mu$ now takes a schematic form
\begin{equation}
\begin{gathered}
\label{dipoleGoldstoneidentity}
    X^\mu = - m^2 \lb \phi^\mu + \vec\xi^\mu \rb + \ldots = 0 \\ 
    \implies \phi_\mu = - \vec\xi_\mu + \frac{1}{m^2}(\ldots)~,
\end{gathered}
\end{equation}
where $\ldots$ denote the potential contributions arising from other terms in the effective action. Provided that we are only interested in the energy scales much smaller than the mass scale $m$ of $\phi_\mu$, we can integrate out the vector Goldstone and work with a simpler effective action $S[\phi] = S[\phi,\phi_\mu = - \vec\xi_\mu + \ldots]$. The variation of $S[\phi]$ is parametrized similar to \cref{eq:PhysicalActionVariation} as 
\begin{align}
    \delta S
    &= \int \df^{d+1}x\sqrt{\gamma}
    \left[
    -{\epsilon}^\mu\delta n_\mu + \biggl(v^{(\mu}\pi^{\nu)}+\frac{1}{2}\tau^{\mu\nu}\biggr)\delta h_{\mu\nu} + J^\mu\delta A_\mu
    + J^{\mu\rho} h_{\rho\lambda}
    \delta A^\lambda_{~\mu}
    + X\delta\phi
    \right],
\end{align}
but with the dipole Goldstone equation of motion $X^\mu$ set to zero.
The description in terms of two Goldstones might still be useful if we want to describe the physics near the mass scale $m$, provided that all other non-conserved microscopic excitations in the theory are sufficiently gapped compared to $m$, in the spirit of ``quasi-hydrodynamics'' with non-conserved degrees of freedom~\cite{Grozdanov:2018fic, Delacretaz:2021qqu, Armas:2021vku, Armas:2023tyx}. See some relevant discussion for ideal s-wave dipole superfluids in~\cite{Armas:2023ouk}.

Demanding that the effective action $S$ is invariant under infinitesimal symmetry variations $\delta_{\hat\scX}$, we can obtain the Ward identities satisfied for arbitrary off-shell configurations of the Goldstones which will be useful later, i.e.
\begin{subequations}
\begin{align}
\begin{split}
  \nabla_\mu'\epsilon^\mu 
    &= -v^\mu f_\mu^X
    - (\tau^{\mu\nu} + \tau^{\mu\nu}_d)
    h_{\lambda\nu}\nabla_\mu v^\lambda~, 
    \\ 
    \nabla_\mu'\lb v^\mu \pi^\nu 
    + \tau^{\mu\nu} + \tau^{\mu\nu}_d \rb
    &= h^{\nu\mu} f_\mu^X 
    - \pi_\mu h^{\nu\lambda}\nabla_\lambda v^\mu~,
    \\ 
    \nabla'_\mu J^\mu &= - X~,\\
    \nabla_\mu'J^{\mu\nu} &= h^\nu_\mu J^\mu
    {\color{Gray}\,-\,X^\nu}~.
\end{split}
\end{align}
The off-shell force term $f^X_\mu$ differs from $f_\mu$  as
\begin{align}
\begin{split}
    f_\mu^X
    &= 
    - F^n_{\mu\nu}\epsilon^\nu 
    - h_{\mu\lambda}A^\lambda_{~\nu}J^\nu 
    + F^\lambda_{\;\;\mu\nu}h_{\rho\lambda}J^{\nu\rho} 
    - n_\mu A^\lambda_{\;\;\rho}
    J^\rho_{\;\;\nu}\nabla_\lambda v^\nu 
    + \xi_\mu X \\
    &\qquad
    {\color{Gray}\,+\, \lb \nabla_\mu \phi^\lambda 
    + A^\lambda_{~\mu}
    \rb X_\lambda 
    - \nabla'_\nu (\phi_\mu X^\nu)}.
    \end{split}
\end{align}
\label{eq:physical-Wardidentity-X}%
\end{subequations}
The gray terms here, and in the rest of the paper, are only present in the two-Goldstone formulation when the dipole Goldstone $\phi_\mu$ has not been integrated out. They are absent in the single-Goldstone formulation where $X^\mu$ is identically zero and $\phi_\mu$ is fixed to its on-shell value $-1/\ell\,\vec\xi_\mu + \ldots$.  
When $\phi$ is taken on-shell as well, $X$ is also set to zero and we recover the Ward identities \eqref{eq:physical-Wardidentity}.

\paragraph*{Dipole-invariant formulation:}

Note that the Ward identities \eqref{eq:physical-Wardidentity-X} are U(1)-invariant but are not manifestly dipole-invariant. For our later discussion of hydrodynamics, it will be convenient to derive a dipole-invariant version of these equations. To this end, we introduce the dipole-invariant versions of background gauge fields 
\begin{align}\begin{split}
    \tilde{A}_\mu 
    &= A_\mu + \Psi_\mu~, \\
    \tilde{a}_{\mu\nu} 
    &= a_{\mu\nu} 
    + h^\rho_\mu h^\sigma_\nu (\nabla_{\rho}\Psi_\sigma + \nabla_\sigma\Psi_\rho)~,
    \label{eq:tilde-background}
\end{split}\end{align}
and accordingly for the composite fields $\tilde F_{\mu\nu}$, $\tilde A^\lambda_{~\mu}$, $\tilde F^\lambda_{~\mu\nu}$, and $\tilde\xi_\mu$, where $\Psi_\mu$ is some field that shifts under dipole transformation as $\Psi_\mu \to \Psi_\mu - \psi_\mu$. Choosing $\Psi_\mu = \phi_\mu$, these definitions reduce to the dipole-invariant quantities defined in~\cite{Jain:2023nbf} for p-wave dipole superfluids. For s-wave dipole superfluids in the U(1)-Goldstone formulation, where the dipole Goldstone $\phi_\mu$ has been integrated out, we need to instead choose $\Psi_\mu = - \vec\xi_\mu$. More generally for s-wave dipole superfluids in the two-Goldstone formulation, we have the option to make either of these choices, as $\phi_\mu$ is only equal to $-\vec\xi_\mu$ up to certain derivative corrections; see~\cref{dipoleGoldstoneidentity}. So, for clarity and to keep the contrast to p-wave dipole superfluids transparent, we will work with an arbitrary $\Psi_\mu$ in the following.  We can also define the dipole-invariant version of the dipole Goldstone itself in the two-Goldstone formulation
\begin{equation}
    \tilde\phi_\mu = \phi_\mu - \Psi_\mu. 
\end{equation}
This object is identically zero if we choose $\Psi_\mu = \phi_\mu$ for defining the dipole-invariants, but is equal to $B_\mu$ for $\Psi_\mu = -\vec\xi_\mu$.

The effective action $S$ of an s-wave dipole superfluid can also be expressed entirely in terms of the U(1)+dipole-invariant objects $n_\mu$, $h_{\mu\nu}$, $\tilde\xi_\mu$, $\tilde A^\lambda_{~\mu}$, and their derivatives. The variation of $S$ with respect to these is given as 
\begin{align}
\label{eq:deltaSdipoleinvariant}
\begin{split}
    \delta S 
    &= \int \df^{d+1}x
    \sqrt{\gamma}\Bigg[
    - \tilde\epsilon^\mu\delta n_\mu 
    + \lb v^{(\mu}\tilde\pi^{\nu)}
    + \frac{1}{2}\tilde\tau^{\mu\nu}\rb \delta h_{\mu\nu} 
    + J^\mu\delta \tilde\xi_\mu 
    + J^{\mu\nu}h_{\nu\lambda}\delta \tilde{A}^{\lambda}_{~\mu}
    \gray{\,\,+\, X^\mu \delta\tilde\phi_\mu} 
    \Bigg]~,
\end{split}
\end{align}
where we have defined the dipole-invariant versions of currents
\begin{align}
\begin{split}
\label{eq:dipoleinvariantcurrents}
    \tilde\epsilon^\mu 
    &= \epsilon^\mu 
    + \lb 2J^{\mu(\rho}\Psi^{\sigma)}
    - J^{\rho\sigma}\Psi^\mu\rb
    \frac{1}{2}\pounds_v h_{\rho\sigma}~,\\
    \tilde\pi^\mu 
    &= \pi^\mu - (J^\nu n_\nu)\Psi^\mu 
    + J^{\rho\mu}F^n_{\rho\sigma}\Psi^\sigma~, \\
    \tilde\tau^{\mu\nu}
    &= \tau^{\mu\nu} 
    - 2
    \Psi^{(\mu}\nabla'_\lambda J^{\nu)\lambda}
    + \nabla_\lambda'\!\lb\Psi^\lambda J^{\mu\nu}\rb~.
\end{split}
\end{align}
 The off-shell Ward identities \eqref{eq:physical-Wardidentity-X} can also be expressed in terms of dipole-invariant objects
\begin{subequations}
\begin{align}
\begin{split}
  \nabla_\mu'\tilde\epsilon^\mu 
    &= -v^\mu \tilde f_\mu^X
    - (\tilde\tau^{\mu\nu} + \tilde\tau^{\mu\nu}_d)
    h_{\lambda\nu}\nabla_\mu v^\lambda~, 
    \\ 
    \nabla_\mu'\lb v^\mu \tilde\pi^\nu 
    + \tilde\tau^{\mu\nu} + \tilde\tau^{\mu\nu}_d \rb
    &= h^{\nu\mu} \tilde f_\mu^X 
    - \tilde\pi_\mu h^{\nu\lambda}\nabla_\lambda v^\mu~,
    \\ 
    \nabla'_\mu J^\mu &= - X~,\\
    \nabla_\mu'J^{\mu\nu} &= h^\nu_\mu J^\mu
    {\color{Gray}\,-\,X^\nu}~,
\end{split}
\end{align}
where the dipole-invariant off-shell force term $\tilde f^X_\mu$ and dipole stress tensor $\tilde\tau^{\mu\nu}_d$ have been accordingly defined as
\begin{align}
\begin{split}
    \tilde f_\mu^X
    &= 
    - F^n_{\mu\nu}\tilde\epsilon^\nu 
    - h_{\mu\lambda} \tilde A^\lambda_{~\nu}J^\nu 
    + \tilde F^\lambda_{\;\;\mu\nu}h_{\rho\lambda}J^{\nu\rho} 
    - n_\mu \tilde A^\lambda_{\;\;\rho}
    J^\rho_{\;\;\nu}\nabla_\lambda v^\nu 
    + \tilde\xi_\mu X \\
    &\qquad
    {\color{Gray}\,+\, \lb \nabla_\mu \tilde\phi^\lambda 
    + \tilde A^\lambda_{~\mu}
    \rb X_\lambda 
    - \nabla'_\nu (\tilde\phi_\mu X^\nu)}, \\
    \tilde\tau_d^{\mu\nu} 
    &= - \tilde{A}^\mu_{\;\;\rho}J^{\rho\nu}.
    \end{split}
\end{align}
\label{eq:physical-Wardidentity-X-inv}%
\end{subequations}
This incarnation of conservation equations will be indispensable for our subsequent discussion of s-wave dipole superfluid hydrodynamics. 

Let us momentarily look at the choice $\Psi_\mu = -\vec\xi_\mu$, which is the only one available to use in the U(1)-Goldstone formulation of s-wave dipole superfluids. In this case, we have
\begin{equation}
    \tilde A_\mu = \Phi\,n_\mu - \dow_\mu\phi, \qquad 
    \tilde\xi_\mu = \Phi\,n_\mu~,
\end{equation}
where $\Phi = v^\mu\xi_\mu = v^\mu A_\mu + v^\mu \dow_\mu\phi$. 
In particular, there is no dipole-invariant version of the spatial components of the U(1) gauge field $A_\mu$ in the U(1)-Goldstone formulation. A corollary to this fact is that the variation of $S$ in \cref{eq:deltaSdipoleinvariant} simplifies to
\begin{align}
\begin{split}
    \delta S
    &= \int \df^{d+1}x
    \sqrt{\gamma}\Bigg[
    - \tilde E^\mu\delta n_\mu 
    + \lb v^{(\mu}\tilde\pi^{\nu)}
    + \frac{1}{2}\tilde\tau^{\mu\nu}\rb 
    \delta h_{\mu\nu} 
    + Q\,\delta \Phi
    + J^{\mu\nu}h_{\nu\lambda}\delta \tilde{A}^{\lambda}_{~\mu} 
    {\color{Gray}\,+\, X^\mu \delta B_\mu}
    \Bigg]~,
\end{split}
\end{align}
where $\tilde E^\mu = {\tilde\epsilon}^\mu - \Phi \lb Q\,v^\mu + \nabla'_\nu J^{\mu\nu} \rb$ and $Q = n_\mu J^\mu$. Note that the spatial components of the U(1) current $h^\mu_\nu J^\nu$ do not appear in \cref{eq:deltaSdipoleinvariant} for the U(1)-Goldstone formulation. 
This is not to say that we have lost information about these components. The full U(1) current $J^\mu$ can still be obtained by varying the action with respect to the true background gauge field $A_\mu$, i.e.
\begin{equation}
    J^\mu = Q\,v^\mu 
    + \nabla'_\nu J^{\mu\nu}
    {\color{Gray}\,+\,X^\nu}~.
    \label{eq:J-def}
\end{equation}
Since $S$ in \cref{eq:deltaSdipoleinvariant} is manifestly parametrized in terms of dipole-invariant objects, the U(1) current identically satisfies the dipole Ward identity.

\section{Hydrostatics}
\label{sec:hydrostatics}

In this section, we discuss the equilibrium aspects of s-wave dipole superfluids with spontaneously broken dipole and U(1) symmetries. Our discussion will be centered around the construction of hydrostatic partition functions~\cite{Banerjee:2012iz, Jensen:2012jh, Bhattacharyya:2012xi, Jensen:2013kka}, from which the conserved currents of an ideal (super)fluid can be obtained. The partition function also allows us to classify possible hydrostatic derivative corrections that characterize equilibrium configurations of the (super)fluid in the presence of time-independent background sources. The formal construction of the hydrostatic partition function for s-wave dipole superfluid is similar to the p-wave case discussed in depth in our previous paper~\cite{Jain:2023nbf}, so we will keep the discussion brief where relevant.

\subsection{The setup}
\label{sec:hydrostatBasics}

We can characterize the state of a (super)fluid in thermal equilibrium in terms of the equilibrium partition function
\begin{align}
    \exp(-\beta_0 W_{\text{eqb}}) 
    = \exp(-\beta_0\mathcal{H}_{\scK})~,
\end{align}
where
$\beta_0 = 1/T_0$ represents the global equilibrium temperature. Furthermore, $\mathcal{H}_\scK$ is the Hamiltonian of the fluid, defined in terms of the conserved densities as
\begin{align}
    \mathcal{H}_{\scK} = \int \df^{d}x
    \Big(\epsilon^t - u_0^i\pi_i - \mu_0 J^t\Big)~,
\end{align}
where $u_0^i$ and $\mu_0$ are the equilibrium fluid velocity and chemical potential respectively. As a consequence of the dipole symmetry algebra, the momentum and chemical potentials shift under a global dipole transformation, $\pi_i \to \pi_i - J^t \psi_i$, $\mu_0 \to \mu_0 + u_0^i \psi_i$; see~\cite{Jain:2023nbf}. In other words, if the equilibrium state is dipole invariant, then a nonzero equilibrium fluid velocity would be in conflict with a well-defined chemical potential and vice-versa. Thus, in general, to have equilibrium states where both charge and momentum are nonzero, the equilibrium state must spontaneously break the dipole symmetry. 

The hydrostatic partition function can be coupled to background sources, including a curved background spacetime, provided that the background is time-independent with respect to the thermal observer. Generically, we can characterize the thermal observer with a set of symmetry data: a time-like vector $K^\mu$, a U(1) parameter $\Lambda_K$, and a dipole parameter $\psi_\mu^K$ (normalized as $v^\mu\psi^K_\mu = 0$), collectively denoted by $\hat\scK = (K^\mu, \Lambda_K, \psi_\mu^K)$. The background sources are required to be invariant with respect to $\hat\scK$, i.e.
\begin{align}
    \delta_{\hat\scK} n_\mu 
    = \delta_{\hat\scK} h_{\mu\nu} 
    = \delta_{\hat\scK} A_\mu
    = \delta_{\hat\scK} A^\lambda_{~\mu}
    =  0~.
    \label{eq:eqb-isometry}
\end{align}
where the symmetry variations are defined as in \cref{eq:back-var}. Under an infinitesimal symmetry transformation $\hat\scX=(\chi^\mu,\Lambda,\psi_\mu)$, the thermal parameters $\hat\scK$ themselves transform according to
\begin{align}
\begin{split}
    \delta_{\hat\scX}K^\mu &= \lie_\chi K^\mu, \\
    \delta_{\hat\scX}\Lambda_K &= \lie_\chi\Lambda_K - \lie_K\Lambda, \\
    \delta_{\hat\scX}\psi^K_\mu &= \lie_\chi\psi^K_\mu 
    - h_\mu^\nu\lie_K\psi_\nu~.
    \label{eq:trans-data}
\end{split}
\end{align}
See~\cite{Jain:2021ibh} for more details.
We can partially fix the symmetries to take $K^\mu = \delta^\mu_i u^i_0/T_0$, $\Lambda_K = \mu_0/T_0$, and $\psi^K_\mu = 0$, but it is clearer to work with the covariant notation. In the presence of background sources, the Hamiltonian modifies to
\begin{align}
\begin{split}
    \beta_0\mathcal{H}_{\scK}
    &= \int\df\Sigma_\mu
  \Big[ K^\nu n_\nu {\epsilon}^\mu
    - K^\nu h_{\nu\rho} \lb v^\mu \pi^{\rho}
    + \tau^{\mu\rho} + \tau^{\mu\rho}_d \rb
    - \lb\Lambda_K + K^\nu A_\nu\rb J^\mu\\
    &\hspace{22em}
    - \lb \psi_\mu^K + h_{\rho\lambda}K^\nu A^\lambda_{~\nu} \rb 
    J^{\mu\rho}
    \Big]~,
    \end{split}
\end{align}
where $\df\Sigma_\mu$ denotes the integral measure on an arbitrary spatial slice transverse to $K^\mu$. It can be checked that the integrand above is conserved due to the Ward identities, hence ${\cal H}_\scK$ is independent of the particular choice of $\df\Sigma_\mu$. It can also be checked that this expression is manifestly invariant under $\hat\scK$-compatible symmetry transformations for which the expressions in \cref{eq:trans-data} vanish.

For s-wave dipole superfluids, with spontaneously broken dipole and U(1) symmetries, the hydrostatic partition function is obtained as a path integral over hydrostatic configurations of the U(1) Goldstone $\phi$, and dipole Goldstone $\phi_\mu$ in the two-Goldstone formulation, satisfying
\begin{equation}
    \delta_{\hat\scK}\phi = 0
    \gray{,\qquad \delta_{\hat\scK}\phi_\mu = 0}~,
    \label{eq:hs-condition-goldstones}
\end{equation}
To wit,\footnote{This should be interpreted as a functional of only the independent components of the covariant background fields.}
\begin{align}
  \exp(-\beta_0 W_\eqb)
  = \int [\df \phi]\gray{[\df\phi_\mu]}
  \exp\Big( {-}\beta_0
  F[n_\mu,h_{\mu\nu},\tilde\xi_\mu,\tilde a_{\mu\nu}]\Big)~.
\end{align}
Here, $F = T_0\int\df\Sigma_\mu K^\mu\mathcal{F}$ is the free energy, which is a scalar under symmetry transformations.
The local equilibrium temperature $T$, fluid velocity $u^\mu$ (normalized as $n_\mu u^\mu = 1$), chemical potential $\mu$, and dipole chemical potential $\varpi_\mu$ get red-shifted in the presence of background sources as

\begin{subequations}
\begin{align}
    T = \frac{1}{n_\nu K^\nu}, \qquad 
    u^{\mu} &= \frac{K^\mu}{n_\nu K^\nu}, \qquad 
    \mu = \frac{\Lambda_K + K^\mu A_\mu}{n_\nu K^\nu}, \qquad 
    \varpi_\mu 
    = \frac{\psi_\mu^K + h_{\mu\lambda}K^\rho A^\lambda_{~\rho}}{n_\nu K^\nu}~.
\end{align}
This particular definition of the thermodynamic parameters is often referred to as the ``thermodynamic frame''; see~\cite{Jensen:2012jh,Kovtun:2022vas} for more discussion.
While $T$ and $u^\mu$ are dipole-invariant, it is easy to see that $\mu$ and $\varpi_\mu$ transform as $\mu \to \mu + u^\mu\psi_\mu$ and $\varpi_\mu \to \varpi_\mu - T\psi_\rho h_\mu^\nu\nabla_\nu (\vec u^\rho/T)$. We can instead define the dipole-invariant version of these fields using $\Psi_\mu$ introduced in \cref{eq:tilde-background}, i.e.
\begin{align}
\begin{alignedat}{3}
    \tmu
    &= \frac{\Lambda_K + K^\mu\tilde A_\mu}
    {n_\nu K^\nu}
    &&= \mu + u^\mu\Psi_\mu
    &&= \Phi
    {\color{Brown} \,-\, \frac{\delta_{\hat\scK}\phi}
    {n_\nu K^\nu}}~, \\\
    \tvarpi_\mu 
    &= \frac{
    \tilde\psi_\mu^K
    + h_{\mu\lambda}K^\rho \tilde A^\lambda_{\;\;\rho}}{n_\nu K^\nu}
    &&= \varpi_\mu - T\Psi_\rho h_\mu^\nu\nabla_\nu \frac{\vec u^\rho}{T}
    &&= \frac{h_{\mu\lambda}K^\rho
    \tilde{A}^\lambda_{\;\;\rho}}
    {n_\nu K^\nu}
    {\color{Brown}\,+\,
    \frac{h_\mu^\rho \delta_{\hat\scK}\Psi_\rho
    }
    {n_\nu K^\nu}}~.
\end{alignedat}
\end{align}
\label{eq:hydrostaticfluidfields-single}%
\end{subequations}
The brown terms above are zero using the hydrostatic condition in \cref{eq:hs-condition-goldstones}. 
We can vary $F$ similar to \cref{eq:deltaSdipoleinvariant} to obtain the \emph{hydrostatic constitutive relations}, i.e. how the conserved currents in hydrostatic equilibrium depend on the time-independent background sources. To wit
\begin{align}
\begin{split}
    \delta F
    &= -\int \df\Sigma_\rho K^\rho \bigg[
    - \tilde E^\mu_{\text{hs}}
    \delta n_\mu 
    + \lb v^{(\mu}\tilde\pi^{\nu)}_{\text{hs}}
    + \frac{1}{2}\tilde\tau^{\mu\nu}_{\text{hs}}
    \rb \delta h_{\mu\nu} 
    + Q_{\text{hs}}\,
    \delta\Phi
    + J^{\mu\nu}_{\text{hs}}
    h_{\nu\lambda}\delta \tilde{A}^{\lambda}_{~\mu}
    \bigg]~~.
    \label{eq:hs-F-1Goldstone}
\end{split}
\end{align}
Therefore, to read out the hydrostatic constitutive relations, we only need to specify the free energy density ${\cal F}$, constructed out of the U(1)+dipole-invariant versions of background fields $n_\mu$, $h_{\mu\nu}$, $\tilde\xi_\mu$, $\tilde a_{\mu\nu}$ and their derivatives, arranged order-by-order in a derivative expansion. 

\subsection{Derivative counting}
\label{sec:derivative_counting}

One of the most elusive features of dipole superfluid hydrodynamics is the underlying derivative counting scheme, which is a crucial ingredient for organizing any low-energy effective field theory. For p-wave dipole superfluids, as we outlined in~\cite{Jain:2023nbf}, the natural derivative counting scheme has dynamical exponent $z=2$, i.e. time-derivatives in the hydrodynamic equations count as double the space-derivatives $v^\mu\partial_\mu \sim {\cal O}(\partial^2)$, $h_\mu^\nu\partial_\nu \sim {\cal O}(\partial^1)$. Among other physical implications, this means that unlike ordinary fluids or superfluids, there are no propagating sound modes in p-wave dipole superfluids, nicely reflecting the immobile nature of charged excitations due to dipole symmetry. On the other hand, s-wave dipole superfluids are more subtle and can seemingly be made compatible with either dynamical exponents $z=1$ or $z=2$, depending on the point of view. We explore this ambiguity in the following.

Since a consistent derivative counting scheme is an integral part of any low-energy effective description, let us look at it carefully in the following. To begin, we associate scaling exponent $1$ with the derivative operator, i.e. $\partial_\mu \sim {\cal O}(\dow^1)$.
We associate the scaling exponent $\varepsilon \equiv z-1$ to $v^\mu$ so that $v^\mu\partial_\mu \sim {\cal O}(\partial^z)$. We are interested in describing thermodynamic systems with nonzero charge density, energy density, and pressure, so we enforce the leading order terms in $n_\mu J^\mu$, $n_\mu\epsilon^\mu$, and $\tau^{\mu\nu}$ to be ${\cal O}(\dow^0)$. In conjunction with the Ward identities \eqref{eq:physical-Wardidentity-X-inv}, this yields the counting scheme for the operators
\begin{subequations}
\begin{gather}
    \tau^{\mu\nu} \sim 
    {\cal O}(\dow^0)~, \qquad 
    J^\mu, \epsilon^\mu \sim 
    {\cal O}(\dow^{\varepsilon})~, \qquad 
    J^{\mu\nu} \sim {\cal O}(\dow^{\varepsilon-1})~, \qquad 
    \pi_\mu \sim {\cal O}(\dow^{-\varepsilon})~, \nn\\ 
    X \sim {\cal O}(\dow^{\varepsilon+1})~{\color{Gray}, \qquad 
    X^\mu \sim {\cal O}(\dow^{\varepsilon})},
\end{gather}
and for the background and Goldstone fields
\begin{gather}
    h_{\mu\nu} \sim 
    {\cal O}(\dow^0)~, \qquad 
    A_\mu,n_\mu \sim {\cal O}(\dow^{-\varepsilon})~, \qquad 
    a_{\mu\nu} \sim {\cal O}(\dow^{-\varepsilon+1}), \qquad 
    v^\mu \sim {\cal O}(\dow^{\varepsilon})~, \nn\\
    \phi \sim {\cal O}(\dow^{-\varepsilon-1})~{\color{Gray}, \qquad 
    \phi_\mu \sim {\cal O}(\dow^{-\varepsilon})}~.
\end{gather} 
Requiring the hydrodynamic fields in \cref{eq:hydrostaticfluidfields-single} to scale homogeneously, we find that 
\begin{equation}
    K^\mu \sim \mathcal{O}(\partial^{\varepsilon})~, \qquad
    \Lambda_K \sim \mathcal{O}(\partial^{0})~, \qquad 
    \psi_\mu^K \sim \mathcal{O}(\partial^{1})~,
\end{equation}
which further results in the counting scheme of the hydrodynamic fields themselves
\begin{equation}
    T, \mu \sim {\cal O}(\dow^0)~, \qquad 
    u^\mu \sim {\cal O}(\dow^{\varepsilon})~, \qquad 
    \varpi_\mu \sim \mathcal{O}(\partial^{1})~.
\end{equation}
\label{eq:scaling-cov}%
\end{subequations}
The scaling rules for the respective dipole-invariant ``tilde'' quantities follow analogously.
In non-covariant notation, these rules imply different scaling behavior or space- and time-components of various tensor structures; see~\cite{Jain:2023nbf} for relevant discussion.

For an ordinary fluid or a U(1) superfluid without dipole symmetry, we know that $\pi_\mu \sim h_{\mu\nu} u^\nu$ at leading order in derivatives, which requires us to take $\varepsilon=0$, $z=1$. For an ordinary U(1) superfluid, we can also use the fact that the effective action (or the free energy at finite temperature) contains a leading order term proportional to $\xi_\mu\xi^\mu$, leading to $J^\mu \sim \xi^\mu$ and again yielding the same counting scheme. On the other hand, the leading term in the effective action for a p-wave dipole superfluid goes as $\tilde a_{\mu\nu}\tilde a^{\mu\nu}$, resulting in $J^{\mu\nu} \sim \tilde a^{\mu\nu}$, setting $\varepsilon=1$, $z=2$ and, in particular, imposing $u^\mu \sim \cO(\dow^{1})$ and $\tilde a_{\mu\nu} \sim \cO(\dow^{0})$. As we found in~\cite{Jain:2023nbf}, there are no equilibrium states with constant nonzero fluid velocity in p-wave dipole superfluids, so it is consistent to assign the scaling ${\cal O}(\dow^1)$ to $u^\mu$. On the other hand, the system does admit equilibrium states with nonzero ``dipole superflow'' $\tilde a_{\mu\nu}\neq 0$, consistent with its $\cO(\dow^0)$ scaling. See~\cite{Jain:2023nbf} for more details.

The situation is slightly trickier for s-wave dipole superfluids. Following our discussion in \cref{sec:dipoleInvariant}, 
after the vector Goldstone $\phi_\mu$ has been taken on-shell, the leading term in the effective action for an s-wave dipole superfluid goes as $\tilde a_{\mu\nu}\tilde a^{\mu\nu}$. 
This results in $J^{\mu\nu} \sim \tilde a^{\mu\nu}$ and $J^\mu \sim 1/\ell\,\partial_\nu \tilde a^{\mu\nu}$, and implies $\varepsilon=1$, $z=2$, same as the p-wave case. This seems natural because, just like their p-wave cousins, s-wave dipole superfluids admit equilibrium states with nonzero dipole superflow $\tilde a_{\mu\nu}\neq 0$.
However, unlike the p-wave case, s-wave dipole superfluids also admit equilibrium states with constant nonzero fluid velocity $u^\mu h_\mu^\nu \neq 0$, so it is natural to set $u^\mu \sim {\cal O}(\dow^0)$, implying $\varepsilon=0$, $z=1$. The latter counting scheme is also supported by the fact that the linearized mode spectrum of s-wave dipole superfluids contains sound modes familiar from ordinary dipole-non-invariant fluid dynamics, characteristic of $z=1$. The seeming incompatibility of these schemes requires us to generalize the derivative counting to include potential anomalous scaling dimensions.

To gain some insight into the problem, consider a rescaling of various fields given as
\begin{gather}
    \epsilon^\mu \equiv \ell^{\varepsilon} \ut\epsilon^\mu~, \qquad
    \pi^\mu \equiv \ell^{-\varepsilon} \ut\pi_\mu~, \qquad
    J^\mu \equiv \ell^{\varepsilon} \ut J^\mu~, \qquad 
    J^{\mu\nu} \equiv \ell^{-1+\varepsilon} \ut J^{\mu\nu}~, \nn\\
    X \equiv \ell^{\varepsilon} \ut X~{\color{Gray}, \qquad 
    X^{\mu}\equiv \ell^{-1+\varepsilon} \ut X^\mu}~, \nn\\
    n_\mu \equiv \ell^{-\varepsilon} \ut n_\mu~, \qquad 
    v^\mu \equiv \ell^{\varepsilon} \ut v^\mu~, \qquad 
    A_\mu \equiv \ell^{-\varepsilon} \ut A_\mu~, \qquad 
    a_{\mu\nu} \equiv \ell^{1-\varepsilon} \ut a_{\mu\nu}~, \nn\\
    \phi \equiv \ell^{-\varepsilon} \ut\phi~{\color{Gray}, \qquad 
    \phi_\mu \equiv \ell^{1-\varepsilon} \ut\phi_\mu}~, \nn\\
    K^\mu \equiv \ell^{\varepsilon} \ut K^\mu~, \qquad 
    \psi_\mu^K \equiv \ell^{1} \ut\psi_\mu^K~, \qquad 
    u^\mu \equiv \ell^{\varepsilon} \ut u^\mu~, \qquad
    \varpi_\mu \equiv \ell^{1}\ut\varpi_\mu~,
    \label{eq:remove-ell}
\end{gather}
and similarly for the respective ``tilde'' quantities, for some inverse length-scale $\ell\sim{\cal O}(\dow^1)$. These scaling transformations are designed such that we have $\ut u^\mu,\ut{\tilde a}_{\mu\nu}\sim{\cal O}(\dow^0)$, as suggested by the allowed equilibrium states, together with the effective dynamical exponent $\ut z = 1$. 
The price we pay is that the dipole Ward identity in \cref{eq:physical-Wardidentity-X,eq:physical-Wardidentity-X-inv} now contains the dimensionful parameter $\ell$, i.e.
\begin{equation}
    \nabla_\mu'\ut J^{\mu\nu} = \ell\,h^\nu_\mu \ut J^\mu~,
\end{equation}
while the remaining Ward identities remain unchanged, for any choice of $\varepsilon$. What we learn from this exercise is that the Ward identities allow for an anomalous dimensionful parameter $\ell$, which is compatible with several derivative counting schemes parametrized by arbitrary $\varepsilon$.
In fact, depending on the equilibrium states under consideration, we can set up the theory of s-wave dipole superfluid hydrodynamics with any dynamical exponent. At the moment, we are not aware of any principle to fix the exponent, except that we must choose $0\leq \varepsilon \leq 1$, and thus $1\leq z\leq 2$, for the hydrodynamic derivative expansion to be well-posed. Outside this bound, one of the gauge-invariant quantities, $u^\mu\sim\cO(\dow^{\varepsilon})$ or ${\tilde a}_{\mu\nu}\sim\cO(\dow^{1-\varepsilon})$, have negative derivative ordering and thus would lead to an inconsistent derivative expansion. In this interpretation, the physical picture is that of a charged fluid deformed by two ``dangerously irrelevant" operators, $u^\mu$ and $\tilde{a}_{\mu\nu}$, whose anomalous dimensions are $\varepsilon$ and $1-\varepsilon$ respectively.

To organize the derivative expansion, then, we take an ``overcomplete'' approach that can safely apply for any $1\leq z\leq 2$. We define the ``[$n$]th derivative order'' as comprised of quantities that appear at $\cO(\dow^{n})$ for some $1\leq z\leq2$ and do not appear at lower derivative orders for any $1\leq z\leq2$. In other words, the ``[$n$]th derivative order'' is comprised of
\begin{equation}
    \cO(\dow^{[n]})
    = \Big\{
    ~\cO(\dow^n)~,~
    \cO(\dow^{n+\varepsilon})~,~
    \cO(\dow^{n+2\varepsilon})~,~ \ldots~,~ 
    \cO(\dow^{n+(1-\varepsilon)})~,~ 
    \cO(\dow^{n+2(1-\varepsilon)})~,~ 
    \ldots~
    \Big\}.
\end{equation}
These are shaded in the following tabular representation
\begin{center}
    \begin{tabular}{ccccccc}
        \cline{1-6}
         \mc{0} & $\varepsilon$ & $\cdots$
         & $n\varepsilon$
         & $(n+1)\varepsilon$
         & $\ldots$
         & [$0$] \\ \cline{2-6}
         \mc{$1 - \varepsilon$}
         & \mc{1} & $\cdots$ &
         $1+(n-1)\varepsilon$
         & $1+n\varepsilon$ & $\cdots$
         & [$1$] \\ \cline{3-6}
         \mc{$\vdots$} & \mc{$\vdots$} & \mc{$\ddots$}
         & $\vdots$ & $\vdots$ & $\vdots$ \\ \hhline{~~~|*3-}
         \mc{$n-n\varepsilon$}
         & \mc{$n-(n-1)\varepsilon$}
         & \mc{$\cdots$}
         & \mc{\cellcolor{RoyalBlue!25} $n$}
         & \cellcolor{RoyalBlue!25} $n+\varepsilon$ 
         & \cellcolor{RoyalBlue!25} $\cdots$
         & [$n$] \\
         \cline{5-6}
         \mc{$(n+1)-(n+1)\varepsilon$}
         & \mc{$(n+1)-n\varepsilon$}
         & \mc{$\cdots$}
         & \mc{\cellcolor{RoyalBlue!25} 
         $n+1-\varepsilon$}
         & \mc{$n+1$} & $\cdots$ \\ \cline{6-6}
         \mc{$\vdots$} & 
         \mc{$\vdots$} & \mc{$\vdots$}
         & \mc{\cellcolor{RoyalBlue!25}
         $\vdots$}
         & \mc{$\vdots$} & \mc{$\ddots$} \\
         $[0]$ & $[1]$ & & $[n]$
    \end{tabular}
\end{center}
We use square brackets in the notation ``[$n$]th derivative order'' to avoid confusion with the naive $\cO(\dow^n)$ counting.
The $n$th row in the table above is $\cO(\dow^n)$ in $z=1$ scheme, while the $n$th column is $\cO(\dow^n)$ in $z=2$ scheme. Let us also note that after integrating out the dipole Goldstone $\phi_\mu$, neither terms like $h^{\mu\nu}{\tilde{A}}_\mu{\tilde{A}}_\nu \sim \mathcal{O}(\partial^{-2\epsilon})$ nor terms involving ${n}_\mu$ without explicit derivatives can appear in the effective action. Hence from an algorithmic point of view, we can think of $\cO(\dow^{n+a\varepsilon})$ terms in the ``rows'' as arising from $n$-explicit covariant derivatives acting on the products of $a$ instances of ${v}^\mu$ and ${u}^\mu$. Similarly, we can think of the $\cO(\dow^{n+b(1-\varepsilon)})$ terms in the ``columns'' as $n$ explicit covariant derivatives acting on the products of $b$ instances of ${\tilde{a}}_{\mu\nu}$, $F_{\mu\nu}$, and $F^n_{\mu\nu}$. Finally, the diagonal terms arise from $n$-explicit covariant derivatives acting on $h_{\mu\nu}$, $\mu$, or $T$. This scheme is overcomplete, i.e. were we to continue down the $n$th column or across the $n$th row \emph{ad infinitum}, we would include terms with derivative ordering greater than or equal to $\mathcal{O}(\partial^{n+1})$. However, this is acceptable as we are sure to capture all terms with derivative ordering less than $\mathcal{O}(\partial^{n+1})$, so that we can trust our results at least to that order. Furthermore, we will mainly consider s-wave dipole superfluids at linear order in fluctuations, in which case only the first few entries around the diagonal will be relevant.

\subsection{Ideal order}
\label{sec:ideal}

At ideal order, the hydrostatic constitutive relations of an s-wave dipole superfluid are characterized by the most generic free energy density ${\cal F}_{\text{ideal}}$ constructed out of [0]th order scalars in the theory. For simplicity, we shall restrict ourselves to terms in $\cal F_{\text{ideal}} $ that are parity-preserving and at most quadratic in $\tilde a_{\mu\nu}$ and $F^n_{\mu\nu}$, which are sufficient to determine the hydrostatic constitutive relations that are at most linear in these fields. Under these assumptions, following the derivative counting scheme outlined in \cref{sec:derivative_counting}, the admissible [0]th order scalars are
\begin{align}
\begin{split}
    \cO(\dow^0):&\qquad 
    T, \quad \tmu, \\
    \cO(\dow^{2\varepsilon}):&\qquad 
    {\vec u}^2 = {u}_\mu{u}^\mu = \ell^{2\varepsilon}\vec{\ut u}^2, \\
    \cO(\dow^{1-\varepsilon}):&\qquad 
    \tr\,{\tilde{a}} = {\tilde a}^\mu_{~\mu} = \ell^{1-\varepsilon}\,\tr\,\tilde{\ut a}, \\
    \cO(\dow^{2-2\varepsilon}):&\qquad 
    {\tilde a}^2 = {\tilde a}_{\mu\nu}
    {\tilde a}^{\mu\nu} = \ell^{2(1-\varepsilon)}\tilde{\ut a}^2, \quad 
    F_n^2 = F^n_{\mu\nu}F_n^{\mu\nu} 
    = \ell^{2(1-\varepsilon)}\ut F_n^2.
\end{split}
\end{align}
Note that there are no $\cO(\dow^\varepsilon)$ scalars in the theory. Higher $\cO(\dow^{\#\varepsilon})$ terms can be obtained by taking higher powers of ${\vec u}^2$. On the other hand, all higher $\cO(\dow^{\#(1-\varepsilon)})$ terms are non-linear in ${\tilde a}_{\mu\nu}$ and $F^n_{\mu\nu}$ and thus have been ignored. In a conventional dipole-non-invariant superfluid, the superfluid velocity $\xi_\mu$ also appears at ideal order via $\xi^2 = h_{\mu\nu}\xi^\mu\xi^\nu$; see e.g. \cite{Bhattacharya:2011eea, Bhattacharyya:2012xi, Jain:2016rlz, Gouteraux:2022qix}. However, due to invariance under dipole symmetry, only the $v^\mu$-contraction of $\xi_\mu$ is allowed and is exactly equal to $\tmu$ in equilibrium. Note also that, unlike p-wave dipole superfluids, we have no terms involving $\tilde F_{\mu\nu}$ because it is equal to $\tmu F^n_{\mu\nu} + 2n_{[\nu}\dow_{\mu]}\tmu$ for the s-wave case in equilibrium.
In the end, we can write down the equilibrium free energy density at ideal order up to quadratic order in fluctuations 
\begin{align}
\begin{split}
    \mathcal{F}_{\text{ideal}} 
    &= - p_f - \frac{p_d}{2}\tr\, {\tilde{a}}
    + \frac{B_d}{8}\tr^2{\tilde{a}} 
    + \frac{G_d}{4}\left({\tilde{a}}^2 
    - \frac{1}{d}\tr^2{\tilde{a}}\right) 
    + \frac{\chi_n}{4}{F}_n^2~, \\
    &= - p_f
    - \frac{\ell^{1-\varepsilon}}{2} p_d\,\tr\,\tilde{\ut a}
    + \frac{\ell^{2-2\varepsilon}}{4} 
    \lb 
    \half B_d\,\tr^2\tilde{\ut a}
    + G_d \lb \tilde{\ut a}^2 - \frac1d \tr^2\tilde{\ut a} \rb \rb
    + \frac{\ell^{-2\varepsilon}}{4} \chi_n \ut F^2_n
    ~.
    \label{eq:ideal-free}
\end{split}
\end{align}
Here $p_f$ is identified as the thermodynamic pressure of the fluid, specified by the equation of state as some function of $T$, $\tmu$, and $\vec u^2$. On the other hand, $p_{d}$, $B_d$, $G_d$ and $\chi_n$ are arbitrary functions of $T$ and $\tmu$, but not of $\vec u^2$. One may check that their respective potential dependence on $\vec u^2$ gets pushed to [1]st and [2]nd derivative orders. 
The thermodynamic derivatives of $p_f$ can be used to define the thermodynamic entropy density $s_f$, charge density $q_f$,  kinetic mass density $\rho_f$, and energy density $\epsilon_f$ according to
\begin{subequations}
\begin{align}
\begin{split}
    \df p_f
    &= s_f \df T + q_f \df \tmu
    + \frac{1}{2}
    \rho_f\df\vec u^2, \\
    \epsilon_f
    &= Ts_f + \tmu q_f + \rho_f\vec u^2 - p_f,
\end{split}
\end{align}
which are all functions of $T$, $\tmu$, and $\vec u^2$ as well.
Similarly, we define $s_d$, $q_d$, and $\epsilon_d$ using $p_d$, i.e.
\begin{align}
    \begin{split}
        \df p_d &= s_d \df T + q_d\df \tmu \\
        \epsilon_d &= Ts_d + \tmu q_d  - p_d,
    \end{split}
\end{align}
\label{eq:thermo}%
\end{subequations}
which are only functions of $T$ and $\tmu$. In the following, it will also be helpful to define a ``total isotropic thermodynamic pressure'' including the dipole contribution 
\begin{equation}
\begin{split}
    p = p_f + \frac{p_d}{2}\tr\, {\tilde{a}}
    - \frac{B_d}{8}\tr^2{\tilde{a}}
    ~,
    \label{eq:total-pressure}
\end{split}
\end{equation}
and similarly $s$, $q$, $\rho$ analogous to \cref{eq:thermo}. Note that $\rho = \rho_f$ because $p_d$, $B_d$ are independent of $\vec u^2$, but we shall update this definition at [1]st order in \cref{sec:hydrostaticCorrections}.

Using \cref{eq:hs-F-1Goldstone}, and noting the relation between the variations of $\tilde A^\lambda_{~\mu}$ and $\tilde a_{\mu\nu}$ from \cref{foot:aAtrans}, we can obtain the constitutive relations for an ideal s-wave dipole superfluid 
\begin{align}
\begin{split}
    \tilde{E}^\mu_{\text{ideal}} 
    &= \lb \epsilon - \tmu q
    \rb u^\mu 
    + p\, \vec u^\mu 
    + p_d \, v^\rho \tilde F_\rho{}^\mu 
    + \nabla'_\nu (\chi_n F^{\mu\nu}_n)~, \\
    \tilde{\pi}_{\text{ideal}}^\mu 
    &= \rho \, \vec{u}^\mu~, \\
    \tilde{\tau}^{\mu\nu}_{\text{ideal}} 
    &= p\, h^{\mu\nu}
    + \rho \, \vec u^\mu\vec u^\nu ~,\\
    Q_{\text{ideal}} 
    &= q~, \\
    J^{\mu\nu}_{\text{ideal}} 
    &= p_d\, h^{\mu\nu}
    - \half B_d\,\tr\,\tilde a\, h^{\mu\nu}
    - G_d\, \tilde a^{\langle\mu\nu\rangle}~,
\end{split}
\label{eq:ideal-consti}
\end{align} 
where the angular brackets in $\tilde a^{\langle\mu\nu\rangle}$ denotes taking the symmetric and traceless part of a tensor with respect to the indices. Here we have not included the terms that are non-linear in $\tilde a_{\mu\nu}$ and $F^n_{\mu\nu}$, though they can be obtained in the supplementary Mathematica notebooks included with the article's submission. However, note that these constitutive relations are not generically complete with regards to such non-linearities because we did not include all the non-linear dependence on $\tilde a_{\mu\nu}$ and $F^n_{\mu\nu}$ in the free energy density ${\cal F}_{\text{ideal}}$ in \cref{eq:ideal-free}. The dipole-invariant energy current and charge current are given as (again, without non-linear terms)
\begin{align}
\begin{split}
    \tilde\epsilon^\mu_{\text{ideal}}
    &= \epsilon\, u^\mu 
    + \lb p -\tmu q \rb \vec u^\mu 
    + J^{\mu\nu}_{\text{ideal}}\, v^\rho \tilde F_{\rho\nu}
    + \nabla'_\nu (\chi_n F^{\mu\nu}_n)
    + \tmu\nabla'_\nu J^{\mu\nu}_{\text{ideal}}~, \\
    J^\mu_{\text{ideal}}
    &= q\, v^\mu 
    + \nabla_\nu J^{\mu\nu}_{\text{ideal}}~.
\end{split}
\end{align}
In particular, note that the fluid velocity does not carry any charge flow, as we would expect in a dipole-invariant theory. However, it does carry energy and momentum flow.

The configuration equation for $\phi$ is given by the U(1) conservation equation
\begin{equation}
\label{U1conservation}
    \nabla'_\mu\!
    \lb q\,\vec u^\mu \rb
    =
    \nabla'_\mu\nabla'_\nu 
    \!\lb p_d\, h^{\mu\nu}
    - \frac{1}{2} B_d\,
    \tr\,\tilde a\, h^{\mu\nu}
    - G_d\,
    \tilde a^{\langle\mu\nu\rangle}
    \rb.
\end{equation}
Here we have used the fact that $v^\mu = TK^\mu - \vec u^\mu$ and $\nabla'_\mu(TQ K^\mu) = 0$ in equilibrium. 

\paragraph*{Equilibrium states:}

Let us specialize to a flat spacetime background in the absence of background sources, and choose the isometry 
\begin{equation}
    K^\mu = \frac{\delta^\mu_t + \delta^\mu_i u^i_0}{T_0}, \qquad 
    \Lambda_K = \frac{\mu_0 - \varpi^{0}_i x^i}{T_0}, \qquad 
    \psi^K_\mu = \delta_\mu^i \frac{\varpi^0_i}{T_0}~,
\end{equation}
where $T_0$, $u^i_0$, $\mu_0$, and $\varpi^0_i$ are constants.
Note that we are allowed to have a spatially linear profile of $\Lambda_K$ because of the dipole shift term in the isometry of $A_\mu$ in \cref{eq:eqb-isometry}. Furthermore, the equilibrium condition $\lie_K \phi = \Lambda_K$ reduces to $\dow_t\phi + u^i_0\dow_i\phi = \mu_0 - \varpi^0_i x^i$. This means that on this background
\begin{equation}
    T = T_0, \qquad 
    u^i = u^i_0, \qquad 
    \tmu = \mu_0 - \ell\varpi^0_i x^i - u^i_0 \dow_i\phi, \qquad 
    \tvarpi_i = \varpi^0_i.
\end{equation}
In particular, $\tmu$ is generically not a constant when $u^i_0\dow_i\phi\neq 0$. 
The non-linear configuration equation in this case reads
\begin{equation}
    u^i_0\dow_i q
    =
    \dow_k\!\lb q_d\, \dow^k \mu \rb
    + \dow_i\dow_j
    \Big(
    B_d\,\delta^{ij}\,\dow_k\dow^k\phi 
    + 2 G_d\,
    \dow^{\langle i}\dow^{j\rangle}\phi
    \Big).
\end{equation}
where the coefficients $q$, $q_d$, $B_d$, and $G_d$ may depend on the non-constant $\mu$. The theory admits the general equilibrium solution
\begin{equation}
    \langle\phi\rangle 
    = \mu_0 t
    + \phi_0
    - \phi^0_{i}
    (x^i - u^i_0 t)
    - \frac{1}{2} \xi^0_{ij} x^i x^j~,
    \label{eq:eqb-state}
\end{equation}
where $\phi_0$, $\phi^0_i$, and $\xi^0_{ij}$ are constants such that $\xi^0_{ij} u^i_0 = \varpi^j_0$. In particular, unlike the p-wave case, this equation admits equilibrium solutions with nonzero $u^i_0$. 
We can use this to find
\begin{align}
    \langle\tmu\rangle &= \mu_0 + \phi^0_{i} u^i_0~, \nn\\
    \langle\tilde a_{ij}\rangle &= 2\xi^0_{ij}~, 
\end{align}
which ultimately yields the equilibrium conserved currents
\begin{align}
\begin{split}
    \langle\epsilon^t_\ideal\rangle 
    &= \epsilon ~, \\ 
    \langle\epsilon^i_\ideal\rangle
    &= \lb \epsilon + p -\tmu q \rb u^i_0~, \\ 
    \langle\pi^i_\ideal\rangle
    &=  q \left(\phi_0^i + \xi^{ij}_{0}x_j \right)
    + \rho\, u^i_0~, \\
    \langle \tau^{ij}_\ideal \rangle 
    &= p\, \delta^{ij}
    + \rho \, u^i_0 u^j_0 
    + \langle J^{ij}_\ideal \rangle\, \tr\, \xi_0 ~, \\
    \langle Q_\ideal \rangle 
    &= q~, \\
    \langle J^i_\ideal \rangle 
    &= 0~, \\
    \langle J^{ij}_\ideal \rangle 
    &= \lb p_d
    - B_d\,\tr\,\xi_0 
    \rb \delta^{ij}
    - 2G_d\, \xi^{\langle ij\rangle}_0 ~.
\end{split}
\end{align}
All the coefficients are understood to be evaluated on the equilibrium configuration.
We can identify $\xi^0_{ij}$ as the equilibrium dipole superflow. 
By restricting to linear constitutive relations with respect to $\tilde a_{\mu\nu}$, we essentially assumed that the equilibrium state has either zero or small dipole superflow.

The equilibrium states are physically characterized by the expectation values of the conserved energy, momentum, and charge densities in equilibrium.  For an ordinary dipole-non-invariant (super)fluid, we can exchange these thermodynamic parameters with the respective chemical potentials in equilibrium $T_0$, $u^i_0$, and $\mu_0$. However, for a dipole superfluid, the homogeneous part of the momentum density splits into the fluid contribution $\rho u^i_0$ and the dipole contribution $q\phi^i_0$, meaning that the fluid is allowed to transition into a state with different $u^i_0$ while conserving its momentum. Assuming the kinetic mass density $\rho$ to be constant, the canonical free energy density of the fluid varies with $\rho^2\vec u^2_0$ as 
\begin{equation}
    \frac{\delta}{\delta(\rho^2\vec u^2)}
    \langle{\cal F}_{\text{ideal}}
    + \mu J^t_{\text{ideal}}
    + u^i \pi_{i\,\text{ideal}}
    \rangle
    =  \frac{1}{\rho}
    \geq 0~.
    \label{eq:freeE-velocity}
\end{equation}
This means that while s-wave dipole superfluids admit equilibrium states with nonzero $u^i_0$, these are thermodynamically unstable and the fluid would prefer into transition to a state with zero $u^i_0$ while dumping all its momentum into $\phi^i_0$. 

\subsection{Derivative corrections}
\label{sec:hydrostaticCorrections}

The derivative corrections to the hydrostatic constitutive relations of an s-wave dipole superfluid can be characterized by the admissible derivative corrections to the free energy density ${\cal F}$. We can generically expand the free energy density order-by-order in the ``over-complete'' derivative expansion as
\begin{equation}
    {\cal F} = {\cal F}_{\text{ideal}}
    + {\cal F}_{[1]}
    + {\cal F}_{[2]}
    + \ldots~,
\end{equation}
where ${\cal F}_{[n]}$ contains all scalars appearing at $[n]$th order in derivatives. For example at [1]-derivative order, the following new scalars can appear in the free energy density
\begin{align}
\begin{split}
   \cO(\dow^{1+\varepsilon})&:\qquad 
   \vec{u}^2\,\tr\,\tilde{a}, \qquad \vec{u}^\mu\partial_\mu \tmu,\quad  
   \vec{u}^\mu\partial_\mu T, \quad 
   \vec{u}^\mu \vec{u}^\nu \tilde{a}_{\mu\nu}, \quad 
   \nabla'_\mu \vec u^\mu~,\\
    \cO(\partial^{1+3\varepsilon})
    &:\qquad \vec{u}^\mu\partial_\mu \vec u^2~.
\end{split}
\label{eq:O1scalars}
\end{align}
We can multiply these terms with $\vec u^2$ to construct higher $\cO(\dow^{1+\#\varepsilon})$ terms. For parity preserving theories, there are no terms of the kind $\cO(\dow^{1+\#(1-\varepsilon)})$. We can omit the total derivative scalar $\nabla'_\mu \vec u^\mu$ from the counting above because it can always be removed from the free energy density by performing integration by parts. Note that $\vec{u}^\mu\tvarpi_{\mu} = T\vec{u}^\mu\partial_\mu (\tmu/T) + \vec{u}^\mu \vec{u}^\nu \tilde{a}_{\mu\nu}/2$, and thus is already accounted for via linear combinations of the scalars in \cref{eq:O1scalars}. This allows us to parametrize the correction to the free energy density as\footnote{The labels on $\lambda_{2,3,4}$ are chosen to coincide with \cite{Jain:2023nbf}.}
\begin{align}
    \mathcal{F}_{[1]} = - \biggl(
    \frac{1}{4} R_d\vec{u}^2 \,\tr\,\tilde{a}
    + \frac{1}{2} M_d\vec{u}^\mu \vec{u}^\nu \tilde{a}_{\langle\mu\nu\rangle}
    + \lambda_2 \vec{u}^\nu\partial_\nu T 
    + \lambda_3 \vec{u}^\nu\partial_\nu \tmu
    + \frac{1}{2}\lambda_4 \vec{u}^\nu\partial_\nu \vec{u}^2
    \biggr)~,
\end{align}
where the coefficients $R_d$, $M_d$, and $\lambda_{2,3,4}$ are allowed to be functions of $\tmu$, $T$ and $\vec{u}^2$, characterizing all possible [1]st order corrections. These result in the hydrostatic constitutive relations
\begin{align}
\begin{split}
    \tilde{E}^\mu_{\hs[1]}
    &= \left(\frac{R_d}{4}\vec{u}^2\tr\,\tilde{a}-\frac{M_d}{2} \vec{u}^\nu\vec{u}^\lambda\tilde{a}_{\langle\nu\lambda\rangle}\right)v^\mu-\nabla_\nu'\left(\left[\lambda_2 T + \lambda_4 \vec{u}^2\right]\vec{u}^\nu\right)u^\mu-\left(\frac{R_d}{2}-\frac{M_d}{d}\right)\tilde{F}^\mu_{\;\;\nu}v^\nu\\
    &\qquad+\left(u^\nu\partial_\nu\left[\lambda_2 T + \lambda_3 \mu + \frac{\lambda_4}{2}\vec{u}^2\right]+\frac{R_d}{2}\vec{u}^2\tr\,\tilde{a} - M_d \vec{u}^\nu\vec{u}^\lambda\tilde{a}_{\langle \nu \lambda \rangle} - M_d \vec{u}^\lambda \tilde{F}_{\lambda\nu}v^\nu \right)\vec{u}^\mu, \\
    \tilde{\pi}^\mu_{\hs[1]}
    &= \partial^\mu\left(\lambda_2 T + \lambda_3  \mu + \frac{\lambda_4}{2} \vec{u}^2\right) + \left(\left[\frac{R_d}{2}-\frac{M_d}{d}\right]\tr\,\tilde{a} -\nabla_\nu' [\lambda_4 \vec{u}^\nu]\right)\vec{u}^\mu + M_d \, \tilde{a}^\mu_{\;\;\nu}\vec{u}^\nu~,
    \\
    \tilde{\tau}^{\mu\nu}_{\hs[1]}
    &= \left(\vec{u}^\rho\partial_\rho\left[\lambda_2 T + \lambda_3 \mu + \frac{\lambda_4}{2}\vec{u}^2\right] + \frac{R_d}{4} \vec{u}^2\tr\,\tilde{a} + \frac{M_d}{2}\left[\vec{u}^\lambda\vec{u}^\rho\tilde{a}_{\langle\lambda\rho\rangle}\right]\right) h^{\mu\nu}+\frac{R_d}{2}\tr\, \tilde{a}\,\vec{u}^\mu\vec{u}^\nu  \\
    &\qquad+M_d\left(\tilde{A}^{(\mu}_{\;\;\lambda}\vec{u}^{\langle \nu)}\vec{u}^{\lambda\rangle}-\frac{\tr\,\tilde{a}}{2}\right)\vec{u}^{\langle \mu}\vec{u}^{\nu\rangle} + \frac{M_d}{2} \vec{u}^2 \tilde{a}^{\langle \mu \nu\rangle}~,  \\
    Q_{\hs[1]} 
    &= -\nabla_\mu'(\lambda_3\vec{u}^\mu)~, \\
    J^{\mu\nu}_{\hs[1]}
    &= \frac{R_d}{2}\vec{u}^2 h^{\mu\nu} + M_d \vec{u}^{\langle \mu}\vec{u}^{\nu \rangle}~.
\end{split}
\end{align}
Here, for simplicity, we have treated the transport coefficients $R_d, M_d, \lambda_{2,3,4}$ as constants. In general, they can be functions of $\mu, T,$ and $\vec{u}^2$, and the ensuing contributions to the constitutive relations are straight-forward to obtain. 

Note that the coefficient $R_d$ can be understood as giving rise to a dipole correction to the kinetic mass density $\rho$. To wit, we can redefine $p_d$ in \cref{eq:ideal-consti} to $p_d^{\text{ideal}} + \half R_d\vec u^2$, which now depends on $\vec u^2$ as well. We can similarly update the definitions of $s_d$, $q_d$, and $\epsilon_d$ as well along the lines of \cref{eq:thermo}, and in addition define a dipole contribution to kinetic mass density $\rho_d = R_d + \vec u^2\dow_{\vec u^2}R_d$. The $R_d$ coefficient then also appears implicitly in $p$ via \cref{eq:total-pressure}, and accordingly in $\epsilon$, $q$, $s$, and $\rho$.

At [2]nd derivative order, we have many terms of both kinds, $\mathcal{O}(\partial^{2 + \#\varepsilon})$ and $\mathcal{O}(\partial^{2 + \#(1-\varepsilon)})$. For instance, considering just the $\mathcal{O}(\partial^2)$ scalars, we have
\begin{align}
\begin{split}
    \cO(\dow^2):&\qquad  
    \nabla'_\mu\nabla^\mu T, \quad 
    \nabla'_\mu\nabla^\mu \mu, \quad \nabla_\mu T \nabla^\mu \tmu, \quad 
    R = R^\lambda_{~\mu\lambda\nu}h^{\mu\nu}, \quad 
    \vec u^\mu \tilde a_{\mu\nu} \tilde a^{\nu\rho}
    \vec u_\rho, \quad \vec{u}^2\tr\,\tilde{a}\\
    &\qquad\vec{u}_\rho\tilde{a}^{\rho\mu}\partial_\mu T, \quad \vec{u}_\rho\tilde{a}^{\rho\mu}\partial_\mu\tmu,\quad \tilde{a}^{\mu\nu}\nabla_{(\mu}\vec{u}_{\nu)},\quad \vec{u}^\mu\partial_\mu (\tr\, \tilde{a}), \quad (F^n)^{\mu\nu}\nabla_{[\mu}\vec{u}_{\nu]}\\
    &\qquad \vec{u}_\rho (F^n)^{\rho\nu}\partial_\nu T, \quad \vec{u}_\rho (F^n)^{\rho\nu}\partial_\nu \mu, \quad v^\mu (F^n)_{\mu\rho}(F^n)^\rho_{\;\;\nu}v^\nu,  \quad \vec{u}^\mu (F^n)_{\mu\rho}(F^n)^\rho_{\;\;\nu}v^\nu, \\
    &\qquad v^\mu(F^n)_{\mu\rho}\tilde{a}^{\rho\nu}\vec{u}_\nu, \quad  \nabla_\mu'(\tilde{a}^{\mu\rho}\vec{u}_\rho) .
   \label{eq:O2scalars}
\end{split}
\end{align}
Here, we used the equilibrium condition $T\partial_\mu(1/T) - F^n_{\mu\nu}u^\nu = 0$ to eliminate many terms. Furthermore, as was the case at [1]-derivative order, the total derivative term in the last line can be written in terms of the others using integration by parts. Due to the proliferation of tensor structures and for ease of reading, we will not write down the complete set of $[2]$-derivative order constitutive relations. As an illustration, however, let us look at the corrections arising from 
\begin{align}
    \mathcal{F}_{[2]} = - \frac{1}{2}\Xi_d\tvarpi^\mu \tvarpi_\mu~,
\end{align}
which consists of many $\mathcal{O}(\partial^2)$ scalars from \cref{eq:O2scalars}. We will take $\Xi_d$, which can be understood as the susceptibility for dipole charge, to be a constant, i.e. $\partial_T\Xi_d = \partial_{\tmu} \Xi_d = \partial_{\vec{u}^2}\Xi_d = 0$. This generates the constitutive relations 
\begin{align}
\begin{split}
    \tilde{E}^\mu_{\hs[2]}
    &= \Xi_d \tvarpi^2 u^\mu 
    - \frac{1}{2} \Xi_d (\tvarpi^\lambda \tvarpi_\lambda) v^\mu-{\color{gray} \Phi X^\mu},
    \\
    \tilde{\pi}^\mu_{\hs[2]}
    &= \Xi_d \tilde{A}^\mu_{\;\;\lambda}\tvarpi^\lambda,
    \\
    \tilde{\tau}^{\mu\nu}_{\hs[2]}
    &= \Xi_d \vec{u}^{(\mu}\tilde{A}^{\nu)}_{\;\;\lambda}\tvarpi^\lambda 
    + \frac{1}{2}\Xi_d (\tvarpi^\lambda\tvarpi_\lambda)h^{\mu\nu}
    - \Xi_d v^\lambda\tilde{F}_\lambda^{\;\;(\mu}\tvarpi^{\nu)}
    ,
    \\
    Q_{\hs[2]} &= \Xi_d\nabla_\mu'(\tilde{A}^\mu_{\;\;\lambda}u^\lambda)
    - \half \Xi_d\tvarpi^\mu F^n_{\mu\nu}(u^\nu+v^\nu) ,
    \\
    J^{\mu\nu}_{\hs[2]} &= \Xi_d \tvarpi^{(\mu}\vec{u}^{\nu)}~.
\end{split}
\end{align}
The generalization to include more non-trivial higher-derivative corrections is straightforward.

\subsection{Discrete symmetries}

 \begin{table}[t]
    \centering
    \begin{tabular}{c|ccc}
         & C & P & T  \\
         \hline 
         $\epsilon^t$, $n_t$, $T$ & $+$ & $+$ & $+$ \\ 
         $\epsilon^i$, $n_i$ & $+$ & $-$ & $-$ \\ 
         $\pi_i$, $v^i$, $u^i$ & $+$ & $-$ & $-$ \\ 
         $\tau^{ij}$, $h_{ij}$ & $+$ & $+$ & $+$ \\ 
         $J^t$, $A_t$, $\mu$ & $-$ & $+$ & $+$ \\ 
         $J^i$, $A_i$ & $-$ & $-$ & $-$ \\ 
         $J^{ij}$, $a_{ij}$, $\phi$ & $-$ & $+$ & $-$
    \end{tabular}
    \caption{Transformation properties of the independent components of symmetry currents under charge conjugation (C), parity (P), and time-reversal (T) discrete transformations. \label{tab:CPT}}
\end{table}

Depending on the physical application in mind, the s-wave dipole superfluid may also possess discrete charge conjugation (C), parity (P), and/or time-reversal (T) symmetries that can impose additional constraints on the constitutive relations. We have listed the transformation properties of  various objects under these discrete symmetries in \cref{tab:CPT}. We have already assumed P symmetry while deriving the hydrostatic constitutive relations above. Furthermore, for instance, if the fluid exhibits T symmetry, the coefficients $p_d$, $M_d$, and their thermodynamic derivatives such as $s_d, \rho_d, q_d$ must vanish.

\section{Hydrodynamics}
\label{sec:dynamics}

Hydrodynamics describes systems near thermal equilibrium, where the relevant physical observables are the conserved currents that slowly relax back to equilibrium. Since conserved quantities cannot be created or destroyed, their equilibration can only occur through transport and takes much longer time compared to non-conserved quantities. Subregions of a hydrodynamic system whose size is greater than the microscopic mean free path are expected to thermalise at timescales much longer than the microscopic collision times.\footnote{This isn't the case when the system is near a finite-temperature critical point, in which case the relaxation time of the non-conserved quantities can be much longer than in typical cases.} These subregions constitute the fluid elements, wherein the conserved currents can be parametrized as a gradient expansion in terms of the background fields and some suitably defined local hydrodynamic variables. 
At each derivative order, the conservation equations can be solved for the hydrodynamic variables in terms of the background fields, yielding the dynamics of the conserved currents. 

\subsection{Hydrodynamic variables and the Second Law}
\label{sec:secondLaw}

One is generally free to choose the hydrodynamic variables to solve for using the conservation equations, however one natural choice is to introduce a set of symmetry parameters $\hat\scB = \{\beta^\mu,\Lambda_\beta, \psi^\beta_\mu\}$. These variables constitute precisely the right number of degrees of freedom corresponding to the energy-momentum, charge, and dipole conservation equations in \cref{eq:physical-Wardidentity-X}. Physically, we can understand $\hat\scB$ as a non-equilibrium generalisation of the thermal isometry $\hat\scK=\{K^\mu, \Lambda_K, \psi^K_\mu\}$ introduced \cref{sec:hydrostatics}, defined such that $\hat\scB\to\hat\scK$ in equilibrium. The fields $\hat\scB$ are required to transform in the same way as $\hat\scK$ as given in \cref{eq:trans-data}. We can also use $\hat\scB$ to generalise the definitions of hydrodynamic variables in \cref{eq:hydrostaticfluidfields-single} to the respective non-equilibrium versions
\begin{subequations}
\begin{align}
    T = \frac{1}{n_\nu\beta^\nu}, \qquad 
    u^{\mu} &= \frac{\beta^\mu}{n_\nu \beta^\nu}, \qquad 
    \mu = \frac{\Lambda_\beta + \beta^\mu A_\mu}{n_\nu \beta^\nu}, \qquad 
    \varpi_\mu 
    = \frac{\psi_\mu^\beta + h_{\mu\lambda}\beta^\rho A^\lambda_{~\rho}}{n_\nu \beta^\nu}~,
\end{align}
together with
\begin{align}
\begin{alignedat}{3}
    \tmu
    &= \frac{\Lambda_\beta + \beta^\mu\tilde A_\mu}
    {n_\nu \beta^\nu}
    &&= \mu + u^\mu\Psi_\mu
    &&= \Phi - \frac{\delta_{\hat\scB}\phi}
    {n_\nu\beta^\nu}~, \\\
    \tvarpi_\mu 
    &= \frac{
    \tilde\psi_\mu^\beta
    + h_{\mu\lambda}\beta^\rho \tilde A^\lambda_{\;\;\rho}}{n_\nu \beta^\nu}
    &&= \varpi_\mu - T\Psi_\rho h_\mu^\nu\nabla_\nu \frac{\vec u^\rho}{T}
    &&= \frac{h_{\mu\lambda}\beta^\rho
    \tilde{A}^\lambda_{\;\;\rho}
    + h_\mu^\rho \delta_{\hat\scB}\Psi_\rho}
    {n_\nu \beta^\nu}~.
\end{alignedat}
\end{align}
\label{hydro variables}%
\end{subequations}
Note that $\delta_{\hat\scB}(\ldots)\neq 0$ outside equilibrium, therefore the second equalities in these equations are different from their respective hydrostatic versions in \cref{eq:hydrostaticfluidfields-single}.  To close the system of equations, we also need the equations of motion for the Goldstones
\begin{equation}
    X = 0~
    \gray{,\qquad X^\mu = 0}~.
\end{equation}
Out of equilibrium, we no longer have the luxury to use a free energy $F$ to obtain $X$ \gray{and $X^\mu$}. These will instead be fixed by imposing a local Second Law of thermodynamics.

\paragraph*{Second Law and the adiabaticity equation:}

At late times, the hydrodynamic description is valid provided that the gradients of the thermodynamic variables are small compared to the inverse mean free path. Over time, the fluid elements exchange heat, momenta, and conserved charges among themselves and ultimately reaching global thermodynamic equilibrium as these gradients relax. Local hydrostatic equilibrium is constrained by the First Law of thermodynamics, which, as we have seen, can be captured by a hydrostatic partition function constructed using the gradients of background fields and dynamical Goldstone fields. However, to describe the relaxation to a global equilibrium state, we must also invoke the Second Law of thermodynamics, which postulates the existence of an entropy current $s^\mu$ with non-negative divergence
\begin{align}
    \label{eq:second-law-definition}
    \nabla_\mu' s^\mu = \Delta \geq 0,
\end{align}
for all solutions of the conservation equations keeping the Goldstone fields off-shell.
In particular, we require that the entropy production rate $\Delta = 0$ in hydrostatic equilibrium.

Traditionally, the entropy current is constructed order-by-order in derivatives using a combination of the First Law and the conservation equations; see for instance~\cite{landaubook}. In the modern treatments of hydrodynamics, see e.g.~\cite{Liu1972MethodOL,Loganayagam:2011mu,Haehl:2015pja}, the Second Law can equivalently be formulated as an off-shell statement by including combinations of the conservation equations \eqref{eq:physical-Wardidentity-X} into the inequality, i.e.
\begin{align} 
\label{eq:Off-shell-2nd}
\begin{split}
    &\nabla_\mu' s^\mu 
    - \beta^\rho n_\rho
    \Big[\nabla_\mu' \epsilon^\mu + \ldots \Big]
    + \beta^\rho h_{\rho\nu}
    \Big[ \nabla'_\mu\!
    \lb v^\mu \pi^\nu + \tau^{\mu\nu} + 
    \tau_d^{\mu\nu}\rb + \ldots \Big] \\
    &\qquad
    + (\Lambda_\beta + \beta^\lambda A_\lambda)
    \Big[\nabla_\mu'J^\mu+ X \Big]
    + (\psi^\beta_\mu + h_{\mu\lambda}\beta^\rho A^\lambda_{\;\;\rho})
    \Big[\nabla_\nu'J^{\nu\mu} - h^{\mu}_{\nu} J^\nu 
    \gray{\,+\, X^\mu}
    \Big]
    = \Delta \geq 0 \;,
\end{split}
\end{align}
known as the \emph{off-shell Second Law}. The $s^\mu$ and $\Delta$ appearing in this equation should be understood as the off-shell generalizations of the respective quantities in \cref{eq:second-law-definition}.
Note that the coefficients multiplying the conservation equations here, typically referred to as Lagrange multipliers, are taken to be of a particular form in terms of the hydrodynamic variables $\hat\scB$. We will later see that this identification is consistent with the hydrostatic requirement of $\hat\scB\to\hat\scK$. Let us define the free energy current $N^\mu$ as
\begin{align}
\begin{split}
    N^\mu 
    &= s^\mu 
    - \beta^\nu n_\nu \epsilon^\mu
    + \beta^\rho h_{\rho\nu}
    \!\lb v^\mu\pi^\nu + \tau^{\mu\nu} + \tau_d^{\mu\nu} \rb
    + (\Lambda_\beta + \beta^\rho A_\rho) J^\mu \\
    &\qquad 
    + \lb\psi^\beta_\nu+\beta^\rho h_{\lambda\nu}A^\lambda_{\;\;\rho}\rb J^{\mu\nu}
    \gray{\,+\,\beta^\nu\phi_\nu X^\mu},
\end{split}
\end{align}
which allows us to recast the off-shell Second Law into the so-called \emph{adiabaticity equation}
\begin{align}
\begin{split}
   \nabla_\mu'N^\mu 
   &=
    - \epsilon^\mu \delta_{\hat\scB} n_\mu 
    + \lb v^\mu \pi^\nu + \frac{1}{2}\tau^{\mu\nu} \rb 
    \delta_{\hat\scB} h_{\mu\nu}
    + J^\mu \delta_{\hat\scB}A_\mu
    + J^{\mu\nu}h_{\nu\lambda} \delta_{\hat\scB} A^{\lambda}_{~\mu} \\
    &\hspace{20em}
    + X \delta_{\hat\scB}\phi 
    \gray{\,+\, X^\mu \delta_{\hat\scB}\phi_\mu}
    + \Delta \;,
\end{split}
\end{align}
where $\Delta \geq 0$. While deriving the adiabaticity equation from the off-shell Second Law, it is useful to note that 
\begin{align}
\begin{split}
    \nabla_\mu(\beta^\rho n_\rho) 
    &= \delta_{\hat\scB} n_\mu + F^n_{\mu\rho}\beta^\rho~, \\
    \nabla_\mu(\Lambda_\beta + \beta^\rho A_\rho) 
    &= \delta_{\hat\scB} A_\mu + F_{\mu\rho}\beta^\rho - \psi^\beta_\mu~, \\
    \nabla_\mu(\beta^\rho h_{\rho \nu}) 
    &= \delta_{\hat\scB} h_{\mu\nu} + 2n_{(\mu}h_{\nu)\lambda}\beta^\rho\nabla_\rho v^\lambda 
    - 2\beta^\rho n_{(\rho}h_{\nu)\lambda}\nabla_\mu v^\lambda
    - h_{\mu\rho}\nabla_\nu \beta^\rho~, \\
    \nabla_\mu(\psi^\beta_\nu + h_{\nu\lambda}\beta^\rho A^\lambda_{~\rho})
    &= h_{\lambda\nu}\delta_{\hat\scB} A^\lambda_{~\mu}
    + (n_\rho\beta^\rho) A^\sigma_{~\mu}h_{\nu\lambda}\nabla_\sigma v^\lambda 
    + h_{\lambda\nu}F^\lambda_{~\mu\rho}\beta^\rho + h_{\lambda\nu}A^\rho_{~\mu}\nabla_\rho\beta^\lambda~, \\
    &\qquad  
    - (\psi^\beta_\sigma + \beta^\rho h_{\lambda\sigma}A^\lambda_{~\rho})n_\nu\nabla_\mu v^\sigma - n_\mu(\psi^\beta_\rho h^{\rho\sigma}+A^\sigma_{~\rho}\beta^\rho)h_{\lambda\nu}\nabla_\sigma v^\lambda~.
\end{split}
\end{align} 

The discussion above can also be expressed in terms of manifestly dipole-invariant fields. The dipole-invariant version of the off-shell Second Law in \cref{eq:Off-shell-2nd} is given as\footnote{Note that the Lagrange multipliers in the dipole-invariant version of the off-shell Second Law \eqref{eq:Dipole-invariant-2nd} are precisely $1/T$, $\vec u_\nu/T$, $\tmu/T$, and $ \tvarpi_\mu/T$, as defined in \cref{hydro variables}, giving \cref{eq:Dipole-invariant-2nd} the interpretation of a non-equilibrium generalization of entropy maximization in equilibrium statistical mechanics.}
\begin{align}\label{eq:Dipole-invariant-2nd}
\begin{split}
    &\nabla_\mu'\tilde{s}^\mu 
    - \frac{1}{T}\Big[
    \nabla_\mu'\tilde{\epsilon}^\mu + \ldots \Big]
    + \frac{\vec u_\nu}{T} \Big[
    \nabla_\nu'\!\lb v^\mu\tilde{\pi}^\nu
    + \tilde{\tau}^{\mu\nu} 
    + \tilde{\tau}_{d}^{\mu\nu} \rb 
    + \ldots \Big] \\
    &\hspace{100pt}
    + \frac{\tmu}{T} \Big[ \nabla_\mu' J^\mu + X \Big]
    + \frac{\tvarpi_\nu}{T} \Big[ 
    \nabla_\mu'J^{\mu\nu}
    - h^\nu_\mu J^\mu 
    \gray{\,+\, X^\nu}
    \Big]
    = \Delta\geq 0~,
\end{split}
\end{align}
where $\tilde{s}^\mu$ is the dipole invariant entropy current defined as
\begin{equation}
    \tilde{s}^\mu = s^\mu 
    + \beta^\rho \Psi_\rho \lb 
    \nabla_\nu'J^{\nu\mu} - h^\mu_\nu J^\nu
    \gray{\,+\, X^\mu}
    \rb.
\end{equation}
We have utilized the definitions of dipole-invariant conserved currents from \cref{eq:dipoleinvariantcurrents}, together with the hydrodynamic variables in \cref{hydro variables}.   We can further define the dipole-invariant free energy current according to
\begin{align}
\label{eq:canonical-free-energy}
    \tilde{N}^\mu = \tilde{s}^\mu 
    - \frac{1}{T} \tilde{\epsilon}^\mu
    + \frac{\vec u_\nu}{T} \Big( v^\mu\tilde{\pi}^\nu 
    + \tilde{\tau}^{\mu\nu} + \tilde{\tau}_d^{\mu\nu} \Big)
    + \frac{\tmu}{T} J^\mu 
    + \frac{\tvarpi_\nu}{T} J^{\mu\nu}
    \gray{\,+\,\beta^\nu \tilde\phi_\nu X^\mu}
    ~,
\end{align}
which satisfies the dipole-invariant version of the adiabaticity equation
\begin{align}
\label{eq:adiabaticity-pinned}
   \nabla_\mu' \tilde N^\mu 
   &=
    - \tilde\epsilon^\mu \delta_{\scB} n_\mu 
    + \lb v^\mu \tilde\pi^\nu 
    + \frac{1}{2}\tilde\tau^{\mu\nu} \rb 
    \delta_{\scB} h_{\mu\nu}
    + J^\mu\!\lb \delta_{\scB} \tilde A_\mu + \tilde\psi^\beta_\mu \rb 
    + J^{\mu\nu}\!\lb h_{\nu\lambda}\delta_{\scB} \tilde A^{\lambda}_{~\mu}
    + \nabla_\mu \tilde \psi_\nu^\beta \rb \nn\\
    &\hspace{20em}
    + X \delta_{\scB}\phi
    \gray{\,+\, X^\mu \lb \delta_{\scB}\phi_\mu - \psi_\mu^\beta \rb}
    + \Delta \;.
\end{align}
Here $\tilde{\psi}^\beta_\mu = \psi^\beta_\mu - h^\nu_\mu\lie_\beta \phi_\nu $ and the operator $\delta_\scB$ denotes an infinitesimal diffeomorphism along $\beta^\mu$ and gauge transformation along $\Lambda_\beta$, but without the dipole transformation along $\psi_\mu^\beta$.

\paragraph*{Eliminating the dipole Goldstone and chemical potential:}

Let us return to the Josephson equation for the dipole Goldstone $\phi_\mu$ given by $X^\mu = 0$, whose form is supposed to be fixed for us by the adiabaticity equation. There is a trivial solution to the adiabaticity equation with the free energy current
\begin{equation}
    \tilde N^\mu 
    \sim -\half  m^2 h^{\rho\sigma}B_\rho B_\sigma\,\beta^\mu~, 
\end{equation}
which results in
\begin{equation}
\begin{split}
    X^\mu 
    &= - m^2 \lb \phi^\mu + h^{\mu\nu} \xi_\nu  \rb
    - T\alpha\, h^{\mu\nu}\lb \delta_\scB\phi_\nu - \psi^\beta_\nu \rb
    + \ldots
    , \\
    \Delta 
    &= T\alpha h^{\mu\nu} \lb \delta_\scB\phi_\mu - \psi^\beta_\mu \rb
    \lb \delta_\scB\phi_\nu - \psi^\beta_\nu \rb
    + \ldots,
\end{split}
\end{equation}
where $\alpha$ is some non-negative transport coefficient. There will also be some $m$-dependent terms in the conserved currents that we have not reported here. Upon imposing $X^\mu = 0$, this implies that
\begin{align}
\begin{split}
    h_\mu^\nu\lb m^2
    + T\alpha\lie_\beta \rb \phi_\nu
    = - m^2\vec\xi_\mu
    + T\alpha \psi^\beta_\mu + \ldots.
\end{split}
\end{align}
In the case of p-wave dipole superfluids, the $m^2$ terms do not appear and the dominant scale in this equation is controlled by $\alpha$, which can then be used to eliminate $\psi^\beta_\mu$, and thus the dipole chemical potential $\varpi_\mu$, from the hydrodynamic description; see~\cite{Jain:2023nbf}. For s-wave dipole superfluids, on the other hand, the dominant scale is $m^2$ and this equation is instead used for eliminating the dipole Goldstone $\phi_\mu$. This ultimately yields the single Goldstone formulation of s-wave dipole superfluids, similar to our discussion in \cref{sec:dipoleInvariant}.

The mechanism for eliminating $\psi_\mu^\beta$, and thus $\tvarpi_\mu$, is slightly different for s-wave dipole superfluids. Note that the adiabaticity equation generically has another solution of the form
\begin{align}
\begin{split}
    J^\mu 
    &= \ldots - T\sigma\, h^{\mu\nu}\lb \delta_\scB \tilde A_\nu 
    + \tilde\psi^\beta_\nu \rb
    + \ldots ~, \\
    \Delta 
    &= \ldots + T\sigma\, h^{\mu\nu}
    \lb \delta_\scB \tilde A_\mu + \tilde\psi^\beta_\mu \rb
    \lb \delta_\scB \tilde A_\nu + \tilde\psi^\beta_\nu \rb
    + \ldots~,
\end{split}
\end{align}
where $\sigma$ is the non-negative U(1) conductivity. With this in place, the dipole Ward identity in \cref{eq:physical-Wardidentity-X} generically implies that 
\begin{align}
\begin{split}
    \psi_\mu^\beta
    &= h_\mu^\nu\delta_\scB A_\nu + \frac{1}{T\sigma}(\ldots) \\
    \implies
    \tvarpi_\mu
    &= h_\mu^\nu\lb T\dow_\nu\frac{\tmu}{T} - \tilde F_{\mu\nu} u^\nu \rb 
    + h_{\nu\lambda}u^\rho \tilde A^\lambda_{~\rho}
    + \frac{1}{\sigma}(\ldots)~.
\end{split}
\end{align}
Broadly speaking, assuming the U(1) conductivity to be nonzero, the dipole chemical potential gets dynamically aligned with the gradient of U(1) chemical potential by the dipole Ward identity.

Having eliminated $\phi_\mu$ and $\tvarpi_\mu$ using $X^\mu = 0$ and the dipole Ward identity, we can write down a cleaner version of the adiabaticity equation \eqref{eq:adiabaticity-pinned}, giving rise to
\begin{align}
\label{eq:adiabaticity}
   \nabla_\mu' \tilde N^\mu 
   &=
    - \tilde E^\mu \delta_{\scB} n_\mu 
    + \lb v^\mu \tilde\pi^\nu 
    + \frac{1}{2}\tilde\tau^{\mu\nu} \rb 
    \delta_{\scB} h_{\mu\nu}
    + Q\, \delta_{\scB} \Phi
    + J^{\mu\nu} h_{\nu\lambda}\delta_{\scB} \tilde A^{\lambda}_{~\mu} 
    + {\cal X} \delta_{\scB}\phi
    + \Delta \;,
\end{align}
where ${\cal X} = X + \nabla'_\mu(Q\,v^\mu + \nabla'_\nu J^{\mu\nu})$ and the dipole-invariant versions of various fields here are defined using $\Psi_\mu = -\vec\xi_\mu$.
We have also redefined the free energy current $\tilde N^\mu \to \tilde N^\mu - \tilde\psi_\nu^\beta J^{\mu\nu} + \delta_\scB\phi\,J^\mu$, thereby removing $\psi^\beta_\mu$ from its definition, i.e.
\begin{align}
    \tilde{N}^\mu = \tilde{s}^\mu 
    - \frac{1}{T} \tilde{\epsilon}^\mu
    + \frac{\vec u_\nu}{T} \Big( v^\mu\tilde{\pi}^\nu 
    + \tilde{\tau}^{\mu\nu} + \tilde{\tau}_d^{\mu\nu} \Big)
    + \frac{\tmu}{T} J^\mu 
    + h_{\nu\lambda} \beta^\rho\tilde A^\lambda_{~\rho} J^{\mu\nu}
    ~.
\end{align}
This will be the version that we will use for the rest of this section to derive the constitutive relations for s-wave dipole superfluid hydrodynamics. 
We need to find the most general expressions for the dipole-invariant observables $\tilde E^\mu$, $\tilde\pi^\mu$, $\tilde\tau^{\mu\nu}$, $Q$, $J^{\mu\nu}$, and $X$ in terms of the hydrodynamic fields $u^\mu$, $T$, $\tmu$ and the U(1)+dipole-invariant versions of background fields $n_\mu$, $h_{\mu\nu}$, $\tilde\xi_\mu = n_\mu\Phi$, $\tilde A^\lambda_{~\mu}$, for some free energy current $\tilde N^\mu$ that satisfy the \eqref{eq:adiabaticity} for some positive semi-definite quadratic form $\Delta \geq 0$. The original adiabaticity equation \eqref{eq:adiabaticity-pinned} might still be useful if we are interested in a hydrodynamic description valid at energy scales comparable to the mass scale $m$ of $\phi_\mu$; see~\cite{Armas:2023ouk}
for the relevant discussion for ideal s-wave dipole superfluids.

Note that, in principle, one could imagine running the same argument to eliminate $\tmu$ from the hydrodynamic description using the Josephson equation of the U(1) Goldstone $\phi$. In detail, the adiabaticity equation admits a solution of the form
\begin{equation}
\begin{split}
    X
    &= - T\sigma_\phi\, \delta_\scB\phi
    + \ldots
    , \\
    \Delta 
    &= \ldots + T\sigma_\phi (\delta_\scB\phi)^2
    + \ldots,
\end{split}
\label{eq:eliminate-dBphi}
\end{equation}
for a non-negative coefficient $\sigma_\phi$. This yields the Josephson equation of the form
\begin{equation}
    u^\mu \tilde\xi_\mu
    = \tmu + \frac{1}{\sigma_\phi}(\ldots)~.
    \label{eq:U1-joseph}
\end{equation}
However, instead of eliminating $\tmu$, in hydrodynamics it is more natural to eliminate $u^\mu\tilde\xi_\mu$ from the hydrodynamic constitutive relations. This choice does, however, have an important consequence: while the naive derivative scaling of $\delta_\scB\phi$ is ${\cal O}(\dow^0)$, the scaling of $X$ together with \cref{eq:eliminate-dBphi} implies that it actually scales as ${\cal O}(\dow^{1+\varepsilon})$ on-shell.

\subsection{Hydrostatic sector}

The constitutive relations allowed by the adiabaticity equation \eqref{eq:adiabaticity} can be broadly classified into hydrostatic and non-hydrostatic sectors. The hydrostatic sector concerns the part of the constitutive relations that remains non-trivial in equilibrium after $\hat\scB$ has been aligned with the isometry $\hat\scK$. The hydrostatic constitutive relations can be characterized by the free energy density ${\cal F}$ introduced in Section \ref{sec:hydrostatics} and does not cause entropy production. On the other hand, the non-hydrostatic sector concerns part of the constitutive relations that vanish in equilibrium. They are further classified into the dissipative and non-hydrostatic non-dissipative sectors, depending on whether they contribute to entropy production or not.

Ideal fluids are part of the hydrostatic sector. It can be checked that the ideal fluid constitutive relations presented in \cref{eq:ideal-consti} satisfy the adiabaticity equation with the free energy current
\begin{align}
    \tilde N^\mu_\ideal &= -{\cal F}_\ideal \beta^\mu + 
    \chi_n F_n^{\mu\nu} \delta_\scB n_\nu 
    {\color{gray} \,-\, \frac{1}{\sqrt{\gamma}}\dow_\nu\bigg( 
    \sqrt{\gamma}\,\frac{1}{T} \chi_n F_n^{\mu\nu}  \bigg)},
\end{align}
and $\Delta_\ideal = 0$.
Note that the additional terms besides ${\cal F}_\ideal\beta^\mu$ vanish in equilibrium. The gray term has been added for convenience; it is a total derivative and trivially drops out of the adiabaticity equation. The corresponding entropy current is given by
\begin{equation}
\begin{split}
    \tilde s^\mu_\ideal 
    &= s\, u^\mu.
\end{split}
\end{equation}
Therefore entropy flow in an ideal fluid is purely in the direction of the fluid velocity.

Following our discussion in \cref{sec:hydrostatics}, more general hydrostatic the most general hydrostatic constitutive relations can be characterized by the hydrostatic free energy density ${\cal F}$, dependent on the dipole-invariant background fields $n_\mu$, $h_{\mu\nu}$, $\Phi$, $\tilde a_{\mu\nu}$ and the thermal isometry $\hat\scK$ that gets promoted to the hydrodynamic fields $\hat\scB$ out of equilibrium. To see that these constitutive relations satisfy the adiabaticity equation, we note the identity
\begin{equation}
\begin{split}
    \nabla'_\mu ({\cal F}\beta^\mu)
    &= {\cal F} v^\mu \delta_\scB n_\mu 
    + \half {\cal F} h^{\mu\nu}\delta_\scB h_{\mu\nu} 
    + \delta_\scB{\cal F} \\ 
    &= \lb \frac{\delta{\cal F}}{\delta n_\mu} 
    + {\cal F} v^\mu \rb \delta_\scB n_\mu 
    + \lb \frac{\delta{\cal F}}{\delta h_{\mu\nu}} 
    + \half {\cal F} h^{\mu\nu} \rb \delta_\scB h_{\mu\nu} 
    + \frac{\delta{\cal F}}{\delta\Phi} \delta_\scB \Phi
    + \frac{\delta{\cal F}}{\delta \tilde a_{\mu\nu}} \delta_\scB \tilde a_{\mu\nu}
    - \nabla'_\mu\Theta_{\cal F}^\mu.
\end{split}
\end{equation}
Here we have utilized the Euler-Lagrange derivatives of ${\cal F}$ and $\nabla'_\mu\Theta_{\cal F}^\mu$ denotes the leftover total derivative term. From here, it follows that the hydrostatic constitutive relations
\begin{align}
\begin{split}
    \tilde E^\mu_\hs 
    &= \frac{\delta{\cal F}}{\delta n_\mu} 
    - h^\mu_\rho \frac{\delta{\cal F}}{\delta \tilde a_{\rho\nu}}
    v^\sigma\tilde F_{\sigma\nu}
    + {\cal F} v^\mu~, \\
    \tilde\pi^{\mu}_\hs 
    &= - n_\nu\frac{\delta{\cal F}}{\delta h_{\mu\nu}}~, \\
    \tilde\tau^{\mu\nu}_\hs 
    &= - 2 h^{\mu}_\rho h^\nu_\sigma \lb
    \frac{\delta{\cal F}}{\delta h_{\rho\sigma}} 
    + 2\frac{\delta{\cal F}}{\delta \tilde a_{\lambda(\rho}}
    \tilde A^{\sigma)}_{~\lambda} \rb
    - {\cal F} h^{\mu\nu}~, \\ 
    Q_\hs 
    &= - \frac{\delta{\cal F}}{\delta \Phi}~, \\ 
    J^{\mu\nu}_\hs 
    &= - 2h^\mu_\rho h^\nu_\sigma 
    \frac{\delta{\cal F}}{\delta \tilde a_{\rho\sigma}}~,
\end{split}
\label{eq:hs-consti-hydro}
\end{align}
identically solve the adiabaticity equation \eqref{eq:adiabaticity}, together with the free energy current 
\begin{equation}
    \tilde N^\mu_\hs = - {\cal F}\beta^\mu - \Theta^\mu_{\cal F}~,
\end{equation}
and $\Delta_\hs = 0$. Therefore, we verify that the transport in the hydrostatic sector is non-dissipative. Note that the hydrostatic constitutive relations in \cref{eq:hs-consti-hydro} reduce to those derived using the hydrostatic partition function in \cref{eq:hs-F-1Goldstone}, upon aligning the hydrodynamic fields $\hat\scB$ with the thermal isometry $\hat\scK$, as promised while setting up the off-shell Second Law near \cref{eq:Off-shell-2nd}.

\subsection{Non-hydrostatic sector}
\label{sec:constitutiveNonHydrostatic}

Outside of hydrostatic equilibrium, one has the freedom to arbitrarily change the definitions of the hydrodynamic variables $T$, $u^\mu$, $\tmu$, and $\tvarpi_\mu$ by suitably redefining the fields $\hat\scB$. In general, two distinct choices of thermodynamic fields are equally valid, as long as they agree with each-other in equilibrium. Different such choices are commonly referred to as \emph{hydrodynamic frames}.\footnote{See, for instance,~\cite{landaubook,Eckart,Kovtun:2019hdm} for more discussion on hydrodynamic frames and some subtle consequences of frame fixing and truncating the derivative expansion in the context of relativistic hydrodynamics.} By requiring that $\hat\scB\to\hat\scK$ in equilibrium, i.e. that the hydrodynamic variables defined in \cref{hydro variables} agree with their hydrostatic definitions in \cref{eq:hydrostaticfluidfields-single}, we have specialized to a class of hydrodynamic frames called \emph{thermodynamic frames}.  Nonetheless, one can still redefine $\hat\scB$ using terms proportional to $\delta_{\hat\scB}(\ldots)$ that vanish in equilibrium and thus preserve the requirement $\hat\scB\to\hat\scK$ in equilibrium. Fixing this residual non-equilibrium redundancy requires us to impose certain constraints on the non-hydrostatic constitutive relations. 
In the following, we shall work in the so-called ``thermodynamic density frame''~\cite{Armas:2021vku}, defined by requiring the non-hydrostatic corrections to the energy density, momentum density, and charge density to vanish, i.e.
\begin{align}
    n_\mu \tilde E_{\text{nhs}}^\mu 
    = Q_{\text{nhs}} 
    = \tilde{\pi}_{\text{nhs}}^\mu=0.
\end{align}

In the thermodynamic density frame, the adiabaticity equation for the non-hydrostatic constitutive relations take the particularly nice form
\begin{align}
\label{eq:framefixedadiabaticity}
    \nabla_\mu'\tilde{N}^\mu_\nhs
    &= -\tilde{E}^\mu_{\text{nhs}}\delta_\scB n_\mu 
    + \frac{1}{2}\tilde{\tau}^{\mu\nu}_{\text{nhs}}\delta_\scB h_{\mu\nu} 
    + J^{\mu\nu}_{\text{nhs}}h_{\nu\lambda}
    \delta_{\scB}\tilde{A}^\lambda_{\;\;\mu}
    + {\cal X}_{\text{nhs}}\delta_{\scB}\phi 
    + \Delta_{\text{nhs}} \nn\\
    \implies 
    \nabla_\mu'\tilde{N}^\mu
    &= 
    \begin{pmatrix}
        {\cal X}_{\text{nhs}} \\
        \tilde{E}^\mu_{\text{nhs}} \\
        \tilde{\tau}^{\mu\nu}_{\text{nhs}} \\
        J^{\mu\nu}_{\text{nhs}}
    \end{pmatrix}^{\!\!\rmT}
    \begin{pmatrix}
        \delta_{\scB}\phi  \\
        - \delta_\scB n_\mu  \\
        \frac{1}{2}h_\mu^\lambda h_\nu^\tau \delta_\scB h_{\lambda\tau} \\
        h_{\lambda(\nu} h^\tau_{\mu)} 
        \delta_{\scB}\tilde{A}^\lambda_{~\tau}
    \end{pmatrix}
    + \Delta_{\text{nhs}}.
\end{align}
The non-hydrostatic constitutive relations can be schematically represented as
\begin{align}
    \begin{pmatrix}
        {\cal X}_{\text{nhs}} \\
        \tilde{E}^\mu_{\text{nhs}} \\
        \tilde{\tau}^{\mu\nu}_{\text{nhs}} \\
        J^{\mu\nu}_{\text{nhs}} 
    \end{pmatrix}
    &= -T
    \begin{pmatrix}
        {\mathfrak D}_{\phi\phi} 
        & {\mathfrak D}^\rho_{\phi\epsilon+}
        & {\mathfrak D}^{\rho\sigma}_{\phi\tau+}
        & {\mathfrak D}^{\rho\sigma}_{\phi d+} \\
        {\mathfrak D}^\mu_{\phi\epsilon-}
        & {\mathfrak D}^{\mu\rho}_{\epsilon\epsilon}
        & {\mathfrak D}^{\mu(\rho\sigma)}_{\epsilon\tau+}
        & {\mathfrak D}^{\mu(\rho\sigma)}_{\epsilon d+} \\
        {\mathfrak D}^{\mu\nu}_{\phi\tau-}
        & {\mathfrak D}^{\rho(\mu\nu)}_{\epsilon\tau-}
        & {\mathfrak D}^{(\mu\nu)(\rho\sigma)}_{\tau\tau}
        & {\mathfrak D}^{(\mu\nu)(\rho\sigma)}_{\tau d+}  \\ 
        {\mathfrak D}^{\mu\nu}_{\phi d-}
        & {\mathfrak D}^{\rho(\mu\nu)}_{\epsilon d-}
        & {\mathfrak D}^{(\rho\sigma)(\mu\nu)}_{\tau d-}
        & {\mathfrak D}^{(\mu\nu)(\rho\sigma)}_{dd} 
    \end{pmatrix}
    \begin{pmatrix}
        \delta_{\scB}\phi \\
        - \delta_\scB n_\rho  \\
        \frac{1}{2}h_\rho^\lambda h_\sigma^\tau \delta_\scB h_{\lambda\tau} \\
        h_{\lambda(\sigma} h^\tau_{\rho)} 
        \delta_{\scB}\tilde{A}^\lambda_{~\tau}
    \end{pmatrix},
    \label{eq:nhs-consti}
\end{align}
where the ${\mathfrak D}$'s are some tensor operators transverse to $n_\mu$ that remain to be specified.

Let us inspect at what derivative order each of the $\fD$ operators contribute to the non-hydrostatic constitutive relations. To this end, it is useful to invoke the dimensionful parameter $\ell$ introduced in \cref{eq:remove-ell}, and express \cref{eq:nhs-consti} as
\begin{equation}
    \begin{pmatrix}
        \ell^{\varepsilon} \ut{\cal X}_{\text{nhs}} \\
        \ut{\tilde{E}}^\mu_{\text{nhs}} \\
        {\tilde{\tau}}^{\mu\nu}_{\text{nhs}} \\
        \ut J^{\mu\nu}_{\text{nhs}} 
    \end{pmatrix}
    = -T
    \begin{pmatrix}
        \ell^{1+\varepsilon} {\mathfrak D}_{\phi\phi} 
        & \ell\, {\mathfrak D}^\rho_{\phi\epsilon+}
        & \ell^{1+\varepsilon} {\mathfrak D}^{\rho\sigma}_{\phi\tau+}
        & \ell^{2} {\mathfrak D}^{\rho\sigma}_{\phi d+} \\
        \ell\, {\mathfrak D}^\mu_{\phi\epsilon-}
        & \ell^{1-\varepsilon} {\mathfrak D}^{\mu\rho}_{\epsilon\epsilon}
        & \ell\, {\mathfrak D}^{\mu(\rho\sigma)}_{\epsilon\tau+}
        & \ell^{2-\varepsilon} {\mathfrak D}^{\mu(\rho\sigma)}_{\epsilon d+} \\
        \ell^{1+\varepsilon} {\mathfrak D}^{\mu\nu}_{\phi\tau-}
        & \ell\, {\mathfrak D}^{\rho(\mu\nu)}_{\epsilon\tau-}
        & \ell^{1+\varepsilon} {\mathfrak D}^{(\mu\nu)(\rho\sigma)}_{\tau\tau}
        & \ell^2\,{\mathfrak D}^{(\mu\nu)(\rho\sigma)}_{\tau d+}  \\ 
        \ell^{2} {\mathfrak D}^{\mu\nu}_{\phi d-}
        & \ell^{2-\varepsilon} {\mathfrak D}^{\rho(\mu\nu)}_{\epsilon d-}
        & \ell^{2}\, {\mathfrak D}^{(\rho\sigma)(\mu\nu)}_{\tau d-}
        & \ell^{3-\varepsilon} {\mathfrak D}^{(\mu\nu)(\rho\sigma)}_{dd} 
    \end{pmatrix}
    \begin{pmatrix}
        \ell^{-\varepsilon}\ut\delta_{\scB}\ut\phi \\
        - \ut\delta_\scB \ut n_\rho  \\
        \half h_\rho^\lambda h_\sigma^\tau \ut\delta_\scB h_{\lambda\tau} \\
        h_{\lambda(\sigma} h^\tau_{\rho)} 
        \ut\delta_{\scB}\tilde{\ut A}^\lambda_{~\tau}
    \end{pmatrix},
    \label{eq:nhs-consti-ell}
\end{equation}
where $\ut\delta_\scB = \ell^{-\varepsilon-1}\delta_\scB \sim \cO(\dow^0)$. In this representation, the column vectors on both sides are explicitly $\cO(\dow^0)$ and all the derivative ordering is dumped within the operator matrix.\footnote{There is an awkward factor of $\ell^\varepsilon$ hanging with ${\cal X}_{\text{nhs}}$ in \cref{eq:nhs-consti-ell}. This could be avoided if we instead worked in a hydrodynamic frame with $Q_{\text{nhs}} \neq 0$ and ${\cal X}_{\text{nhs}}=0$, though it is more straightforward in the context of the adiabaticity equation work in a frame with $\cal{X}_{\text{nhs}}$.} Note that $\delta_\scB\phi$ counts as ${\cal O}(\dow^{1+\varepsilon})$, as discussed near \cref{eq:U1-joseph}, so $\ut\delta_{\scB}\ut\phi \sim {\cal O}(\dow^\varepsilon)$. Furthermore, assuming the theory to be parity-preserving, the odd-rank ${\mathfrak D}$ operators must have at least one occurrence of $\vec u^\mu$ or $\dow_\mu$, meaning that the leading terms in these operators must scale as $\cO(\dow^\varepsilon)$. On the other hand, the even-rank ${\mathfrak D}$ operators can scale as $\cO(\dow^0)$. 
Taking note of this, the respective leading order contributions of the ${\mathfrak D}$ operators to the non-hydrostatic constitutive relations can be summarised by the leading derivative ordering of the operator matrix in \cref{eq:nhs-consti-ell}. We find
\begin{align}
    \cO\begin{pmatrix}
        \dow^{1+\varepsilon}
        & \dow^{1+\varepsilon}
        & \dow^{1+\varepsilon}
        & {\color{BrickRed}\dow^{2}} \\
        \dow^{1+\varepsilon}
        & {\color{Green}\dow^{1-\varepsilon}}
        & \dow^{1+\varepsilon} 
        & {\color{BrickRed}\dow^{2}} \\
        \dow^{1+\varepsilon}
        & \dow^{1+\varepsilon}
        & \dow^{1+\varepsilon}
        & {\color{BrickRed}\dow^2} \\ 
        {\color{BrickRed}\dow^{2}}
        & {\color{BrickRed}\dow^{2}}
        & {\color{BrickRed}\dow^2}
        & {\color{BrickRed}\dow^{2+(1-\varepsilon)}}
    \end{pmatrix}
    \sim \cO\begin{pmatrix}
        \dow^{[1]}
        & \dow^{[1]}
        & \dow^{[1]}
        & {\color{BrickRed}\dow^{[2]}} \\
        \dow^{[1]}
        & {\color{Green}\dow^{[0]}}
        & \dow^{[1]} 
        & {\color{BrickRed}\dow^{[2]}} \\
        \dow^{[1]}
        & \dow^{[1]}
        & \dow^{[1]}
        & {\color{BrickRed}\dow^{[2]}} \\ 
        {\color{BrickRed}\dow^{[2]}}
        & {\color{BrickRed}\dow^{[2]}}
        & {\color{BrickRed}\dow^{[2]}}
        & {\color{BrickRed}\dow^{[2]}}
    \end{pmatrix}.
\end{align}
Interestingly, we see that the non-hydrostatic contributions arising from $\fD_{\epsilon\epsilon}$ can appear at [0]th order in the constitutive relations. In particular, they can contribute at 0-derivative order in $z_\eff=2$ counting scheme. Similar results were obtained for p-wave dipole superfluids in~\cite{Jain:2023nbf}. We will look at these in detail in \cref{sec:nhs-0}. The contributions arising from $\fD_{\phi\phi}$, $\fD_{\phi\epsilon\pm}$, $\fD_{\phi\tau\pm}$, $\fD_{\epsilon\tau\pm}$, and $\fD_{\tau\tau}$ appear at [1]st order onward, which we discuss in \cref{sec:nhs-1}. Finally, the contributions arising from $\fD_{\phi d\pm}$, $\fD_{\epsilon d\pm}$, $\fD_{\tau d\pm}$, and $\fD_{dd}$ in the last row and column can only start contributing from [2]nd order, which we will not explore in this work. We have summarized the derivative counting of contributions from various terms in the $\fD$ operators in \cref{tab:nhs-counting}.

\begin{table}[th]
    \centering
    \begin{tabular}{c|ccc}
         & $\dow^n$ & $\dow^{n+\#\varepsilon}$
         & $\dow^{n+\#(1-\varepsilon)}$ \\
         \hline 
         
         $\fD_{\epsilon\epsilon}$ 
         & {\color{Green}$\dow^{n+(1-\varepsilon)} \sim \dow^{[n]}$}
         & $\dow^{(n+1)+(\#-1)\varepsilon} \sim \dow^{[n+1]}$
         & {\color{Green}$\dow^{n+(\#+1)(1-\varepsilon)} \sim \dow^{[n]}$}
         \\

         $\fD_{\phi\epsilon\pm}$,
         $\fD_{\epsilon\tau\pm}$ 
         & $\dow^{n+1} \sim \dow^{[n+1]}$
         & $\dow^{(n+1)+\#\varepsilon} \sim \dow^{[n+1]}$
         & $\dow^{(n+1)+\#(1-\varepsilon)} \sim \dow^{[n+1]}$ \\

         $\fD_{\phi\phi}$, 
         $\fD_{\phi\tau\pm}$,
         $\fD_{\tau\tau}$ 
         & $\dow^{(n+1)+\varepsilon} \sim \dow^{[n+1]}$
         & $\dow^{(n+1)+(\#+1)\varepsilon} \sim \dow^{[n+1]}$
         & {\color{BrickRed}$\dow^{(n+2)+(\#-1)(1-\varepsilon)} \sim \dow^{[n+2]}$} \\

         $\fD_{\phi d\pm}$, $\fD_{\tau d\pm}$
         & {\color{BrickRed}$\dow^{n+2} \sim \dow^{[n+2]}$}
         & {\color{BrickRed}$\dow^{(n+2)+\#\varepsilon} \sim \dow^{[n+2]}$}
         & {\color{BrickRed}$\dow^{(n+2)+\#(1-\varepsilon)} \sim \dow^{[n+2]}$} \\ 

         $\fD_{\epsilon d\pm}$
         & $\dow^{(n+1)+(1-\varepsilon)} \sim \dow^{[n+1]}$
         & {\color{BrickRed}$\dow^{(n+2)+(\#-1)\varepsilon} \sim \dow^{[n+2]}$}
         & $\dow^{(n+1)+(\#+1)(1-\varepsilon)} \sim \dow^{[n+1]}$ \\ 

         $\fD_{dd}$
         & {\color{BrickRed}$\dow^{(n+2)+(1-\varepsilon)} \sim \dow^{[n+2]}$}
         & {\color{Brown}$\dow^{(n+3)+(\#-1)\varepsilon} \sim \dow^{[n+3]}$}
         & {\color{BrickRed}$\dow^{(n+2)+(\#+1)(1-\varepsilon)} \sim \dow^{[n+2]}$} \\ 

         \hline

         Parity-preserving
         & $(r+n)$ even
         & $(r+n+\#)$ even
         & $(r+n)$ even
    \end{tabular}
    \caption{Derivative ordering of non-hydrostatic constitutive relations arising from various $\fD$ operators in \cref{eq:nhs-consti-ell}. Here $n$ is a non-negative integer, while $\#$ represents a positive integer. Certain terms in the table above are not allowed by parity symmetry depending on the rank $r$ of the tensors $\fD$'s, summarised in the last row.}
    \label{tab:nhs-counting}
\end{table}

By inspecting this table, we observe that the [0]th order transport in a parity-preserving theory is sensitive to only dipole superflow and [1]st order to only the fluid velocity. They can mix at [2]nd order onward. This is a consequence of the fact that in the presence of a velocity $\vec{u}^\mu \neq 0$ but no dipole superflow, SO($d$) rotation symmetry is broken to SO($d-1$). While the [1]st order terms are classified by their irreducible representations of SO($d-1$) in the standard way, a consequence of the derivative scheme we present and parity-invariance is that the [0]th order contributions to $\epsilon^\mu$ are vectors under SO($d$). 

\subsubsection{[0]-derivative order}
\label{sec:nhs-0}

From \cref{tab:nhs-counting}, we can see that the non-hydrostatic constitutive relations at [0]th order can appear via $\cO(\dow^0)$ and $\cO(\dow^{\#(1-\varepsilon)})$ terms in $\fD_{\epsilon\epsilon}$. Note that $\cO(\dow^{\#\varepsilon})$ terms in $\fD_{\epsilon\epsilon}$ combine with $\ell^{1-\varepsilon}$ in \cref{eq:nhs-consti-ell} and get pushed to [1]-derivative order. To wit, we find
\begin{align}
\begin{split}
    {\mathfrak D}^{\mu\nu}_{\epsilon\epsilon[0]}
    &= T\kappa\,h^{\mu\nu}
    + \ldots,
\end{split}
\end{align}
where the transport coefficients can be arbitrary functions of $T$ and $\tmu$, as well as of the scalars made out of $\tilde a_{\mu\nu}$ and $F^n_{\mu\nu}$ such as $\tr\,\tilde a$, $\tilde a^2$, and $F_n^2$. The ellipsis denote higher ${\cal O}(\dow^{\#(1-\varepsilon)})$ symmetric-traceless tensors constructed using various mutual contractions of $\tilde{a}_{\mu\nu}$ and $F^n_{\mu\nu}$.
The transport coefficients coupled to these tensors can be independently constrained, but they appear at ${\cal O}(\dow^{2(1-\varepsilon)})$ and higher, and thus are less important than the terms included above. 
The entropy production quadratic form is given as
\begin{align}
\begin{split}
    \frac{1}{T}\Delta_{\text{nhs}}^{[0]} 
    &= T\kappa
    \lb h^{\mu\nu}\delta_{\scB}n_\mu \delta_\scB n_\nu\rb 
    + \ldots~,
\end{split}
\end{align}
with $\tilde N^\mu_{\nhs[0]} = 0$.
Requiring the positivity of entropy production results in the constraints
\begin{align}
    \kappa \geq 0~.
\end{align}

This results in the dissipative constitutive relations for energy flux 
\begin{equation}
    \tilde E^\mu_{\nhs[0]}
    = \ell^{-\varepsilon}\,T^2\,\kappa\, h^{\mu\nu}\,\delta_\scB n_\nu~.
\end{equation}
The transport coefficient $\kappa$ can be identified as the thermal conductivity. Note that the background fluid velocity plays no role here and that $\tilde{E}^{\mu}_{\text{nhs}[0]}$ is a vector under SO$(d)$.

\subsubsection{[1]-derivative order}
\label{sec:nhs-1}

There are two kinds of contributions to the non-hydrostatic constitutive relations at [1]st order. The first kind are $\cO(\dow^{\#\varepsilon})$ terms in $\fD_{\epsilon\epsilon}$, leading to 
\begin{align}
    \fD_{\epsilon\epsilon[1]}^{\mu\nu}
    = T\,{\kappa}^{(2)}_\perp 
    (\vec u^2 h^{\mu\nu} - \vec u^\mu\vec u^\nu)
    + T\,\kappa^{(2)}_\| \vec u^\mu \vec u^\nu,
\end{align}
where the transport coefficients $\kappa^{(2)}_\perp$, $\kappa^{(2)}_\|$ can be functions of $T$, $\tmu$, and $\vec{u}^2$, while remaining at [1]st order. Note, however, that both of these coefficients only contribute to linearized analysis when fluctuating around a state with nonzero fluid velocity.
In principle, $\cO(\dow^{1})$ and $\cO(\dow^{1+\#(1-\varepsilon)})$ terms in $\fD_{\epsilon\epsilon}$ also contribute at this order, but no such terms can be constructed for a parity-preserving theory. The second kind of contributions are $\cO(\dow^{0})$ and $\cO(\dow^{\#\varepsilon})$ terms appearing in $\fD_{\phi\phi}$, $\fD_{\phi\epsilon\pm}$, $\fD_{\phi\tau\pm}$, $\fD_{\epsilon\tau\pm}$, and $\fD_{\tau\tau}$, i.e. 
\begin{align}
    {\mathfrak D}_{\phi\phi[1]} 
    &= \sigma_\phi, \nn\\
    {\mathfrak D}^\mu_{\phi\epsilon\pm[1]}
    &= \bigg[
    \lb \gamma_\epsilon^{(1)} 
    \pm \bar\gamma_\epsilon^{(1)} \rb  \vec u^\mu \bigg]_{\varepsilon}, \nn\\
    {\mathfrak D}^{\mu\nu}_{\phi\tau\pm[1]}
    &= \lb \gamma_\tau \pm \bar\gamma_\tau \rb h^{\mu\nu}
    + \bigg[
    \lb \gamma^{(2)}_{\tau} 
    \pm \bar\gamma^{(2)}_{\tau} \rb \vec u^\mu \vec u^\nu 
    \bigg]_{2\varepsilon},
    \nn\\
    {\mathfrak D}^{\mu(\rho\sigma)}_{\epsilon\tau\pm[1]}
    &= \bigg[
    2\lb \eta_\epsilon^{(1)} \pm \bar\eta^{(1)}_\epsilon\rb  h^{\mu(\rho}\vec u^{\sigma)}
    + \lb \zeta_{\epsilon}^{(1)} \pm \bar\zeta_{\epsilon}^{(1)} \rb \vec u^\mu h^{\rho\sigma} \bigg]_{\varepsilon}
    + \bigg[
    \lb \zeta_{\epsilon}^{(3)} \pm \bar\zeta_{\epsilon}^{(3)} \rb \vec u^\mu \vec u^\rho \vec u^\sigma \bigg]_{3\varepsilon}, \nn\\
    {\mathfrak D}^{(\mu\nu)(\rho\sigma)}_{\tau\tau[1]}
    &= 2\eta\, h^{\rho\langle\mu}h^{\nu\rangle\sigma} 
    + \zeta\, h^{\mu\nu}h^{\rho\sigma} \nn\\
    &\qquad 
    + \bigg[
    4\eta^{(2)}\vec{u}^{(\mu} h^{\nu)(\rho}\vec{u}^{\sigma)}
    + \lb\zeta^{(2)}+\bar\zeta^{(2)} \rb
    h^{\mu\nu}\vec{u}^\rho\vec{u}^\sigma
    + \lb\zeta^{(2)}-\bar\zeta^{(2)}\rb
    \vec{u}^\mu\vec{u}^\nu h^{\rho\sigma}
    \bigg]_{2\varepsilon} \nn\\
    &\qquad 
    + \bigg[
    \zeta^{(4)}
    \vec{u}^\mu\vec{u}^\nu\vec{u}^\rho\vec{u}^\sigma\bigg]_{4\varepsilon},
\end{align}
The subscript under the square brackets denote the leading derivative order of the contained terms.
All the transport coefficients appearing here are arbitrary functions of all $T$, $\tmu$, and $\vec u^2$, with each dependence on $\vec u^2$ resulting in an additional $\dow^{2\varepsilon}$ to the derivative counting. Any dependence on $\tilde{a}^{\mu\nu}$ gets pushed to [2]nd order or higher. The barred transport coefficients identically drop out of the entropy production and are thus classified as ``non-hydrostatic non-dissipative''.

To obtain the positivity constraints implied by the Second Law, let us first look at the entropy production quadratic form in the isotropic limit that $\vec u^\mu \to 0$. We find
\begin{align}
    \frac{1}{T}\Delta_{\text{nhs}}^{[1]}\Big|_{\vec u \to 0}
    &= \begin{pmatrix}
        \delta_\scB \phi \\ \frac{1}{2} h^{\rho\sigma}\delta_{\scB}h_{\rho\sigma}
    \end{pmatrix}^{\mathrm T}
    \begin{pmatrix}
        \sigma_\phi & \gamma_\tau \\
        \gamma_\tau & \zeta
    \end{pmatrix}
    \begin{pmatrix}
        \delta_\scB \phi \\ \frac{1}{2} h^{\rho\sigma}\delta_{\scB}h_{\rho\sigma}
    \end{pmatrix} \nn\\
    &\qquad 
    + 2\eta \,h^{\mu\langle \rho}h^{\sigma\rangle \nu}
    \left(\frac{1}{2}\delta_\scB h_{\mu\nu}\right)
    \left(\frac{1}{2}\delta_\scB h_{\rho\sigma}\right)~,
\end{align}
together with $\tilde N^\mu_{\nhs[1]} = 0$,
which leads to the constraints
\begin{align}
    \sigma_\phi \geq \gamma_\tau^2/\zeta~, \qquad 
    \zeta \geq 0~, \qquad
    \eta \geq 0~.
\end{align}
These constraints are sufficient for linearized hydrodynamics near a zero velocity state.

However, more generally for non-zero $\vec{u}^\mu$, there are two subtleties  that we must address. Firstly, note that we are able to independently constrain the non-hydrostatic transport coefficients at [0]th and [1]st orders for zero $\vec u^\mu$ because they only involve to the vector and scalar/tensor sectors respectively. This is no longer true for nonzero $\vec u^\mu$ and we must look at the combined entropy production $\Delta_{\text{nhs}}$. Secondly, nonzero $\vec u^\mu$ breaks the SO($d$) rotation symmetry to SO($d-1$), so $\Delta_{\text{nhs}}$ should be decomposed using the irreducible representations of SO($d-1$) to read out the correct positivity constraints, i.e.
\begin{align}
    \mathbb{S} \equiv 
    \begin{pmatrix}
    \vec{u}^2\delta_\scB \phi \\ 
    - \vec{u}^2\vec u^\mu\delta_\scB n_\mu \\
    \half \Delta^{\mu\nu}\delta_\scB h_{\mu\nu} \\  
    \half \vec{u}^\mu \vec{u}^{\nu} \delta_\scB h_{\mu\nu}
    \end{pmatrix}~,~~
    {\mathbb V}_\mu \equiv 
    \begin{pmatrix}
        - \vec{u}^2 \Delta_{\mu}{}^\nu \delta_\scB n_\nu \\ 
        \Delta_\mu^{~\nu} \vec u^\sigma \delta_\scB h_{\nu\sigma}
    \end{pmatrix}~,~~
    {\mathbb T}_{\mu\nu}
    = \frac{1}{2}
    \lb \Delta_{\mu}^{~~\rho} \Delta_{\nu}^{~\sigma}
    - \frac{\Delta_{\mu\nu}\Delta^{\rho\sigma}}{d-1}
    \rb \delta_\scB h_{\rho\sigma}~,
    \label{eq:SOd-1tensors}
\end{align}
where $\Delta^{\mu\nu} = \vec{u}^2h^{\mu\nu} - \vec{u}^\mu\vec{u}^\nu$. In terms of these, the entropy production is given as
\begin{align}
    \frac{\vec{u}^8}{T} \Delta_\nhs 
    &= \vec{u}^4{\mathbb S}\,{\mathfrak C}_{\mathbb S}\,{\mathbb S} 
    + {\mathbb V}_\mu\,{\mathfrak C}_{\mathbb V}\,{\mathbb V}^\mu 
    + 2\eta {\mathbb T}_{\mu\nu}{\mathbb T}^{\mu\nu},
    \label{eq:Delta-1}
\end{align}
together with $\tilde N^\mu_{\nhs[1]} = 0$, 
where we have identified the transport coefficient matrices
\begin{align}
    {\mathfrak C}_{\mathbb S}
    &= \begin{pmatrix}
        \sigma_\phi &  \gamma_\epsilon^{(1)} & \gamma_\tau 
        & \gamma_\tau
        + \vec{u}^2\gamma_\tau^{(2)} \\ 
         \gamma_\epsilon^{(1)} 
        & \frac{T\kappa}{\vec{u}^2} + T\kappa_\|^{(2)}
        &  \zeta_\epsilon^{(1)}
        &  \zeta_\epsilon^{(1)} + 2\eta_\epsilon^{(1)}
        + \vec{u}^2\zeta^{(3)}_\epsilon
        \\
        \gamma_\tau
        &  \zeta_\epsilon^{(1)}
        & \zeta + {\textstyle\frac{2}{d(d-1)}}\eta
        & \zeta - \frac{2}{d}\eta
        + \vec{u}^2\zeta^{(2)} \\
        \gamma_\tau+\vec{u}^2\gamma_\tau^{(2)}
        & \zeta_\epsilon^{(1)} + 2\eta_\epsilon^{(1)} + \vec{u}^2\zeta^{(3)}_\epsilon
        & \zeta - \frac{2}{d}\eta
        + \vec{u}^2\zeta^{(2)} 
        & \zeta + {\textstyle 2\frac{d-1}{d}}\eta
        + \vec{u}^2(2\zeta^{(2)} + 4\eta^{(2)})
        + \vec{u}^4\zeta^{(4)}
    \end{pmatrix}, \nn\\
    {\mathfrak C}_{\mathbb V}
    &= 
    \begin{pmatrix}
         T\kappa + T\kappa^{(2)}_\perp \vec u^2 
        &  \vec{u}^2\eta_\epsilon^{(1)} \\
        \vec{u}^2\eta_\epsilon^{(1)} & 
        \vec{u}^2\eta + \ell^{4\varepsilon}\vec{u}^4\eta^{(2)}
    \end{pmatrix}.
\label{eq:SOd-1matrices}
\end{align}
The positivity of entropy production implies that
\begin{equation}
    {\mathfrak C}_{\mathbb S} \geq 0~, \qquad 
    {\mathfrak C}_{\mathbb V}  \geq 0~, \qquad 
    \eta \geq 0~.
\end{equation}
where a positive semi-definite matrix means that all its eigenvalues are non-negative. These constraints are improvements on the [0]th order and [1]st order anisotropic constraints. 

While correct as written, these matrices are slightly deceptive. For a consistent gradient expansion, we must have $\vec{u}_0^2 \sim \cO(\dow^{2\varepsilon}) \ll 1$ and that the transport coefficients are $\mathcal{O}(1)$. This causes mixed-order entries in the matrices ${\mathfrak C}_{\mathbb S}$ and ${\mathfrak C}_{\mathbb V}$ to very weakly constrain terms with explicit $\vec{u}$ dependence. For instance, the Second Law requires that
\begin{align}
\kappa\geq 0, \quad T\kappa + \ell^{2\varepsilon}T\kappa^{(2)}_\perp \vec u^2 
     \geq 0.
\end{align}
Considering small $\vec{u}^2$ tells us that $\kappa^{(2)}$ is unconstrained except when $\kappa = 0$. The only case where this does not apply is for $\varepsilon = 0$ (or $z=1$), in which case we can consider explicit SO$(d-1)$ representations from the beginning. Therefore to correctly identify the Second Law constraints, we should isolate those sources which lead to mixed entries in ${\mathfrak C}_{\mathbb S}$ and ${\mathfrak C}_{\mathbb V}$. In fact, this is simple to do: these terms arise simply from the SO$(d-1)$ decomposition of $\delta_\scB n_\mu$. We learned from the [0]th order non-hydrostatic terms that this is inappropriate to do when $\kappa \neq 0$. Instead, the subleading $\vec{u}$ dependence corrects the eigenvector corresponding to $\kappa$ at [1]st derivative order,
The reason for this is that we can define a vector
\begin{align}
\begin{split}
   -\delta_\scB \mathcal{N}_\mu 
   &\equiv -\left(1 + \vec{u}^2\frac{\kappa^{(2)}_\perp}{2\kappa}\right)\delta_\scB n_\mu
    + \frac{\eta_\epsilon^{(1)}}{T\kappa}
    \half\vec{u}^\nu\delta_{\scB}h_{\mu\nu}
    + \frac{\zeta_\epsilon^{(1)}}{T\kappa}\vec{u}_\mu
    \half h^{\rho\sigma}\delta_\scB h_{\rho\sigma}\\
   &\qquad
   + \frac{\zeta_\epsilon^{(3)}}{T\kappa}\vec{u}_\mu
   \half \vec{u}^\rho\vec{u}^\sigma \delta_\scB h_{\rho\sigma}
   - \frac{\kappa_{\|}^{(2)}-\kappa_\perp^{(2)}}{2\kappa}\vec{u}_\mu
   \vec{u}^\nu\delta_\scB n_\nu
  + \frac{\gamma_\epsilon^{(1)}}{T\kappa} \vec{u}_\mu
  \delta_\scB\phi
  + \ldots~,
\end{split}
\label{eq:NewVector}
\end{align}
and a modified set of scalar sources and scalar transport matrix
\begin{align}
    \hat{\mathbb{S}}\equiv\left(\begin{array}{c}
    \vec{u}^2\delta_\scB \phi\\
    \frac{1}{2}\Delta^{\mu\nu}\delta_{\scB}h_{\mu\nu}\\
    \frac{1}{2}\vec{u}^\mu\vec{u}^\nu\delta_\scB h_{\mu\nu}
    \end{array}\right), \qquad 
    \hat{\mathfrak{C}}_{\mathbb{S}} \equiv \left(\begin{array}{ccc}
    \sigma_\phi & \gamma_\tau & \gamma_\tau\\
    \gamma_\tau & \zeta+\frac{2}{d(d-1)}\eta & \zeta-\frac{2}{d}\eta\\
    \gamma_\tau & \zeta-\frac{2}{d}\eta & \zeta+\frac{2(d-1)}{d}\eta 
    \end{array}\right)+\mathcal{O}(\dow^{2\varepsilon}).
\end{align}
Then, the positivity of entropy production can be written
\begin{align}
    \frac{\vec{u}^8}{T}\Delta_{\text{nhs}}
    = \vec{u}^8T\kappa\, h^{\mu\nu}\delta_\scB \mathcal{N}_\mu\delta_\scB \mathcal{N}_\nu 
    + \vec{u}^4 \hat{\mathbb{S}}\hat{\mathfrak{C}}_{\mathbb{S}}\hat{\mathbb{S}}
    + 2\eta\mathbb{T}_{\mu\nu}\mathbb{T}^{\mu\nu}~.
\end{align}
Hence, when $\kappa \neq 0$ and $\varepsilon\neq 0$, the complete set of positivity conditions up to [1]st order are, 
\begin{align}
    \kappa \geq 0, \quad \eta \geq 0, \quad 
    \hat{\mathfrak{C}}_{\mathbb{S}}\geq 0
\end{align}
with all other transport coefficients at [1]st order being unconstrained, particularly those appearing in \cref{eq:NewVector}. For example, consider $\delta_\scB\phi \neq 0$ but all other non-hydrostatic tensors vanishing. Then
\begin{align}
    \Delta_{\text{nhs}} 
    = \left(T\sigma_\phi + \vec{u}^2\frac{(\gamma_\epsilon^{(1)})^2}{\kappa}+\ldots\right)(\delta_\scB\phi)^2 \geq 0~,
\end{align}
leading to $\sigma_\phi\geq 0$ but no constraints on $\gamma_{\epsilon}^{(1)}$ as a consequence of $\vec{u}^2\ll 1$.

\subsubsection{[2]-derivative order}

The spectrum of non-hydrostatic constitutive relations at [2]nd order is extremely rich, as we can infer from \cref{tab:nhs-counting}. As an illustration, let us only look at the terms in $\fD$ operators that contribute to the linearized constitutive relations at [2]nd order around a state with zero $\vec u^\mu$ and $\tilde a_{\mu\nu}$. In other words, we only look at terms in $\fD$ constructed out of $T$, $\mu$, and $h_{\mu\nu}$, with any explicit derivatives appearing at the right
\begin{align}
\begin{split}
    \fD_{\epsilon\epsilon[2]}^{\mu\nu}
    &= - \sigma_\epsilon\nabla^\mu \nabla^\nu
    - \gamma_\epsilon h^{\mu\nu} \nabla'_\lambda \nabla^\lambda, \\
    \fD_{\phi\epsilon\pm[2]}^\mu 
    &= \lb \bar\sigma_{\phi\epsilon} \pm {\sigma}_{\phi\epsilon} \rb \nabla^\mu, \\
    \fD^{\mu(\rho\sigma)}_{\epsilon\tau\pm[2]}
    &= \lb \bar\sigma_{\epsilon\tau} \pm {\sigma}_{\epsilon\tau} \rb h^{\rho\sigma} \nabla^\mu
    + 2\lb \bar\gamma_{\epsilon\tau} \pm {\gamma}_{\epsilon\tau} \rb h^{\mu\langle\rho} \nabla^{\sigma\rangle}, \\
    \fD^{\mu(\rho\sigma)}_{\epsilon d\pm[2]}
    &= \lb \bar\sigma_{\epsilon d} \pm {\sigma}_{\epsilon d} \rb h^{\rho\sigma} \nabla^\mu
    + 2\lb \bar\gamma_{\epsilon d} \pm {\gamma}_{\epsilon d} \rb h^{\mu\langle\rho} \nabla^{\sigma\rangle}, \\
    \fD_{\phi d\pm[2]}^{(\mu\nu)}
    &= \lb \gamma_d \pm \bar \gamma_d \rb h^{\mu\nu}, \\
    \fD^{(\mu\nu)(\rho\sigma)}_{\tau d\pm[2]}
    &= 2\lb\eta_{\tau d} \pm \bar\eta_{\tau d} \rb
    h^{\rho\langle\mu}h^{\nu\rangle\sigma} 
    + \lb\zeta_{\tau d} \pm \bar\zeta_{\tau d} \rb
    h^{\mu\nu}h^{\rho\sigma}, \\
    \fD^{(\mu\nu)(\rho\sigma)}_{dd[2]}
    &= 2\eta_d\, h^{\rho\langle\mu}h^{\nu\rangle\sigma} 
    + \zeta_d\, h^{\mu\nu}h^{\rho\sigma}.
    \end{split}
\end{align}
In the presence of a background $\tilde{a}_{\mu\nu}$ or $\vec{u}^\mu$, many more terms can be written down, including terms that do not contain explicit derivatives. For instance, we can construct rank-3 tensors out of $\vec{u}^\mu$ in addition to $h^{\mu\nu}$ which will contribute to $\fD^{\mu(\rho\sigma)}_{\epsilon d\pm}$. All these terms have been considered at $\mathcal{O}(\partial^2)$ in our previous work on p-wave dipole superfluids~\cite{Jain:2023nbf}. 

We can borrow the calculation of the Second Law constraints at [2]nd derivative order also from~\cite{Jain:2023nbf}. In particular, we identify a vector 
\begin{align}
\begin{split}
   {\cal V}_\mu &\equiv 
   T\delta_\scB n_\mu 
   - \frac{\sigma_\epsilon}{2\kappa}\nabla_\mu\nabla^\rho\delta_{\scB}n_\rho
   - \frac{\sigma_{\epsilon\tau}-\frac{2}{d}\gamma_{\epsilon\tau}}{\kappa}\nabla_\mu\left(\frac{1}{2}h^{\rho\sigma}\delta_\scB h_{\rho\sigma}\right)
   - \frac{\sigma_{\epsilon d}-\frac{2}{d}\gamma_{\epsilon d}}{\kappa}\nabla_\mu\left(h^\rho_\lambda \delta_\scB \tilde{A}^\lambda_{\;\;\rho}\right) \\
   &\qquad 
   - \frac{\gamma_\epsilon}{2\kappa}\nabla_\rho' \nabla^\rho \delta_\scB n_\mu
   - \frac{\gamma_{\epsilon\tau}}{\kappa}\nabla^\rho \delta_\scB h_{\rho\mu}
   - \frac{\gamma_{\epsilon d}}{\kappa}
   \nabla^\rho\left(2h_{\lambda(\mu} \delta_\scB\tilde{A}^\lambda_{\;\;\rho)}\right)
   - \frac{\sigma_{\phi\epsilon}}{\kappa}
   \nabla_\mu\delta_\scB\phi~.
\end{split}
\end{align}
Since we have specialized around a state with zero $\vec u^\mu$, we may use the $\SO(d)$ decomposition of the entropy production quadratic form. We find 
\begin{align}
\begin{split}
    \frac{1}{T}\Delta_{\text{nhs}}
    &= \frac{\kappa}{T} {\cal V}_\mu {\cal V}^\mu 
    + \begin{pmatrix}
        \delta_\scB \phi \\ \ell^{\varepsilon}\frac{1}{2} h^{\rho\sigma}\delta_{\scB}h_{\rho\sigma} \\
        \ell h^\rho_\sigma \delta_\scB\tilde A^\sigma_{~\rho}
    \end{pmatrix}^{\mathrm T}
    \begin{pmatrix}
        \sigma_\phi & \gamma_\tau & \gamma_d \\
        \gamma_\tau & \zeta & \zeta_{\tau d} \\
        \gamma_d & \zeta_{\tau d} & \zeta_d
    \end{pmatrix}
    \begin{pmatrix}
        \delta_\scB \phi \\ \ell^{\varepsilon}\frac{1}{2} h^{\rho\sigma}\delta_{\scB}h_{\rho\sigma} \\
        \ell h^\rho_\sigma \delta_\scB\tilde A^\sigma_{~\rho}
    \end{pmatrix} \nn\\
    &\qquad 
    + 2 h^{\rho\langle\mu} h^{\nu\rangle\sigma}
    \begin{pmatrix}
        \ell^{\varepsilon}\half\delta_\scB h_{\mu\nu} \\ 
        \ell h_{\mu\lambda} \delta_\scB\tilde A^\lambda_{~\nu}
    \end{pmatrix}^\rmT
    \begin{pmatrix}
        \eta & \eta_{\tau d} \\ 
        \eta_{\tau d} & \eta_d
    \end{pmatrix}
    \begin{pmatrix} 
        \ell^\varepsilon\half\delta_\scB h_{\rho\sigma} \\ 
        \ell h_{\rho\lambda} \delta_\scB\tilde A^\lambda_{~\sigma}
    \end{pmatrix},
\end{split}
\end{align}
Hence $\kappa$, the $3\times 3$ coefficient matrix in the scalar sector, and the $2\times 2$ coefficient matrix in the tensor sector are required to be positive semi-definite. Note that the ``non-hydrostatic non-dissipative'' barred coefficients do not appear in the entropy production and thus are left entirely unconstrained by the Second Law of thermodynamics.

\section{Dispersion relations and response functions}
\label{sec:dispersionResponse}

In this section, we consider the linearized perturbations of our hydrodynamic model, which can be used to obtain the hydrodynamic predictions for the dispersion relations and linear response functions of an s-wave dipole superfluid.
In particular, we consider plane wave fluctuations of the fluid variables around a generic equilibrium state
\begin{align}
    \begin{split}
    T &= \,T_0 + \delta T\, e^{-i\omega t + k_ix^i}~, \\
    u^i &= u^i_0 + \delta u^i \,e^{-i\omega t+ik_i x^i}~, 
    \\
    \tmu &= \mu_0
    + \delta \tmu \, e^{-i\omega t+ik_i x^i}~, 
    \\
    \phi 
    &=  \mu_0 t
    - \frac{\ell}{2} \xi^0_{ij} x^i x^j  
    + \delta\phi\, e^{-i\omega t+ik_i x^i}~.
    \label{eq:dynamical-fluctuations}
\end{split}
\end{align}
We have set $\phi_0$ and $\phi^0_i$ appearing in \cref{eq:eqb-state} to zero, which we can always reinstate by redefining the equilibrium chemical potential $\mu_0 \to \mu_0 + \phi^0_i u^i_0$ and performing a global U(1)+dipole transformation $\Lambda = - \phi_0 + \phi^0_i x^i$, $\psi_i = - \phi^0_i$ that leaves the flat background fields invariant.

\subsection{Dispersion relations}
\label{sec:dispersion}
 
We can express the linearized equations of motion in the absence of sources in the following schematic way
\begin{align}
\label{eq:schematic-linear}
        \mathcal{M}(\omega,k)
    \begin{pmatrix}
        \delta T\\ 
        \delta u_{\|}\\ 
        \delta \tmu \\ 
        \delta \phi \\ 
        \delta u_\perp
    \end{pmatrix} 
    = 0~,
\end{align}
where ${\cal M}(\omega,k)$ is the dispersion matrix and we have denoted $\delta u_{\|} = \hat{k}_i\delta u^i$ and $\delta u_\perp^i = (\delta^{i}_j -\hat{k}^i\hat{k}_j)\delta u^j$ to be the longitudinal and transverse projections of $\delta u^i$ in the Fourier basis, where $\hat k^i = k^i/|k|$. The linearized equations only admit nontrivial solutions for hydrodynamic dispersion relations $\omega(k)$ that satisfy
\begin{align}
    \det\mathcal{M}\biggl(\omega(k),k\biggr)=0.
\end{align}

In the following, it will be helpful to identify various thermodynamic derivatives 
\begin{align}
    \begin{gathered}
    \chi_{ss} = \frac{\partial s}{\partial T}~, \qquad 
    \chi_{sq} = \frac{\partial s}{\partial \tmu} = \frac{\partial q}{\partial T}~, \qquad 
    \chi_{qq} = \frac{\partial q}{\partial \tmu}~,\\
    \chi_{s\rho} = 2\frac{\partial s}{\partial \vec u^2} = \frac{\partial \rho}{\partial T}~, \qquad
    \chi_{q\rho} = 2\frac{\partial q}{\partial \vec u^2} = \frac{\partial \rho}{\partial \tmu}~,
    \qquad
    \chi_{\rho\rho} = 2\frac{\partial \rho}{\partial \vec u^2}~.
    \end{gathered}
\end{align}
Note that, despite the notation, $\chi_{s\rho},\chi_{q\rho},\chi_{\rho\rho}$ are technically not thermodynamic susceptibilities. In particular, note that $\chi_{\rho\rho}$ is not the momentum susceptibility, but we will use this notation since it simplifies future expressions. 

\paragraph*{Zero dipole superflow:}

Let us first consider the linearized mode spectrum around an equilibrium state with vanishing fluid velocity and dipole superflow, $u^i_0 = \xi^0_{ij} = 0$. We find a pair of ``normal'' sound modes and $d-1$ copies of shear modes familiar from ordinary hydrodynamics, with the dispersion relations
\begin{subequations}
\begin{align}
\begin{split}
    \omega_{s,\pm}
    &= \pm v_s k - \frac{i}{2} \Gamma_s k^2
    + \mathcal{O}(k^3)~, \\
    \omega_D 
    &= -iD k^2+\mathcal{O}(k^4)~,
\end{split}
\end{align}
where sound speed, attenuation, and shear diffusion coefficients are 
\begin{align}
\label{eq:sound-velocity}
\begin{split}
    v_s^2
    &= \frac{Ts}{\rho} \frac{\dow(p-\mu q)}{\dow\epsilon}
    \bigg|_{q,\rho\vec u} =
    \frac{s^2}{\rho} \frac{\chi_{qq}}{\chi_{qq}\chi_{ss}-\chi_{sq}^2}~,  \\
    \Gamma_s 
    &=  \frac{\zeta+2\frac{d-1}{d}\eta}{\rho}
    + \frac{\rho\, v_s^2}{Ts^2} \kappa
    + \frac{(q+\gamma_\tau-\bar\gamma_\tau)
    (q-\gamma_\tau-\bar\gamma_\tau)}{\rho\,\sigma_\phi} ~, \\
    D &= \frac{\eta}{\rho}.
\end{split}
\end{align}
\label{eq:linear-spectrum}%
\end{subequations}
All coefficients are understood to be evaluated on the equilibrium configuration. In an ordinary dipole-non-invariant U(1) superfluid, we also find a pair of second-sound modes $\omega_{\pm} \sim\pm v_{\text{sf}}\, k$, where $v_{\text{sf}}^2 = f_s/\chi_{qq}$, arising from a term like $\half f_s \xi_i \xi^i$ in the free energy density. However, the superfluid density term is disallowed by dipole symmetry, causing the second-sound mode in an s-wave dipole superfluid to instead feature a ``magnon-like" dispersion relation with subdiffusive attenuation, i.e.
\begin{align}
\label{eq:magnon-dispersions}
\begin{split}
    \omega_{m,\pm} 
    &= \pm
    v_m k^2 
    + \mathcal{O}(k^4)~, \qquad 
    v_m^2 = \frac{B_d+2\frac{d-1}{d}G_d}{\chi_{qq}}~.
\end{split}
\end{align}
This mode spectrum agrees with the analysis of~\cite{Jain:2023nbf} at leading order in derivatives.

\paragraph*{Nonzero dipole superflow:}

We can turn on the dipole superflow $\xi^0_{ij}\neq 0$ while keeping the fluid velocity $u^i_0=0$. Assuming isotropic dipole superflow $\xi^0_{ij} = \xi_0\delta_{ij}$, all the modes remain qualitatively the same, except that the thermodynamic parameters are comprised of both the fluid and dipole contributions according to \cref{eq:total-pressure}. The spectrum is considerably more involved in the case of anisotropic superflow. In particular, we find that the shear and sound modes couple among each other non-trivially, and the resultant shear modes carry different diffusion constants in different transverse spatial directions.
On the other hand, we find that the magnon-like second-sound mode remains largely unchanged and the superflow effects only show up at $\cO(k^4)$. More details can be found in the supplementary Mathematica notebook.

An interesting aspect of nonzero superflow states in ordinary dipole-non-invariant superfluids is the Landau instability at large superflow~\cite{landauBook2, Gouteraux:2022qix, Arean:2023nnn}. It is natural to ask whether there is an analogue of this instability for dipole superflow states as well.\footnote{The original argument by Landau for the instability requires Galilean boost symmetry and the Goldstone spectrum to have $\lim_{k\to 0}\omega/k \neq 0$ and thus fails for dipole superfluids.} Let us assume isotropic superflow for simplicity. Since the dipole contributions to thermodynamic quantities as defined via \cref{eq:total-pressure} are not necessarily sign constrained, one way such instabilities might arise is when the coefficients $v_s^2$, $v_m^2$, $\Gamma_s$ or $D$ in \cref{eq:linear-spectrum,eq:magnon-dispersions} change signs at large enough values of the superflow. In particular, the total kinematic mass density at nonzero superflow is given as $\rho = \rho_f + d\,\rho_d\,\xi_0$, which becomes negative and causes the sound and shear modes to become unstable for $\xi_0 < \rho_d/(d\,\rho_d)$. Similarly, the total charge susceptibility is given as $\chi_{qq} = \chi_{qq}^f + d\,\chi^d_{qq}\,\xi_0 - d^2/2(\dow^2B_d/\dow\tilde\mu^2)\xi_0^2$, which becomes negative for sufficiently large magnitude of $\xi_0$ and causes the magnon-like second-sound mode to become unstable. More intricate stability criteria exist for anisotropic dipole superflows. It will be interesting to see whether these instabilities are realized in microscopic models with dipole symmetry. 

\paragraph*{Nonzero fluid velocity and dynamical instability:}

As we discussed around \cref{eq:freeE-velocity}, an s-wave dipole superfluid admits equilibrium states with nonzero fluid velocity, $u^i_0\neq 0$, but these states are thermodynamically unfavorable. Accordingly, we find that linearized perturbations around an equilibrium state with  $u^i_0\neq 0$ feature a dynamical instability in the magnon-like second-sound mode.
We expect that this instability is responsible for driving the system back to the $u^i_0 = 0$ state. The general form of the magnon mode at $u^i_0\neq 0$ is quite complicated, so let us look at a simple case with $q_0 = \mu_0 = 0$ and turn off all non-hydrostatic transport coefficients except $\sigma_\phi$, together with all non-hydrostatic coefficients except $B_d$, $G_d$, the diagonal susceptibilities $\chi_{ss}$, $\chi_{qq}$, and $\rho_0$. In this simple scenario, charge fluctuations and fluctuations of the Goldstone only couple to each other and satisfy
\begin{align}
\begin{split}
    \partial_t\delta q
    &= f_s\partial^2\delta\phi
    - \left(B_d+2\frac{d-1}{d}G_d\right)\partial^4\delta\phi~, \\
    \partial_t\delta\phi
    &= \frac{1}{\chi_{qq}}\delta q + \frac{1}{\sigma_\phi}\left(u_0^i\partial_i\delta q 
    + f_s\partial^2\delta\phi
    - \left[B_d+2\frac{d-1}{d}G_d\right]\partial^4\delta\phi
    \right)~,
\end{split}
\end{align}
where we have used \cref{U1conservation} for $X_{\text{nhs}}$ and $\delta q = \chi_{qq}\delta\tilde{\mu}$.
We have included the dipole-non-invariant superfluid density $f_s$ for comparison. Defining $u_\| = u^i_0 k_i/|k|$, for a dipole-non-invariant superfluid with $f_s\neq 0$, this results in the dispersion relations
\begin{equation}
    \omega_{m\pm}
    = \pm v_{\text{sf}} k
    - \frac{i}{2} k^2 \frac{f_s}{\sigma_\phi} \left(1 \mp \frac{u_\|}{v_{\text{sf}}}\right)
    + \cO(k^3)~,
\end{equation}
which is stable for $|u_\|| \leq v_{\text{sf}}$. Since dipole symmetry sets $f_s =0$ in an s-wave dipole superfluid, this bound is violated for any $u_\| \neq 0$. Indeed, for $f_s=0$, the dispersion relations yield 
\begin{equation}
    \omega_{m,\pm}
    = \pm v_m k^2
    \lb 1 + \frac{i}{2}\frac{\chi_{qq}}{\sigma_{\phi}} u_\| k \rb
    + {\cal O}(k^4)~.
\end{equation}
As a consequence, a state with nonzero fluid velocity in a dissipative s-wave dipole superfluid is destabilized by the magnon-like second-sound mode propagating along the fluid flow. Beyond the simple case discussed above, various other transport coefficients such as viscosities and thermal conductivity can also contribute to this instability. As a nice final check in this simple case, setting $u_\| = f_s = 0$ indeed leads to two stable modes with 
\begin{align}
    \omega_{m,\pm} = \pm v_m k^2
    -\frac{i}{2} \frac{\chi_{qq}}{\sigma_\phi}v_m^2  k^4
    + \mathcal{O}(k^5)~.
\end{align}

\subsection{Response functions}
\label{sec:response}

We can now turn on small variations of the background fields and study the induced variations of the hydrodynamic observables to read out the respective response functions~\cite{Kovtun:2012rj}. This method allows us to obtain the response functions of hydrodynamic densities as well as fluxes, which are not available in the traditional Kadanoff-Martin approach to response functions~\cite{1963AnPhy..24..419K}. To wit, consider the fluctuations of the background sources parametrized as
\begin{align}
\begin{gathered}
    n_t = 1 + \delta n_t\, e^{-i\omega t + ik_i x^i}, \qquad 
    n_i = \delta n_i\, e^{-i\omega t + ik_i x^i}, \\
    g_{ij} = \delta_{ij} + \delta g_{ij}\,
    e^{-i\omega t + ik_i x^i}, \qquad 
    v^{i} = \delta v^i\, e^{-i\omega t + ik_i x^i}, \\
    A_{t}
    = \delta A_t\, e^{-i\omega t + ik_i x^i}, \qquad 
    A_{i}
    = \delta A_i\,
    e^{-i\omega t + ik_i x^i}, \qquad 
    a_{ij}
    = \delta a_{ij}\, e^{-i\omega t + ik_i x^i}.
    \label{eq:background-fluctuations}
 \end{gathered}
\end{align}
The response functions of an operator $O(t,\vec x)$ can be found by varying the one-point function in the presence of a source $s(t,\vec{x})$, denoted as $\langle O(t,\vec x)\rangle_s$, in the following way,
\begin{align}
G^{R}_{OO}(t-t',\vec{x}-\vec{x}') = -i\theta(t-t')\langle [O(t,\vec{x}),O(t',\vec{x}')]\rangle 
= -\left.\frac{\delta}{\delta s(t',\vec x')}{\delta\langle O(t,\vec{x})\rangle_s}\right|_{s=0}.
\end{align}
In momentum-space, we can schematically represent the hydrodynamic equations and operators in the presence of background source fluctuations as
\begin{subequations}
\begin{align}
\label{eq:schematic-linear-source}
        \mathcal{M}(\omega,k)
    \begin{pmatrix}
        \delta T\\ 
        \ell^\varepsilon \delta u_{\|}\\ 
        \delta \mu \\ 
        \ell^{-\varepsilon} \delta \phi \\ 
        \ell^{\varepsilon}\delta u_\perp
    \end{pmatrix} 
    = \mathcal{S}(\omega, k)\,\delta s~, \qquad 
    O(\omega,k)
    = {\cal X}(\omega,k)
    \begin{pmatrix}
        \delta T\\ 
        \ell^\varepsilon \delta u_{\|}\\ 
        \delta \mu \\ 
        \ell^{-\varepsilon} \delta \phi \\ 
        \ell^{\varepsilon}\delta u_\perp
    \end{pmatrix} 
    + {\cal C}(\omega,k)\,\delta s~,
\end{align}
which, in turn, allows us to read out the response functions 
\begin{equation}
    G^R_{OO}(\omega,k) 
    = {\cal X}(\omega,k)\,{\cal M}(\omega,k)^{-1}\,{\cal S}(\omega,k)
    + {\cal C}(\omega,k)~.
\end{equation}
\label{eq:general-response}%
\end{subequations}
Since the hydrodynamic dispersion relations $\omega(k)$ are the solutions of $\det {\cal M}(\omega(k),k) = 0$, they can equivalently be defined as the poles of the response functions.

Because of the constraints imposed by the Aristotelian structure of the background geometry, we must be careful to vary only the independent components of $n_\mu$, $h_{\mu\nu}$, and $a_{\mu\nu}$. In non-covariant notation, these are
\begin{equation}
\begin{gathered}
    n_\mu =
    \begin{pmatrix}
         n_t \\ n_i
    \end{pmatrix}, \qquad 
    h_{\mu\nu} = 
    \begin{pmatrix}
       v^i v_i/(v^t)^2 & - v_j/v^t \\
       - v_i/v^t & g_{ij}
    \end{pmatrix}, \\ 
    v^\mu =
    \begin{pmatrix}
       (1 - v^i n_i)/n_t \\ v^i
    \end{pmatrix}, \qquad 
    h^{\mu\nu} =
    \begin{pmatrix}
        n^kn_k (v^t)^2 & -v^t(n^j-n^kn_k v^j) \\ 
        -v^t(n^i-n^kn_k v^i) & 
        n^k n_k v^i v^j - 2v^{(i}n^{j)}  + g^{ij}
    \end{pmatrix}, \\ 
    A_\mu =
    \begin{pmatrix}
         A_t \\ A_i
    \end{pmatrix}, \qquad 
    a_{\mu\nu} = 
    \begin{pmatrix}
        a_{ij}v^iv^j/(v^t)^2 & - a_{ij}v^i/v^t \\
       - a_{ij}v^j/v^t & a_{ij}
    \end{pmatrix}.
\end{gathered}
\label{eq:non-cov-sources}
\end{equation}
Here, $g_{ij}$ is the spatial metric, used to lower non-covariant spatial indices, and $g^{ij}$ is the inverse of the spatial metric $g_{ij}$ and is used to raise non-covariant indices. Similarly we have the independent currents
\begin{equation}
\begin{gathered}
    \epsilon^\mu = 
    \begin{pmatrix}
    \epsilon^t \\ \epsilon^i
    \end{pmatrix}, \qquad 
    \pi^\mu = 
    \begin{pmatrix}
       - \pi^i n_i/n_t \\ \pi^i
    \end{pmatrix}, \qquad 
    \tau^{\mu\nu} =
    \begin{pmatrix}
        \tau^{ij}n_in_j/n_t^2 & - \tau^{ij}n_i/n_t \\ 
        - \tau^{ij}n_j/n_t & \tau^{ij}
    \end{pmatrix} \\ 
    J^\mu = 
    \begin{pmatrix}
    J^t \\ J^i
    \end{pmatrix}, \qquad 
    J^{\mu\nu} =
    \begin{pmatrix}
        J^{ij}n_in_j/n_t^2 & - J^{ij}n_i/n_t \\ 
        - J^{ij}n_j/n_t & J^{ij}
    \end{pmatrix}.
\end{gathered}
\label{eq:non-cov-currents}
\end{equation}
Finally, though the generating functional is written more succinctly in terms of $A^\lambda_{\;\;\mu}$, this field is a combination of the U(1) gauge field $A_\mu$ and the dipole gauge field $a_{\mu\nu}$ and we should account for the conversion as given in \cref{foot:aAtrans}. 
After all of this has been accounted for, the set of independent sources and associated operators are
\begin{equation}
    \begin{split}
        \delta s
        &= \begin{pmatrix}
            {- \delta n_t} \\
            {- \delta n_i} \\
            {- \delta v^i} \\
            \half \delta g_{ij} \\
            \delta A_t \\
            \delta  A_i \\
            \half\delta a_{ij} 
        \end{pmatrix}, \qquad
        \langle {O} \rangle_s
        = \sqrt{\gamma}\,
        \begin{pmatrix}
            \epsilon^t + \pi_k v^k \\
            \epsilon^i - \ell^{-1}J^{ik} F_{tk} \\
            \pi_i - \tau_{ik}n^k \\
            \tau^{ij} - 2A^{(i}_{~k}J^{j)k} \\
            J^t \\
            J^i \\
            J^{ij}
        \end{pmatrix}~,
    \end{split}
    \label{eq:physical-operators}
\end{equation}
where we have ignored terms that are non-linear in background fields.
Note that we can always use a dipole transformation to set $\delta A_i$ to zero. By doing this, we lose direct access to the response functions involving 
the flux $J^i$, which can instead be obtained using the dipole Ward identity.

Since states with nonzero equilibrium fluid velocity are thermodynamically and dynamically unstable, we will take $u^i_0 = 0$ in the following.

\subsubsection{Zero superflow}

In this section, we present the response functions around an equilibrium state with zero dipole superflow, $\xi^0_{ij} = 0$. We start by presenting response functions in the optical ($k_i \to 0$) limit, where there is no distinction between transverse and longitudinal responses. Note that, up to contact terms, all response functions involving the conserved densities $\epsilon^t$, $J^t$, and $\pi_i$ and the U(1) flux $J^i$ vanish. For simplicity of expressions, we will define
\begin{align}
\begin{split}
    \bar{v}_u^2 
    &= \frac{q}{\rho}\frac{\partial p}{\partial q} 
    + \frac{Ts+\mu q}{\rho}\frac{\partial p}{\partial \epsilon}~,\\
    \bar{v}_d^2 
    &= \frac{q_d}{\rho}\frac{\partial p_d}{\partial q} 
    + \frac{T s_d + \mu q_d}{\rho}\frac{\partial p_d}{\partial \epsilon}~,\\
    \bar{v}_{ud}^2 
    &= \frac{q_d}{\rho}\frac{\partial p}{\partial q}
    + \frac{T s_d + \mu q_d}{\rho}\frac{\partial p}{\partial \epsilon} 
    = \frac{q}{\rho}\frac{\partial p_d}{\partial q}
    + \frac{Ts + \mu q}{\rho}\frac{\partial p_d}{\partial \epsilon}~.
\end{split}
\end{align}
The first of these matches the expression for the sound velocity in boost agnostic fluids at zero background fluid velocity; see~\cite{Armas:2020mpr}. The second is the equivalent expression when $p\to p_d$ and the final expression mixes these two. Using these, the optical response functions are given as
\begin{align}
\begin{split}
    \ell^{2\varepsilon}
    G^R_{\epsilon^i\epsilon^j}(\omega) 
    &= \lb \frac{T^2s^2}{\rho}
    - i\omega T\kappa \rb \delta^{ij}~, \\
    G^R_{\tau^{ij}\tau^{kl}}(\omega) 
    &= \left(\rho\bar{v}_u^2 -\frac{d-2}{d}p_f- i\omega \left(\zeta-\frac{\gamma_\tau^2-\bar{\gamma}_\tau^2}{\sigma_\phi}\right) \right) \delta^{ij}\delta^{kl}
    + \bigg( p_f - i\omega\eta\bigg) 2\delta^{k\langle i}\delta^{j\rangle l}~, \\
    \ell^{2\varepsilon-2}G^R_{J^{ij}J^{kl}}(\omega) 
    &= \biggl(\rho \bar{v}_d^2
    + B_d -i\omega
    \lb\zeta_d-\frac{\gamma_d^2-\bar{\gamma}_d^2}{\sigma_\phi}\rb \biggr) \delta^{ij}\delta^{kl}
    + \bigg(G_d - i\omega\eta_d\bigg) 2\delta^{k\langle i}\delta^{j\rangle l}~, \\
    \ell^{\varepsilon-1}G^{R\pm}_{\tau^{ij}J^{kl}}(\omega) 
    &= \lb \rho\bar{v}_{ud}^2-\frac{d-2}{d}p_d
    - i\omega\lb 
    \zeta_{\tau d} \pm \bar{\zeta}_{\tau d}
    - \frac{(\gamma_d\pm\bar{\gamma}_d)(\gamma_\tau\mp\bar{\gamma}_\tau)}{\sigma_\phi}
    \rb\rb \delta^{ij}\delta^{kl} \\
    &\qquad
    + \bigg( 
    p_d - i\omega\lb\eta_{\tau d}\pm\bar\eta_{\tau d}\rb
    \bigg)
    2\delta^{k\langle i}\delta^{j\rangle l}~.
    \label{eq:optical_response}
\end{split}
\end{align}
For brevity, we have used the following notation for off-diagonal response functions $G^{R+}_{O_1O_2} = G^{R}_{O_1O_2}$, $G^{R-}_{O_1O_2} = G^{R}_{O_2O_1}$.
We can use these to find the Kubo formulae
\begin{align}
\begin{split}
    \kappa
    &=
    \lim_{\omega\to 0}\lim_{k\to 0}
    \frac{-1}{T\omega} \Im G^R_{\epsilon_\perp\epsilon_\perp}~, \\
    \eta
    &=
    \lim_{\omega\to 0}\lim_{k\to 0}
    \frac{-1}{\omega} \Im G^R_{\tau_{\perp\|}\tau_{\perp\|}}~, \\
    \zeta -\frac{\gamma_\tau^2-\bar{\gamma}_\tau^2}{\sigma_\phi}+ 2\frac{d-1}{d}\eta
    &=
    \lim_{\omega\to 0}\lim_{k\to 0}
    \frac{-1}{\omega} \Im G^R_{\tau_{\|\|}\tau_{\|\|}}~, \\ 
    \eta_d
    &=
    \lim_{\omega\to 0}\lim_{k\to 0}
    \frac{-1}{\omega} \Im G^R_{J_{\perp\|}J_{\perp\|}}~,  \\
    \zeta_d - \frac{\gamma_d^2-\bar{\gamma}_d^2}{\sigma_\phi} + 2\frac{d-1}{d}\eta_d
    &=
    \lim_{\omega\to 0}\lim_{k\to 0}
    \frac{-1}{\omega} \Im G^R_{J_{\|\|}J_{\|\|}}~, \\
    \eta_{\tau d}
    &=
    \lim_{\omega\to 0}\lim_{k\to 0}
    \frac{-1}{2\omega} \Im \lb G^R_{\tau_{\perp\|}J_{\perp\|}}
    + G^R_{J_{\perp\|}\tau_{\perp\|}}
    \rb~, \\
    \zeta_{\tau d}  -\frac{\gamma_\tau\gamma_d - \bar{\gamma}_\tau\bar{\gamma}_d}{\sigma_\phi}+ 2\frac{d-1}{d}\eta_{\tau d}
    &=
    \lim_{\omega\to 0}\lim_{k\to 0}
    \frac{-1}{2\omega} \Im \lb G^R_{\tau_{\|\|}J_{\|\|}}
    + G^R_{J_{\|\|}\tau_{\|\|}}
    \rb~, \\
    \bar\eta_{\tau d}
    &=
    \lim_{\omega\to 0}\lim_{k\to 0}
    \frac{-1}{2\omega} \Im \lb G^R_{\tau_{\perp\|}J_{\perp\|}}
    - G^R_{J_{\perp\|}\tau_{\perp\|}}
    \rb~, \\
    \bar\zeta_{\tau d} -\frac{\gamma_\tau\bar{\gamma}_d-\bar{\gamma}_\tau\gamma_d}{\sigma_\phi}+ 2\frac{d-1}{d}\bar\eta_{\tau d}
    &=
    \lim_{\omega\to 0}\lim_{k\to 0}
    \frac{-1}{2\omega} \Im \lb G^R_{\tau_{\|\|}J_{\|\|}}
    - G^R_{J_{\|\|}\tau_{\|\|}}
    \rb~.
    \label{eq:Kubo_formulae}
\end{split}
\end{align}
The $\|$ and $\perp$ components of various operators are defined to be along and transverse to $k_i$ respectively.
Other transport coefficients are accessible via Kubo formulae using finite wavevector responses. We will not report those here.

We can also compute the correlation functions at nonzero wavevector, $k_i\neq 0$. The general form of these is quite complicated, so we only look at the transverse operators $\pi_\perp$, $J_\perp$, $\epsilon_\perp$. The correlation functions involving $\tau_{\|\perp}$, $J_{\|\perp}$ are related to these using Ward identities and have been presented in the supplementary Mathematica notebook. Defining
\begin{align}
    Q_\perp \equiv i\omega - \frac{\eta}{\rho}k^2~,
\end{align}
we find
\begin{align}
\label{eq:non-polarized-response}
\begin{split}
    G^R_{\pi_\perp \pi_\perp}
    &= -\rho_\perp + \frac{i\omega\rho_\perp + \mathcal{O}(k^2)}{Q_\perp}~, \\
    G^R_{J_\perp J_\perp} 
    &= k^2G_d - \frac{\omega^2 k^2 \eta_d + \mathcal{O}(k^4)}{Q_\perp},\\
    G^R_{\epsilon_\perp\epsilon_\perp}
    &= k^2\lb\mu^2G_d+\chi_n\rb
    - \frac{i\omega T^2s^2/\rho_\perp + \omega^2 T\kappa + \cO(k^2)}{Q_\perp}~, \\
     G^{R\pm}_{\pi_\perp J_\perp}
     &= q - \frac{i\omega k^2(\bar\eta_{\tau d}\pm\eta_{\tau d}) + {\cal O}(k^4)}{Q_\perp}~, \\
     G^{R\pm}_{\pi_\perp \epsilon_\perp}
     &= - p + q\mu + \frac{i\omega Ts
     + i\omega k^2(\bar\gamma_{\epsilon\tau}\mp \gamma_{\epsilon\tau}
     - \mu\bar\eta_{\tau d} \mp \mu \eta_{\tau d}) + \cO(k^4)}{Q_\perp}~, \\
     G^{R\pm}_{J_\perp \epsilon_\perp}
     &=  k^2 \mu G_d \pm i\ut\omega p_d
     - \frac{i\omega k^2 Ts(\bar\eta_{\tau d}\mp\eta_{\tau d})/\rho_\perp 
     - \omega^2 k^2
     (\gamma_{\epsilon d}\mp \bar\gamma_{\epsilon d} + \mu_0 \eta_d)
     + \mathcal{O}(k^4)}{Q_\perp}~.
\end{split}
\end{align}
In the above expressions, we have used $\rho_\perp = \rho$. This notation will become more useful in the next subsection when working with nonzero dipole superflow.

\subsubsection{Non-zero superflow}

Finally, we look at the response functions around an equilibrium state with nonzero dipole superflow, $\xi^0_{ij} \neq 0$. One hiccup we immediately encounter is that $\tilde a_{\mu\nu}$ using \cref{eq:background-fluctuations,eq:dynamical-fluctuations} becomes spatially-inhomogeneous when $\xi^0_{ij} \neq 0$, and thus the resulting linearized hydrodynamic equations do not admit plane wave solutions. Therefore, instead of using the parametrization of $\phi$ and $a_{ij}$ from \cref{eq:background-fluctuations,eq:dynamical-fluctuations}, it is helpful to perform a U(1)+dipole transformation on the equilibrium state with $\Lambda = - \half \xi^0_{ij}x^i x^j$, $\psi_i = \xi^0_{ij}x^j$, and parametrize the respective fluctuations as
\begin{equation}
    \phi 
    = \mu_0 t
    + \delta\phi\, e^{-i\omega t+ik_i x^i}, \qquad 
    a_{ij}
    = 2\xi^0_{ij}
    + \lb \delta a_{ij}
    + 2\xi_0{}_{(i}{}^k\delta g_{j)k}
    \rb e^{-i\omega t+ik_i x^i}~,
\end{equation}
The $\delta g_{ij}$ term in the parametrization of $a_{ij}$ ensures that the operator-source relations in \cref{eq:physical-operators} remain unmodified. The rest of the story follows along the lines of our general discussion of response functions given around \cref{eq:general-response}. 

The response functions for generic superflow are quite complicated, so we focus on a simpler case of a diagonal superflow $\xi^0_{ij} = \text{diag}(\xi_\|,\xi_\perp,\xi_\perp,\ldots)$. Similar to the observation made in the mode spectrum, we find that the isotropic superflow $\xi_0 = \xi_\| = \xi_\perp$ does not do anything interesting besides appearing implicitly in various thermodynamic quantities via \cref{eq:total-pressure}. More generally for states with $\Omega = \xi_\perp - \xi_\| \neq 0$, the transverse pole gets replaced by
\begin{align}
    Q_\perp \equiv i\omega 
    - \frac{\eta + 2\Omega\eta_{\tau d} + \Omega^2\eta_d}{\rho_\perp}k^2~,
\end{align}
where $\rho_\perp = \rho + 2/d\,M_d\Omega$. With these updated definitions, the diagonal transverse response functions $G^R_{\pi_\perp\pi_\perp}$, $G^R_{J_\perp J_\perp}$, and $G^R_{\epsilon_\perp\epsilon_\perp}$ remain the same as \cref{eq:non-polarized-response}. The off-diagonal transverse response functions get additional contributions 
\begin{align}
\begin{split}
    G^{R\pm}_{\pi_\perp J_\perp}
     &= (\ldots) - \frac{\pm i\omega k^2\Omega\eta_d + {\cal O}(k^4)}{Q_\perp}~, \\
     G^{R\pm}_{\pi_\perp \epsilon_\perp}
     &= (\ldots)
     + \frac{d-1}{d}G_d\Omega^2
     + \frac{i\omega k^2\Omega (
     \bar\gamma_{\epsilon d} 
     \mp\gamma_{\epsilon d}
     \mp \mu\eta_d
     ) + \cO(k^4)}{Q_\perp}~, \\
     G^{R\pm}_{J_\perp \epsilon_\perp}
     &=  (\ldots) \mp  \frac{2}{d}i\omega \Omega G_d
     - \frac{\mp i\omega k^2 Ts\Omega\eta_d + \mathcal{O}(k^4)}{Q_\perp}~.
\end{split}
\end{align}
The optical response functions reported in \cref{eq:optical_response} remain the same up to some new $\Omega$-dependent contact terms. The detailed expressions are provided in the supplementary Mathematica notebook.

\subsubsection{Discrete symmetries and Onsager's relations}

If we require the underlying microscopic description of the theory to feature certain discrete time-reversal symmetry, e.g. T, CT, PT, or CPT, the response functions satisfy the Onsager's reciprocity relations
\begin{equation}
    G^{R}_{OO}(\omega,\vec k) 
    = \Theta \cdot \Big[G^{R}_{OO}(\omega,-\vec k)\Big]^\rmT \cdot \Theta~,
    \label{eq:onsager}
\end{equation}
where $\Theta$ is a diagonal matrix denoting the eigenvalues of operators under the said discrete transformation given in table \ref{tab:CPT}; see \cite{Kovtun:2012rj} for more details. Imposing Onsager's relations for T or PT symmetries, the coefficients
\begin{equation}
    p_d, \qquad M_d, \qquad 
    \gamma_\tau, \qquad 
    \bar\gamma_d, \qquad 
    \gamma_{\epsilon\tau}, \qquad 
    \bar\gamma_{\epsilon d}~, \qquad 
    \eta_{\tau d}, \qquad 
    \zeta_{\tau d}~,
\end{equation}
must vanish. On the other hand, for CT or CPT symmetries, the coefficients
\begin{equation}
    \gamma_\tau, \qquad 
    \gamma_d, \qquad 
    \gamma_{\epsilon\tau}, \qquad 
    \gamma_{\epsilon d}~, \qquad 
    \bar\eta_{\tau d}, \qquad 
    \bar\zeta_{\tau d}~,
\end{equation}
must be odd functions of $\tmu$, while all other coefficients must be even functions of $\tmu$.

\section{Discussion}
\label{sec:discussion}

Dipole-symmetric and translation-invariant systems admit two known phases with hydrodynamic low-energy descriptions, an s-wave phase and a p-wave phase. In this work, we have developed a dissipative hydrodynamic theory of the s-wave phase, while in~\cite{Jain:2023nbf} we constructed the description of the p-wave phase. See also~\cite{Armas:2023ouk}. The s-wave phase is a dipole superfluid whose global U(1) and dipole symmetries are spontaneously broken, in addition to unbroken spacetime translation and spatial rotation symmetries. Note that dipole symmetry is not compatible with spacetime boost symmetry. The symmetry breaking pattern implies a scalar U(1) Goldstone $\phi$ and a vector dipole Goldstone $\phi_i$. However the vector Goldstone is massive and can be integrated out from the fluid description~\cite{Jain:2023nbf, Armas:2023ouk}. We carefully analyzed the derivative counting scheme appropriate for these fluids and used it to construct the hydrodynamic constitutive relations up to leading hydrostatic and dissipative derivative corrections consistent with the Second Law of thermodynamics. Our hydrodynamic framework includes a fully non-linear coupling to a curved spacetime background, together with background U(1) and dipole gauge fields~\cite{Jain:2021ibh, Bidussi:2021nmp}. In addition to studying the linearized mode spectrum of these fluids, this enabled us to also obtain the response functions of hydrodynamic operators and Kubo formulae for dissipative transport coefficients.

The most salient feature of our construction is a consistent gradient expansion scheme to organize the hydrodynamic constitutive relations, which is complicated by the fact that the fluid velocity $u^i$ and dipole superflow $\tilde a_{ij}$ have non-trivial scaling dimensions and compete to control the gradient expansion. In particular, near an equilibrium state where one of these operators is non-vanishing, we can attempt to formulate a derivative counting scheme in which $\partial_t\sim\cO(\dow^z)$, where $z=1$ for $\langle u^i\rangle =u^i_0\neq 0$ and $z=2$ for $\langle\tilde a_{ij}\rangle = 2\xi^0_{ij}\neq 0$. However, in either case, the vanishing operator is dangerously irrelevant and its fluctuations lead to derivative mixing in the spectrum and spoil the derivative counting scheme. Nonetheless, this annoyance only arises in a truncation scheme for the constitutive relations, where one assigns some terms to be more important than the others in the low-energy description. In the end, the physical results such as dispersion relations, response functions, or Kubo formulae, are agnostic of the choice of $z$, provided that we work at sufficiently high derivative orders. To accommodate this subtlety, we introduced a dimensionful parameter $\ell\sim\cO(\partial)$ and an anomalous scaling dimension $\varepsilon=z-1$. The hydrodynamic theory is consistent when $u^i \sim \ell^{\varepsilon}$ and $\xi_{ij} \sim \ell^{1-\varepsilon}$, for any scaling dimension in the range $0\leq\varepsilon\leq 1$. Equipped with these, we introduced a generalized ``overcomplete'' gradient expansion organized in terms of $\ell$ and $\partial_i$, which allows for a consistent hydrodynamic theory irrespective of the choice of $\varepsilon$.

We then utilized this novel derivative counting scheme to write down the hydrodynamic constitutive relations including leading derivative corrections. We undertook this task in two steps. First, in \cref{sec:hydrostatics}, we constructed the most general hydrostatic constitutive relations that characterize the fluid in equilibrium when coupled to arbitrary time-independent background sources. To this end, we used the machinery of hydrostatic effective actions developed in~\cite{Jensen:2012jh, Banerjee:2012iz, Bhattacharyya:2012xi, Jensen:2013kka}; see~\cite{Jensen:2014ama, Banerjee:2015uta, Armas:2020mpr} for relevant discussion concerning non-relativistic and boost-agnostic fluids. Second, in \cref{sec:dynamics}, we constructed the non-hydrostatic constitutive relations compatible with the Second Law of thermodynamics and the ensuing inequality constraints on the transport coefficients. Here we invoked the adiabaticity equation and the off-shell formalism of hydrodynamics from~\cite{Loganayagam:2011mu}, appropriately generalized to fluids with dipole symmetry as developed in~\cite{Jain:2023nbf, Armas:2023ouk}. In this section, we also formulated an all-order argument for the low-energy equivalence between the two-Goldstone and single-Goldstone formulations of s-wave dipole superfluid dynamics.

We then utilized our hydrodynamic framework to study the linearized mode spectrum of s-wave dipole superfluids in \cref{sec:dispersion}. We found a pair of linearly propagating sound modes $\omega\sim\pm k$, shear diffusion modes $\omega\sim -ik^2$ in transverse spatial directions, and a pair of magnon-like modes $\omega\sim \pm k^2$, which we explained are analogue of second sound in systems with dipole symmetry. The qualitative nature of the mode spectrum persists with or without an equilibrium dipole superflow, while the explicit parameters appearing therein are sensitive to the superflow. Naively, the theory also admits equilibrium states with nonzero fluid velocity, but the magnon-like modes in these states are plagued by an unstable sign-indefinite imaginary part at order $\sim \pm i (u_0\cdot k) k^2$. In fact, there are zero-velocity states in the theory that carry the same conserved momentum as the nonzero velocity states, but have lower canonical free energy and thus are thermodynamically preferred. Depending on the parameters of the model, the nonzero dipole superflow states might also feature instabilities are large enough values of the superflow, akin to the Landau instability in ordinary dipole-non-invariant superfluid dynamics~\cite{landauBook2} (see recent discussions in~\cite{Gouteraux:2022qix,Arean:2023nnn}). 

We emphasize that the instability at nonzero velocity exists even when our description includes the massive vector dipole Goldstone $\phi_i$. In that case, the normal component of the charge flux along $u^i$ is opposed by two superfluid components, one along $\phi_i$ and another along $\dow_i\phi$, so that the net charge flux vanishes on account of dipole symmetry. However, the equation of motion for $\phi_i$ fixes it to be along $-\partial_i\phi$ in equilibrium. Ultimately, the nonzero velocity states are subdominant to states with nonzero Goldstone profiles $\phi_i = -\rho/q\,u_i$, $\phi = \rho/q\,u_ix^i$. 

In \cref{sec:response} we used the hydrodynamic equations coupled to background sources to obtain the linear response functions of hydrodynamic operators, together with Kubo formulae for dissipative transport coefficients. Some illustrative cases of these results are presented in the main text, while more general results have been provided in a supplementary Mathematica notebook. Special attention is needed to compute response functions in states with nonzero dipole superflow. If we work in a gauge where the U(1) Goldstone attains a spatially-quadratic profile, the hydrodynamic equations coupled to background sources are non-homogeneous and thus do not admit a plane-wave decomposition, making them harder to solve. On the other hand, we find that we can equivalently work in a gauge where the background dipole gauge field is fluctuating around a nonzero value, but the hydrodynamic equations remain homogeneous. Finally, we use the off-diagonal response functions to derive the Onsager's constraints on dissipative transport coefficients for systems with a discrete time-reversal symmetry.

We presented qualitative results for simple choices of superflow, but not the most general one. While technically tedious, it would be interesting to look at these more closely for qualitatively distinct signatures that might be relevant for experimental realizations. These explorations will also pave a more comprehensive stability analysis of states with large dipole superflow. We have investigated the thermodynamic and dynamical instability of states with nonzero fluid velocity. We predicted the endpoint of these instabilities to be states with zero velocity but a spatially-linear profile for the U(1) Goldstone, based on the fact that they have the same conserved momentum density but lower canonical free energy. However, we have not explored the precise mechanism for such a transition, which is an interesting endeavor for the future. There is also a possibility that this instability drives the system to an entirely different phase, e.g. one where the translation symmetries are also spontaneously broken giving rise to a crystalline structure; see e.g.~\cite{Delacretaz:2017zxd, Armas:2019sbe, Armas:2020bmo}.

With this long-desired hydrodynamics in hand, we would like to make contact with transport in real-world systems with particles of restricted mobility. Combined with our results in~\cite{Jain:2023nbf}, we have found ``smoking-gun'' signatures of dipole symmetry in gapless phases with dipole symmetry.  Namely, the existence of magnon-like propagating modes with subdiffusive attenuation, characterized by the dispersion relations $\omega\sim \pm k^2 - ik^4$. These are the only propagating modes in the p-wave phase, while in the s-wave phase these are accompanied by the normal sound modes, $\omega \sim \pm k - ik^2$, familiar from ordinary fluids. Another distinction between the two phases of dipole superfluids are the shear modes, which in the p-wave phase are subdiffusive, $\omega\sim -ik^4$, and diffusive in the s-wave phase, $\omega\sim -ik^2$.

The most promising relevant experimental results to date come from tilted optical lattices~\cite{tilted}, where cold atoms were prepared in a spatially modulated state with wavelength $\lambda$ and subjected to a constant linear external U(1) potential. The atoms were observed to thermally relax with an approximate timescale $\tau\sim \lambda^4$ corresponding to subdiffusive behavior, with striking similarities to both s-wave and p-wave dipole superfluids. In these systems, translation symmetry is broken, so it is perhaps not surprising that only the dissipative part of the spectrum would be observed.
Other particularly interesting avenues for experimental exploration are topological defects, e.g. vortices in superfluids or dislocations/disclinations in crystals. While not strictly immobile, it is energetically costly to move these topological defects and thus their low-energy behavior is well-approximated by dipole, or higher-multipole, symmetry; see~\cite{Pretko:2017kvd, 2018PhRvL.120s5301P, Nguyen:2020yve, Gromov:2017vir, Doshi:2020jso}. Finally, from a theoretical aspect, it will also be interesting to compare this approach to another recent take on topological defects using approximate higher-form symmetries~\cite{Armas:2023tyx, Armas:2022vpf}. 

While this paper was nearing completion, the work of~\cite{Glodkowski:2024ova} appeared on arXiv with a different perspective on s-wave superfluid dynamics. The authors included new degrees of freedom corresponding to ``internal dipole moment density'' into their hydrodynamic description, however these turn out to generically be gapped and do not affect the low-energy dynamics that is dictated purely by the gapless degrees of freedom. Similar phenomenology has also been observed in the context of ``internal spin density'' degrees of freedom in hydrodynamics~\cite{Gallegos:2021bzp, Hongo:2021ona, Gallegos:2022jow}. The authors also implement a mixed derivative counting scheme wherein time-derivatives $\dow_t$ count as $\cO(\dow^2)$ when acting on the charge density, internal dipole density, and the Goldstones, and as $\cO(\dow^1)$ when acting on the energy density and momentum density. It is not immediately obvious to us how to import this counting scheme into our framework. In particular, $\dow_t$ on curved spacetime gets promoted to $\dow_t + v^i\dow_i$, with $v^i$ being the background frame velocity, and it is unclear what the derivative scaling of $v^i$ should carry. 
Nonetheless, as we emphasized earlier in the discussion, physical results are agnostic to the particular counting scheme being implemented so long as we work to sufficiently high derivative orders in the constitutive relations. We have explicitly verified this by comparing our results with that of~\cite{Glodkowski:2024ova} and found agreement where they overlap. 

\subsection*{Acknowledgments}

We are grateful to Jay Armas, Andrew Lucas,  Matthew Roberts, and Charles Stahl for helpful discussions. The work of KJ, RL, and EM~was supported in part by the NSERC Discovery Grant program of Canada. The work of AJ was funded by the European Union’s Horizon 2020 research and innovation programme under the Marie Skłodowska-Curie grant agreement NonEqbSK No. 101027527. AJ is also partly supported by the Netherlands Organization for Scientific Research (NWO) and by the Dutch Institute for Emergent Phenomena (DIEP) cluster at the University of Amsterdam. Part of this project was carried out during ``The Many Faces of Relativistic Fluid Dynamics'' and the ``Quantum Materials With and Without Quasiparticles" programs at KITP, Santa Barbara, supported by the National Science Foundation under Grant No. NSF PHY-1748958 and PHY-2309135.

\bibliographystyle{JHEP}
\bibliography{refs2}

\end{document}